\documentclass[11pt,a4paper]{article}
\usepackage[utf8]{inputenc}
\usepackage[english]{babel}
\usepackage{amsmath,bm}
\usepackage{slashed}
\usepackage{amsfonts}
\usepackage{amssymb}
\usepackage{multirow}
\usepackage{cancel}
\usepackage[numbers,sort&compress]{natbib}
\usepackage[colorlinks = true,
            linkcolor = blue,
            urlcolor  = blue,
            citecolor = blue,
            anchorcolor = blue]{hyperref}
\usepackage{makeidx}
\usepackage{authblk}
\usepackage{float}
\usepackage{mathtools}
\usepackage{tabu}
\usepackage{graphicx}
\usepackage{titlesec}
\usepackage{authblk}
\usepackage[subrefformat=parens,labelformat=parens]{subfig}
\newcommand{\bea}{\begin{eqnarray}}
\newcommand{\bec}{\begin{columns}}
\newcommand{\eec}{\end{columns}}
\newcommand{\eea}{\end{eqnarray}}
\newcommand{\beq}{\begin{equation}}
\newcommand{\eeq}{\end{equation}}
\newcommand{\ec}{\end{center}}

\newcommand{\bsym}{\boldsymbol}
\newcommand{\bc}{\begin{center}}
\usepackage[left=2cm,right=2cm,top=2cm,bottom=2cm]{geometry}

\title{{\Large {\bf 
Exploring  maverick top partner decays at the LHC
}}}

\author[a,\thanks{\href{mailto:shivam.59910103@gm.rkmvu.ac.in}
{shivam.59910103@gm.rkmvu.ac.in}}]{Shivam Verma}

\author[a,\thanks{\href{mailto:sanjoy.phy@gm.rkmvu.ac.in}{sanjoy.phy@gm.rkmvu.ac.in}}]{Sanjoy Biswas}

\affil[a]{{\it
Department of Physics, Ramakrishna Mission Vivekananda Educational and Research Institute, 
         Belur Math, Howrah 711202, India}}

\author[b,\thanks{\href{mailto:anirbanc@iitk.ac.in}{anirbanc@iitk.ac.in}}]{Anirban Chatterjee}

\affil[b]{{\it Department of Physics, Indian Institute of Technology, Kanpur 208016, India} }

\author[c,\thanks{\href{mailto:joyganguly.hep2022@gmail.com}{joyganguly.hep2022@gmail.com}}]{Joy Ganguly}

\affil[c]{{\it
  Department of Physics, Indian Institute of Technology, Hyderabad 502285, India}}

\date{}

\begin{document}

\maketitle
\thispagestyle{empty}

\begin{abstract}

In this work, we have considered an extension of the standard model (SM) with a $SU(2)_L$ singlet vectorlike quark (VLQ) with electric charge $Q=+2/3$. The model also contains an additional local $U(1)_d$ symmetry group and the corresponding gauge boson is the dark photon. The VLQ is charged while all the SM particles are neutral under the new $U(1)_d$ gauge group. Even though in this model the VLQ possesses many properties qualitatively similar to that of the traditional top partner ($T_p$), there are some compelling differences as well. In particular, its branching ratio to the traditional modes ($T_p \to bW, tZ, th$) are suppressed which in turn helps to evade many of the existing bounds, mainly coming from the LHC experiments. In an earlier work, such a VLQ is referred to as ``maverick top partner". It has been shown that the top partner in this model predominantly decays to a top quark and a dark photon/dark Higgs pair ($T_p \to t\gamma_d ,~th_d$) over a large region of the parameter space. The dark photon can be made invisible and consequently, it gives rise to the missing transverse energy ($\cancel{E} _{T}$) signature at the LHC detector. We have mainly focused on the LHC signatures and future prospects of such top partners. In particular, we have studied the $t\bar{t}+\cancel{E}_{T}$ and $t+\cancel{E}_{T}$ signatures in the context of the LHC via pair and single productions of the top partner, respectively at 13 and 14 TeV LHC center of mass energies assuming that the dark photon either decays into an invisible mode or it is invisible at the length scale of the detector. We have shown that one can exclude $\sin\theta_L \sim 0.025$ (0.05) for $m_{_{T_p}} \leq $ 2.0 (2.6) TeV at $\sqrt{s}=14$ TeV with an integrated luminosity of 3 ab$^{-1}$ using the single top partner production channel.
 
\end{abstract}

\clearpage

\pagenumbering{arabic}

\pagestyle{plain}
\section{I{\small NTRODUCTION}}

Search for new resonances or particle-like states predicted by the physics beyond the Standard Model (BSM) is an important goal for the Large Hadron Collider (LHC). 
One of the foremost among them is search for vectorlike fermions (VLFs), in particular, vectorlike quarks (VLQs). VLQs are omnipresent in many extensions 
of the SM such as SM with extra dimension \cite{Allanchach:2019wrx,Randall:1999ee, Carena:2007ua}, composite Higgs model \cite{Chivukula:2000mb, Kaplan:1983sm, Agashe:2004rs}, primarily motivated to solve the Higgs mass problem in the SM 
\cite{Arkani-Hamed:2002iiv, Chacko:2005pe,Arkani-Hamed:2002ikv}. A lot of studies has been done on the phenomenological 
aspects of VLQs in a general setup \cite{Aguilar-Saavedra:2009xmz, Berger:2012ec, Buchkremer:2013bha, Kim:2018mks, Alhazmi:2018whk, Anandakrishnan:2015yfa, Dolan:2016eki, Chala:2018qdf, Gopalakrishna:2013hua, Dermisek:2019vkc, Han:2018hcu, Colucci:2018vxz, Dobrescu:2016pda, Gopalakrishna:2015wwa, Dasgupta:2021fzw, Corcella:2021mdl, Dermisek:2021zjd, duPlessis:2021xuc}.
However, the existence of VLQ (or VLF in general) can also be motivated in a class of theories where the SM is augmented with a 
dark sector (DS) and the VLFs act as mediators between the visible and the DS (often referred as portal matter)\cite{PhysRevD.99.115024, Rizzo:2022qan, Qin:2021cxl, Wojcik:2022rtk}. The DS in its
most simple form can contain a dark photon \cite{Fabbrichesi:2020wbt} corresponding to a dark $U(1)_d$ gauge symmetry. 
In many extensions of the SM motivated by the solution to the dark matter (DM) problem, dark photon is introduced  in order to explain 
the small scale structure of the universe \cite{Foot:2016wvj}.

The theoretical framework considered in this study closely follows that of Ref.~\cite{Kim:2019oyh}. We have considered a $SU(2)_L$ singlet vectorlike quark 
carrying $+{2}/{3}$ unit of electric charge. The VLQ is also charged under an additional  $U(1)_d$ dark gauge symmetry whereas all the SM fields are
neutral under this new symmetry. The $U(1)_d$ symmetry is spontaneously broken by the vacuum expectation value (VEV) of a complex scalar field (dark Higgs)
and the gauge boson (dark photon) corresponding to the local $U(1)_d$  symmetry becomes massive. 
The dark Higgs field is not only responsible for giving mass to the dark photon in this framework but it also plays an important role, namely, it 
induces a mixing between the top-quark and the VLQ. We will only consider the VLQ mixing with the 3rd generation quark with same charge. 
However, as we will see later on, the vectorlike quark being charged under the additional $U(1)_d$ 
gauge interaction, its branching ratio into the traditional decay modes ($T_p \to bW, tZ, th$)  are all suppressed. 
In this framework, it predominantly decays to a top-quark and a dark photon/dark Higgs ($T_p \to t\gamma_d,~ t h_d$). 
Such a VLQ having same electric charge and color quantum number as that of SM top-quark with traditional decay modes suppressed is referred to 
as ``maverick top partner" in \cite{Kim:2019oyh}. Similar extensions with VLQs carrying $-{1}/{3}$ unit of electric charge was earlier proposed 
in \cite{PhysRevD.99.115024, Rizzo:2022qan} which also predicts an enhanced branching ratio of the {bottom} partner to the nonstandard modes.  
The origin of this enhancement in branching ratio to the nonstandard modes can be traced back to the hierarchies between
{\it (i) the top and the maverick top partner masses, (ii) the vacuum expectation values of the two Higgs fields
and (iii) the fermion mass ratio and the fermion mixing angle}.
The existence of such a top partner, not only gives rise to rich collider phenomenology but also helps to evade the constraints coming from the LHC
data depending on their transformation properties under the SM gauge group. The collider constraints come from pair production 
of top partner or single production of top partner in association with a forward jet at the LHC. The current LHC data is mostly sensitive to VLQ searches in 
the traditional decay modes ($T_{p} \to bW, tZ, t h$) \cite{ATLAS:2018dyh,ATLAS:2018uky,CMS:2016lel,CMS:2019too}. 
Nonstandard or exotic decays of the vectorlike quarks in different set-ups with different collider signatures have been considered in the literature \cite{Bizot:2018tds, Xie:2019gya, Cacciapaglia:2019zmj, Chala:2017xgc, Aguilar-Saavedra:2017giu, Benbrik:2019zdp,Bhardwaj:2022nko,Das:2018gcr,Banerjee:2016wls,Banerjee:2022izw,Banerjee:2022xmu}.

In the present work, we will assume that the dark photon always decays to a pair of DM particles. Consequently, 
in a collider experiment the signature of the dark photon
production will be {\it missing momentum}. Our motivation is to constrain the decay of the maverick top partner into a {\it top-quark and dark photon pair} 
at the LHC in the $t\bar{t}$+{\it missing transverse energy} ($\cancel{E}_{T}$) and $t+\cancel{E}_{T}$ channels and for that it is sufficient to treat the dark photon and its subsequent decay products 
to be invisible at the length scale of the detector. The details of the DM sector and its phenomenology is not considered here. We emphasize that the allowed region of the parameter space that is consistent with the perturbative unitarity and electroweak precision (EWP) measurements \cite{Ciuchini:2013pca, deBlas:2016ojx, Chen:2017hak} can be excluded at the {high-luminosity (HL) LHC}. We have also set the exclusion limit on the  relevant parameter space using the latest LHC data in the $t\bar{t}+\cancel{E}_{T}$ and $t+\cancel{E}_{T}$ channels \cite{CMS:2021beq, ATLAS:2022vff, Kraml:2016eti}.

The rest of the paper is organized as follows: In Sec. \ref{sec:model}, we briefly outline the theoretical framework considered for
the present analysis including various constraints on the model parameters. The event selection for both the single and pair production 
of maverick top partner(s) is detailed in Sec. \ref{Tp@LHC}. In Sec. \ref{results}, we present the results of our numerical simulation and discuss their 
significances. Finally, we conclude in Sec. \ref{conclusions}.

\section{Theoretical framework}\label{sec:model}

In this section, we briefly summarize the theoretical framework  proposed in \cite{Kim:2019oyh}. The model contains a $SU(2)_L$ singlet vectorlike quark
having electric charge $+{2}/{3}$ unit. The top partner is also charged under an additional  $U(1)_d$ dark gauge symmetry. The gauge boson  corresponding
to the local $U(1)_d$  symmetry is the dark photon  (in the limit of very tiny kinetic mixing). The model also contains a complex  scalar field $\Phi_d$ charged under 
the $U(1)_d$ and singlet under the SM gauge group. The dark photon becomes massive as $\Phi_d$ develops a nonzero vacuum expectation value, $\langle \Phi_d \rangle= {v_d}/{\sqrt{2}}$.
The corresponding Higgs mode in the dark sector after mixing with the SM Higgs gives a dark Higgs in the mass eigenbasis. All the SM fields are neutral under the $U(1)_d$.
The full Lagrangian of the model is detailed below:

The relevant fields content and their representations under $SU(3)_C \times SU(2)_L\times U(1)_Y\times U(1)_d$ symmetry groups are summarized in Table.~\ref{table:charges}.

\begin{table}[H]
\begin{center}
\begin{tabular}{lcccc}
\hline
Fields & $SU(3)_C$ & $SU(2)_L$ & $Y$ & $Y_d$\\
			\hline \hline
			$t^{'}_{R}$ & 3 & 1 & 2/3 & 0 \\
			
			$b_R$ & 3 & 1 & -1/3 &  0 \\
			
			$Q_L=\begin{pmatrix}
			t^{'}_{L}\\
			b_L
			\end{pmatrix}$ & 3 & 2 & 1/6 & 0 \\
			
			$\Phi$ & 1 & 2 & 1/2 & 0 \\
			
			$T^{'}_{L}$ & 3 & 1 & 2/3 & 1 \\
			
			$T^{'}_{R}$ & 3 & 1 & 2/3 & 1 \\
			
			$\Phi_d$ & 1 & 1 & 0 & 1 \\
			\hline\hline
			
\end{tabular}
\caption{{ \scriptsize Representations of the relevant field content under the full symmetry group of the theory. The remaining SM fields are all neutral under the new $U(1)_d$.}}
\label{table:charges}
\end{center}
\end{table} 

The relevant sectors of the Lagrangian invariant under $SU(3)_C \times SU(2)_L\times U(1)_Y\times U(1)_d$ symmetry group are given below:

\begin{eqnarray}\label{eq:L_gauge}
{\mathcal{L}}_{\rm Gauge}=-\frac{1}{4}G_{\mu\nu}^a G^{a,\mu\nu}-\frac{1}{4}W_{\mu\nu}^i W^{i,\mu\nu}-\frac{1}{4}B_{\mu\nu}^{\prime} B^{\prime,\mu\nu}+\frac{\varepsilon^{\prime}}{2\cos\theta_W}B_{d,\mu\nu}^{\prime}B^{\prime\mu\nu}-\frac{1}{4}B_{d,\mu\nu}^{\prime} B_d^{\prime,\mu\nu},
\end{eqnarray}
where $G_{\mu\nu}^a$ are the $SU(3)_C$ field strength tensor with $a=1,\cdots,8$, $W_{\mu\nu}^i$ are $SU(2)_L$ field strength tensor with $i=1,2,3$,  $B_{\mu\nu}^{\prime}$ is that of the $U(1)_Y$  and $B_{d,\mu\nu}^{\prime}$ corresponds the field strength tensor of the additional $U(1)_d$. The kinetic mixing term is parametrized by $\varepsilon^{\prime}$.

\begin{eqnarray}
{\mathcal{L}}_{\rm Scalar}=|D_{\mu}\Phi|^2+|D_{\mu}\Phi_d|^2 - V(\Phi,\Phi_d)
\end{eqnarray}
where $\Phi$ and $\Phi_d$ are the SM and dark sector Higgs fields, respectively and the corresponding scalar potential is
given by 

\begin{eqnarray}\label{eq:scalar potential}
V(\Phi,\Phi_d)=-\mu^2|\Phi|^2+\lambda |\Phi|^4-\mu_{h_d}^2 |\Phi_d|^2+\lambda_{h_d}|\Phi_d|^4+\lambda_{h h_d}|\Phi|^2||\Phi_d|^2.
\end{eqnarray}

The gauge covariant derivative has the form 
\begin{eqnarray}
D_{\mu}=\partial_{\mu}-ig_{_S} t^a G_{\mu}^a
-i g T^i W_{\mu}^i-i g^{\prime}Y B_{\mu}^{\prime}-i g_d^{\prime}Y_d B_{d,\mu}^{\prime}.
\end{eqnarray}
Here, $g_{_{S}}$, $g$, $g^{\prime}$, $g_d^{\prime}$ are the $SU(3)_C$, $SU(2)_L$, $U(1)_Y$ and $U(1)_d$ couplings constants, respectively and $t^{a}$'s and $T^{i}$'s are the generators of the $SU(3)_C$ and $SU(2)_L$ groups, respectively.

 When the $SU(3)_C \times SU(2)_L\times U(1)_Y\times U(1)_d$ symmetry is spontaneously broken to $SU(3)_C \times U(1)_{\rm em}$
 by the choice of minimum value configurations of $\Phi$ and $\Phi_d$ which in unitary gauge can be written as 
 \begin{eqnarray}
 \Phi=\begin{pmatrix}
 0 \\
 \frac{v_{_{\rm EW}}+h'}{\sqrt{2}}
 \end{pmatrix},\quad \Phi_d=\frac{1}{\sqrt{2}}(v_d+h^{\prime}_d).
 \end{eqnarray}
  one ends up with two massive scalar modes (and massive gauge bosons of the corresponding broken gauge groups). Here, $v_{\rm EW}$ 
  has value 246 GeV.
   
 To find the mass basis we rotate $h^{\prime}$ and $h^{\prime}_d$ as
 
 \begin{eqnarray}
 \begin{pmatrix}
 h \\
 h_d
 \end{pmatrix}
 =\begin{pmatrix}
 \cos\theta_{S} & -\sin\theta_{S} \\
 \sin\theta_{S} & \cos\theta_{S}
 \end{pmatrix} \begin{pmatrix}
 h^{\prime} \\
 h^{\prime}_d
 \end{pmatrix}
 \end{eqnarray}
 
 The $h$ can be identified as the observed Higgs boson  with mass $m_h=125$ GeV while $h_d$ can be taken as BSM Higgs with mass $m_{h_d}$. Hence, the free parameters are $\theta_S$, $m_{h_d}$ and $v_d$ as all the other parameters in the potential of Eq.~(\ref{eq:scalar potential}) can be expressed in terms of the above-mentioned free parameters.
 

The full lagrangian for the fermion sector involving 3rd generation quarks and VLQ is given by
\bea\label{eq:Fermion lagrangian}
\mathcal{L}_{\rm Fermion}=\bar{Q}_Li\slashed{D}Q_L+\bar{t^{\prime}}_{R}i\slashed{D}t^{\prime}_{R}+\bar{b}_Ri\slashed{D}b_R+ {\overline{T}^{\prime}i\slashed{D}T^{\prime}}+\mathcal{L}_{\rm Yuk}
\eea 
 
 where, $\mathcal{L}_{\rm Yuk}$ is given by
 \begin{eqnarray}
 \mathcal{L}_{\rm Yuk}=-y_{_b} \bar{Q}_L\Phi b_R-y_{_t}\bar{Q}_L \tilde{\Phi}t^{\prime}_{R}-\lambda_T \Phi_d \overline{T}^{\prime}_{L}t^{\prime}_{R}-m_{_{T}}\overline{T}^{\prime}_{L}T^{\prime}_{R}+{\rm H.c.}.
 \end{eqnarray}
 
when $\Phi$ and $\Phi_d$ get VEVs, mass matrix in $t^{\prime}$ and $T^{\prime}$ basis can be written in a form as
 \begin{eqnarray}
 \mathcal{L}_{u3-{\rm mass}}=-\overline{\chi}_L \mathcal{M}\chi_R + {\rm H.c.},
  \end{eqnarray}
  where,
  
  \bea
  \chi_{\tau}=\begin{pmatrix}
  t^{\prime}_{\tau} \\
  T^{\prime}_{\tau}
  \end{pmatrix},\quad \mathcal{M}=\begin{pmatrix}
  \frac{y_t\, v_{_{\rm EW}}}{\sqrt{2}} & 0 \\
  \frac{\lambda_T \, v_d}{\sqrt{2}} & m_{_{T}}
  \end{pmatrix},
  \eea

  Here $\tau=L,R$. 
  
  One requires the biunitary transformation of the following kind in order to diagonalize the above mass matrix
  \begin{eqnarray}
  \begin{pmatrix}
  t_L \\
  T_{p_{_L}}
  \end{pmatrix}=\begin{pmatrix}
  \cos\theta_L & -\sin\theta_L \\
  \sin\theta_L & \cos\theta_L
  \end{pmatrix}
  \begin{pmatrix}
  t^{\prime}_{L} \\
  T^{\prime}_{L}
  \end{pmatrix},\quad
  \begin{pmatrix}
  t_R \\
  T_{p_{_R}}
  \end{pmatrix}=\begin{pmatrix}
  \cos\theta_R & -\sin\theta_R \\
  \sin\theta_R & \cos\theta_R
  \end{pmatrix}
  \begin{pmatrix}
  t^{\prime}_{R} \\
  T^{\prime}_{R}
  \end{pmatrix}
  \end{eqnarray}
  
 Here $t$ can be identified as the SM top quark having mass {$m_t=173.2$} GeV. The fermion sector contains two free parameters: the mass of top partner, $m_{_{T_p}}$ and the mixing angle involving the left-handed fields, $\theta_L$. We take $\sin \theta_{L}$, $m_t$ and $m_{_{T_p}}$ as independent parameters for the rest of the discussion.

 To verify the allowed ranges of $\sin \theta_{L}$ consistent with perturbative unitarity we trade Eq.~(\ref{eq:lam_T}) to write $\sin\theta_L$ as 
  
 \bea
\left| \sin \theta_{L}\right| =\frac{1}{2}\sqrt{\frac{2 m_{_{T_p}}^{2}-2m_{t}^{2}-\lambda_{T}^{2}v_{d}^{2}}{m_{_{T_p}}^{2}-m_{t}^{2}}\left(1-\sqrt{1-\frac{8\lambda_{T}^{2}v_{d}^{2}m_{t}^{2}}{\left(2 m_{_{T_p}}^{2}-2m_{t}^{2}-\lambda_{T}^{2}v_{d}^{2}\right)^{2}}}\right)}
\eea

  To have real solution for $\sin\theta_{L}$ one needs $|\lambda_T| < \sqrt{2}(m_{_{T_{p}}}- m_t)/v_d$. On the other hand the perturbative unitarity bound on $\lambda_{T}$ requires $\lambda_{T} < 4\sqrt{2\pi}$ \footnote{The perturbative unitarity bound has been estimated by studying the $h_d t\to h_d t$ scattering process in \cite{Kim:2019oyh}}. We can combine these two conditions and write
  
  \bea
  |\lambda_T| < \sqrt{2}\min{\left(\frac{m_{_{T_{p}}}- m_t}{v_d}, 4\sqrt{\pi}\right)}\label{eq:bound_lamT}
  \eea
  
\begin{figure}[H]
\centering
\subfloat[\label{fig:lamt_mtp_st_vd_1_gev}]{\includegraphics[width=0.5\columnwidth]{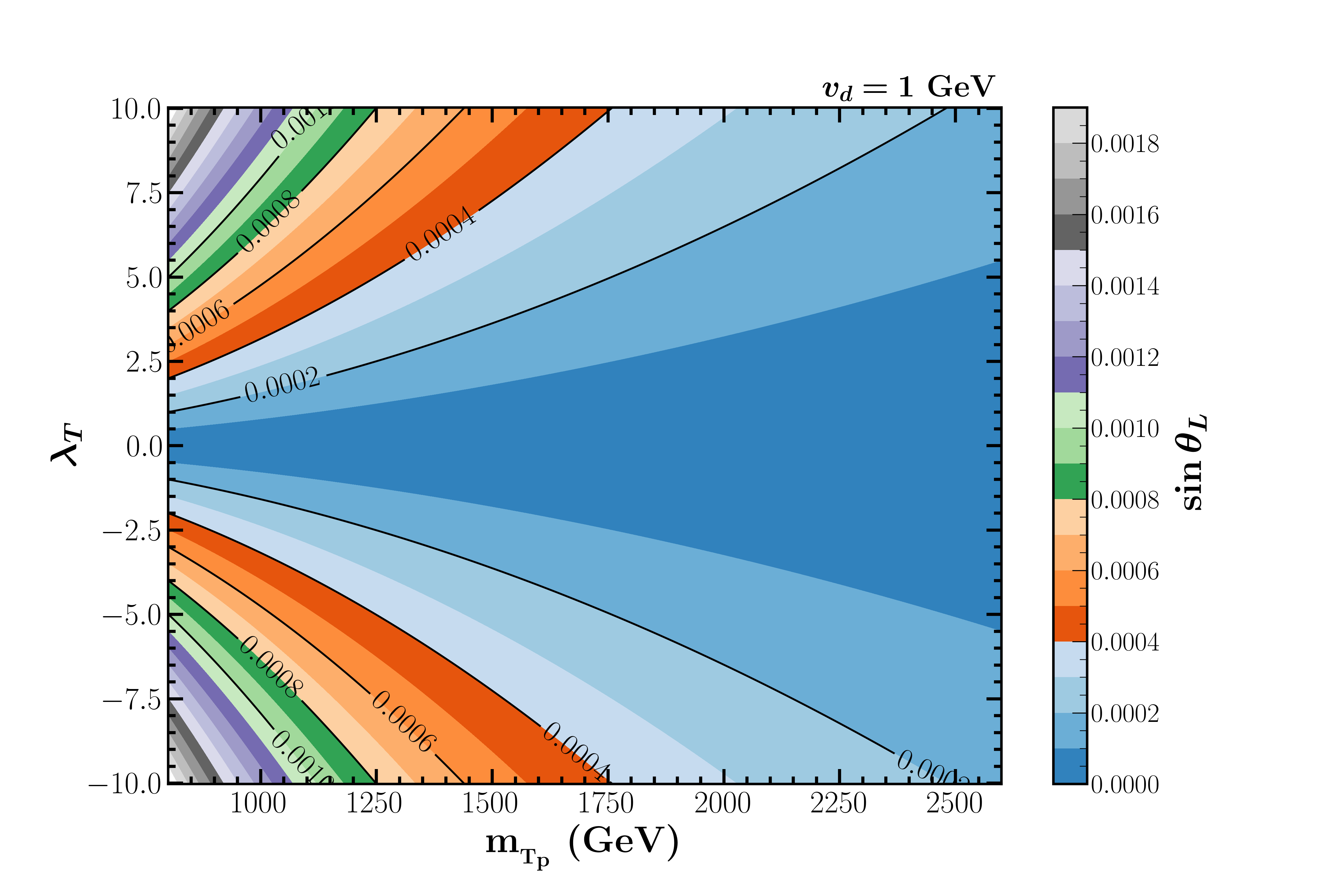}}
\subfloat[\label{fig:lamt_mtp_st_vd_10_gev}]{\includegraphics[width=0.5\columnwidth]{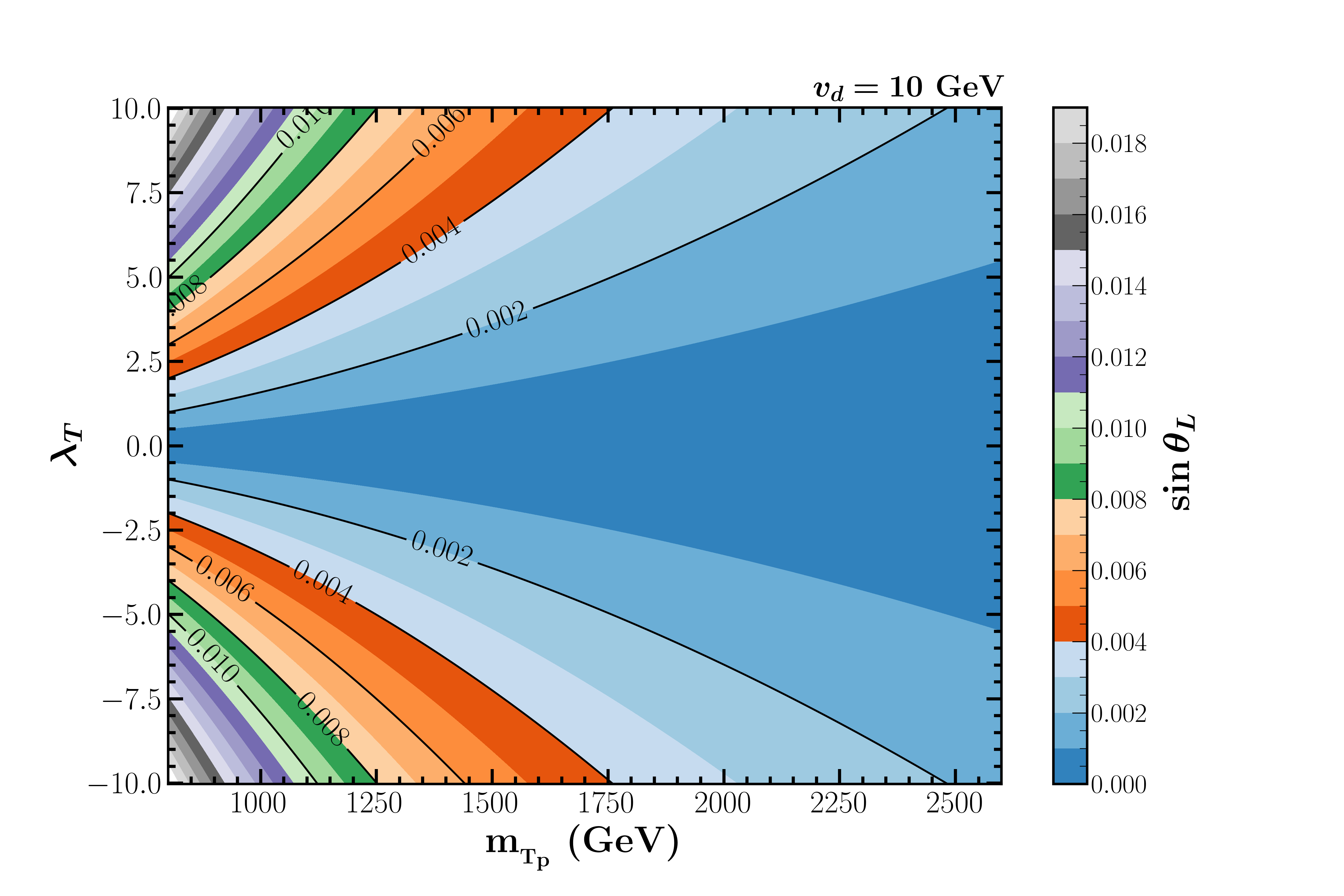}}

\subfloat[\label{fig:lamt_mtp_st_vd_100_gev}]{\includegraphics[width=0.5\columnwidth]{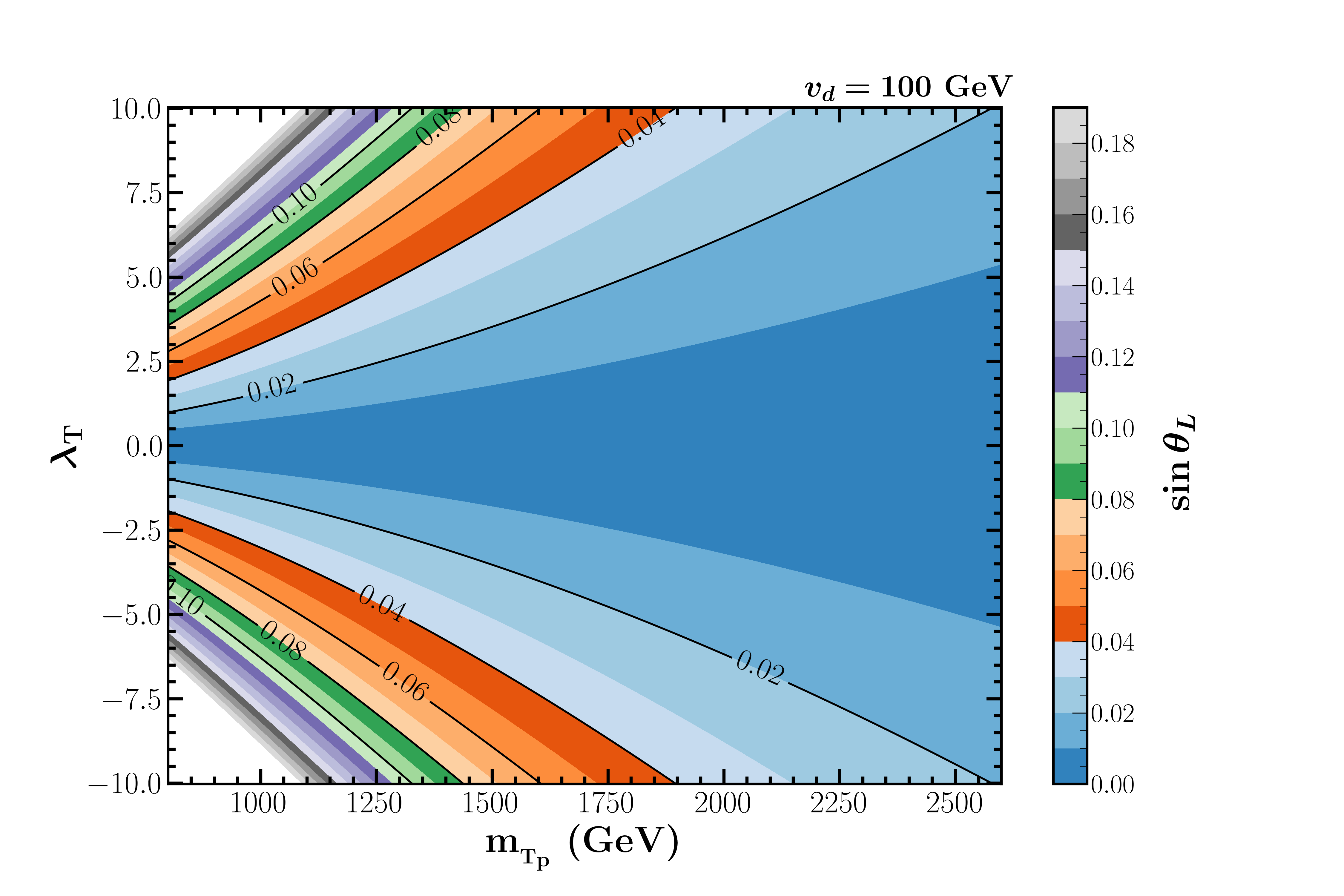}}
\subfloat[\label{fig:lamt_mtp_st_vd_200_gev}]{\includegraphics[width=0.5\columnwidth]{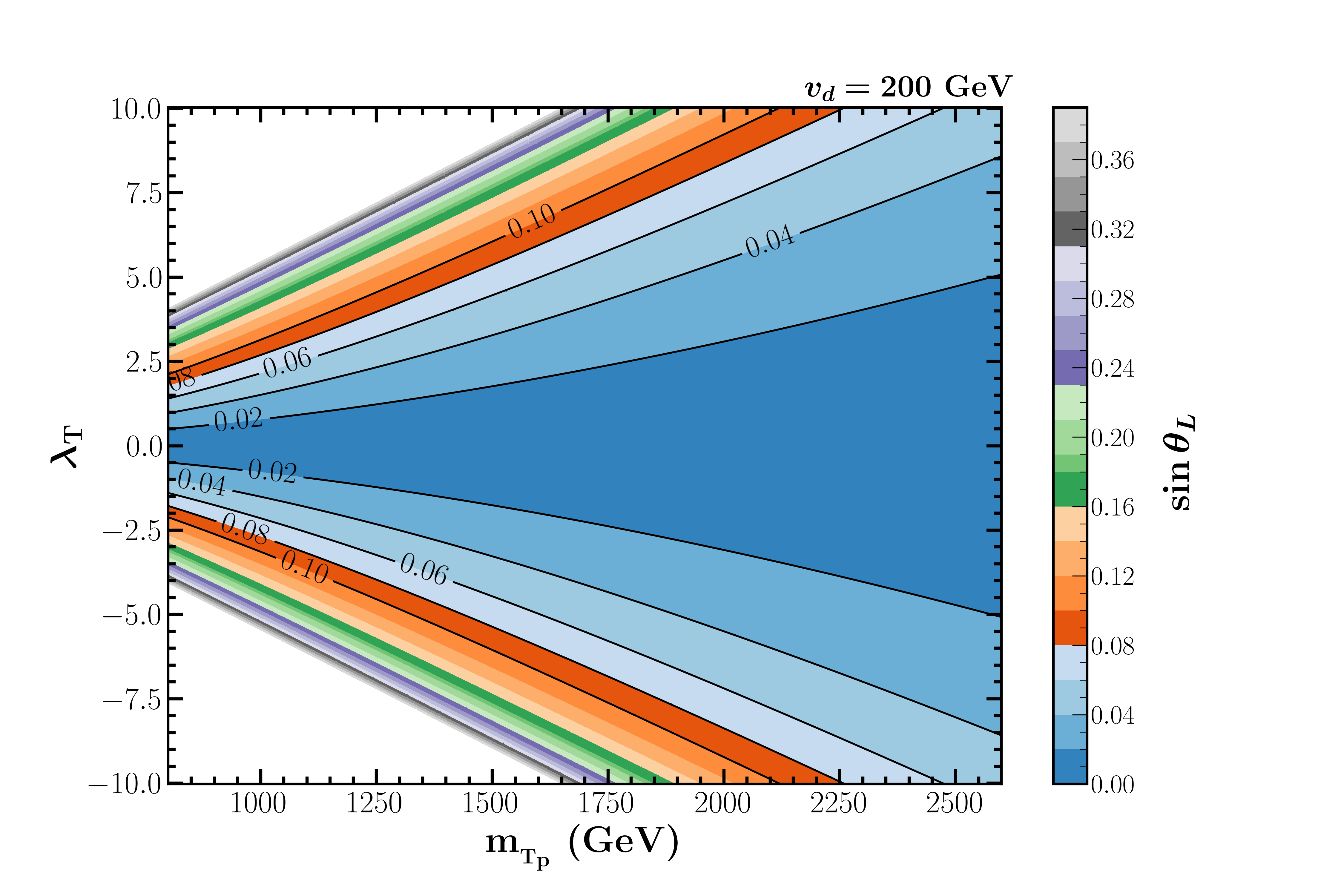}}
\caption{{ \scriptsize Allowed ranges of $\sin\theta_{L}$ in the $\lambda_{T} - m_{_{T_{p}}}$ plane for different choices of $v_{d}$ consistent with perturbative unitarity bound.}}
\label{fig:lamT-mTp}
\end{figure}

  In Fig.~\ref{fig:lamT-mTp}, we illustrate the effect of Eq.~(\ref{eq:bound_lamT}) on the allowed ranges of $\sin\theta_{L}$ as a function of the top partner mass.


Finally, the relevant parameters which play a crucial role in the discussion that follows are  $\sin\theta_{L}$, $v_d$, $m_{\gamma_d}$, $m_{h_d}$, and 
$m_{_{T_{p}}}$. These parameters can be considered as free, albeit with certain restrictions in their allowed ranges coming from various 
constraints such as perturbative unitarity bound, constraints coming from electroweak precision observables (EWPO)\cite{deBlas:2016ojx, Chen:2017hak} and Higgs signal strength measurements. 

{The mixing angle between the SM Higgs and the dark Higgs is constrained by various observations \cite{Robens:2015gla, Robens:2019ynf,Robens:2022oue,Adhikari:2022yaa,Robens:2022cun}. For low mass region ($10 ~{\rm GeV} \leq m_{h_d} \leq 100$ GeV) there is a stringent lower limit ($0.9-0.998$) on the allowed values of $|\sin\theta_{S}|$ for $v_{d}\sim v_{_{EW}}$ coming from LEP experiment. 

{
For $m_{h_d}>100$ GeV, the constraint coming from LHC heavy Higgs searches and Higgs signal strength measurement are relevant and it depends on the mass of the dark Higgs.
For $m_{h_d}\sim 200-1000$ GeV, the Higgs signal strength measurement sets a limit on $|\sin\theta_{S}|< 0.24$ \cite{Robens:2022cun}. However, heavy Higgs searches at the LHC give more stringent bound on $|\sin\theta_{S}|$ in the range $200-600$ GeV and the present upper limit on the allowed values of $|\sin\theta_{S}|$ is $0.2$  \cite{Adhikari:2022yaa,Robens:2022cun}. 

For higher masses, the constraint coming from the correction to W boson mass and the perturbative unitarity requirement is more stringent, for example, $|\sin\theta_{S}|$ as low as $0.14$ can be ruled out for $m_{h_d} \sim 1$ TeV \cite{Robens:2022cun}.
}

Exotic decays of the SM Higgs via the mixing of the SM Higgs with the dark Higgs also provides strong constraint on the scalar mixing angle which is sometimes more constraining than the bound coming from the Higgs signal strengths measurements. For example, Higgs to invisible searches can be used to set limits on $|\sin\theta_S|$  using the formula quoted in Ref. \cite{Kim:2019oyh},
\beq
|\sin\theta_S | \leq 4.6\times 10^{-4}~ \left(\frac{v_d}{\rm GeV}\right) \sqrt{{\rm BR}_{\rm lim}}
\eeq
where ${\rm BR}_{\rm lim}$ is the upper limit on the branching ratio of the SM Higgs into invisible decay modes. 
The latest CMS  ~\cite{CMS:2022ofi} and ATLAS ~\cite{ATLAS:2023tkt} searches for the Higgs $\to$ invisible decay modes provide,
\beq
{\rm BR}_{\rm lim} = 
\begin{cases}
0.15& {\rm CMS}  \\
0.107 & {\rm ATLAS }  \\
\end{cases}
\eeq

\noindent
For $v_d = 100$ GeV, this translates to $|\sin\theta_S| \leq 0.018$ (CMS) and $0.015$ (ATLAS). 
}

We define the ratio (${\rm R}_{_{\Gamma}}$) of the decay widths in the nonstandard and the standard (or traditional) modes as,

\bea\label{eq:RG_1}
{\rm R}_{_{\Gamma}}&=&\frac{\Gamma(T_p \to t+h_{d}/\gamma_{d})}{\Gamma(T_p \to t/b+W/Z/h)}
\eea

In the limit $|\sin\theta_d | ,\, |\sin\theta_{S}| ,\, \varepsilon$ \footnote{$\varepsilon$ and $\varepsilon^{\prime}$ are related via $\varepsilon = \varepsilon^{\prime} /\sqrt{1-\varepsilon^{\prime 2}/\cos^{2}\hat{\theta}_{W}}$, where $\cos\hat{\theta}_{W} = \cos\theta_{W}+\mathcal{O}(\varepsilon^{2})$} $,\, ~{\rm and}~ m_t/ m_{_{T_p}} \ll 1$ this ratio is given by Eq.~(\ref{eq:RG_2}). It 
illustrates that the enhancement in the decay width in the nonstandard mode compared to that for the standard one in two different region of 
the parameter space in the $\sin\theta_{L} - m_{_{T_{p}}}$ plane significant for the {HE-HL LHC} analysis. In particular, at the {HE-HL
LHC} both large $m_{_{T_{p}}}$ and small $|\sin\theta_{L}|$ will be probed. Hence, it is important to consider the following two scenarios: (i) $|\sin\theta_{L}| \ll m_t/m_{_{T_{p}}} \ll 1$  and (ii) $m_t/m_{_{T_{p}}} \ll |\sin\theta_{L}| \ll 1$, relevant for LHC.

\bea\label{eq:RG_2}
{\rm R}_{_{\Gamma}}&\approx &\frac{1}{2}
	\begin{cases}
		\left(\frac{m_{_{T_{p}}}}{m_{t}}\right)^{2} \left(\frac{v_{_{\rm EW}}}{v_d}\right)^{2}& {\rm for, }~ |\sin\theta_{L}|\ll \frac{m_{t}}{m_{_{T_{p}}}}\ll 1 \\
		\left(\frac{m_{t}}{m_{_{T_{p}}} \sin^{2}\theta_{L}}\right)^{2} \left(\frac{v_{_{\rm EW}}}{v_d}\right)^{2} & {\rm for, }~  \frac{m_{t}}{m_{_{T_{p}}}} \ll | \sin\theta_{L}|\ll 1
	\end{cases}
\eea

Eq.~(\ref{eq:RG_2}) explains when $|\sin\theta_{L}| \ll m_t/m_{_{T_{p}}} \ll 1$, the enhancement in ${\rm R}_{_{\Gamma}}$ has its origin in the hierarchy between
the fermion masses ($m_{_{T_{p}}}/m_t$) and the two VEVs ($v_{_{\rm EW}}/v_d$). 

Whereas, in the other limit, {\it i.e.}, $m_t/m_{_{T_{p}}} \ll |\sin\theta_{L}| \ll 1$ \footnote{However, this region of parameter space is disfavored by perturbative unitarity constraint.} the enhancement can be traced back to the hierarchy between the two VEVs, {\it i.e.}, ($v_{_{\rm EW}}/v_d$) and that of $m_t/({m_{_{T_{p}}} \sin^2\theta_{L}})$. It shows even when $v_d \sim v_{_{\rm EW}}$ one can have a small enhancement in ${\rm R}_{_{\Gamma}}$ if, $m_t/({m_{_{T_{p}}} \sin^2\theta_{L} }) > 1$ , {\it i.e.}, when $\sin^{2}\theta_{L}< m_t/m_{_{T_p}}<|\sin\theta_L| << 1$.

\begin{figure}[H]
\centering
\subfloat[\label{fig:RG_vd_10_gev}]{\includegraphics[width=0.5\columnwidth]{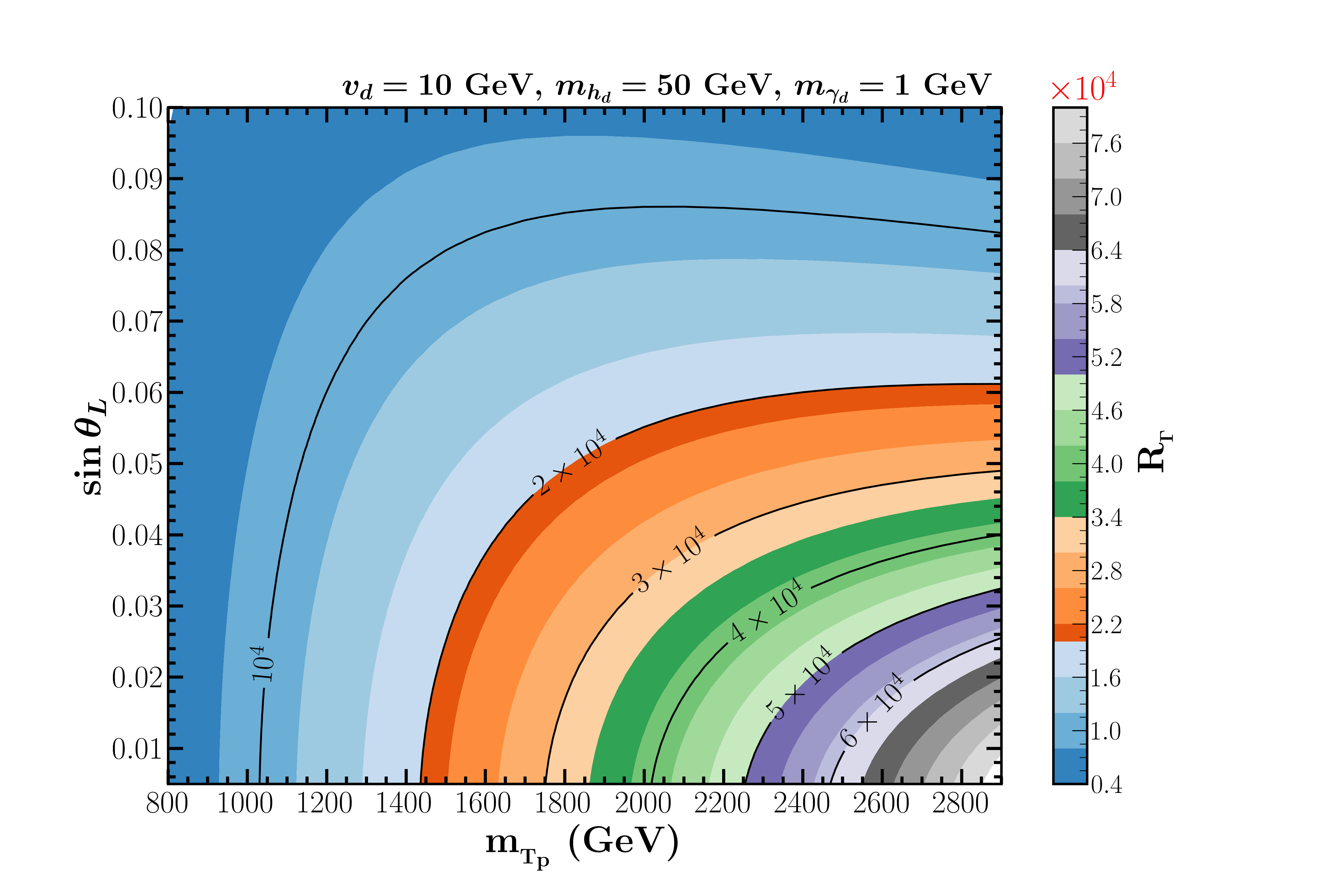}}
\subfloat[\label{fig:RG_vd_100_gev}]{\includegraphics[width=0.5\columnwidth]{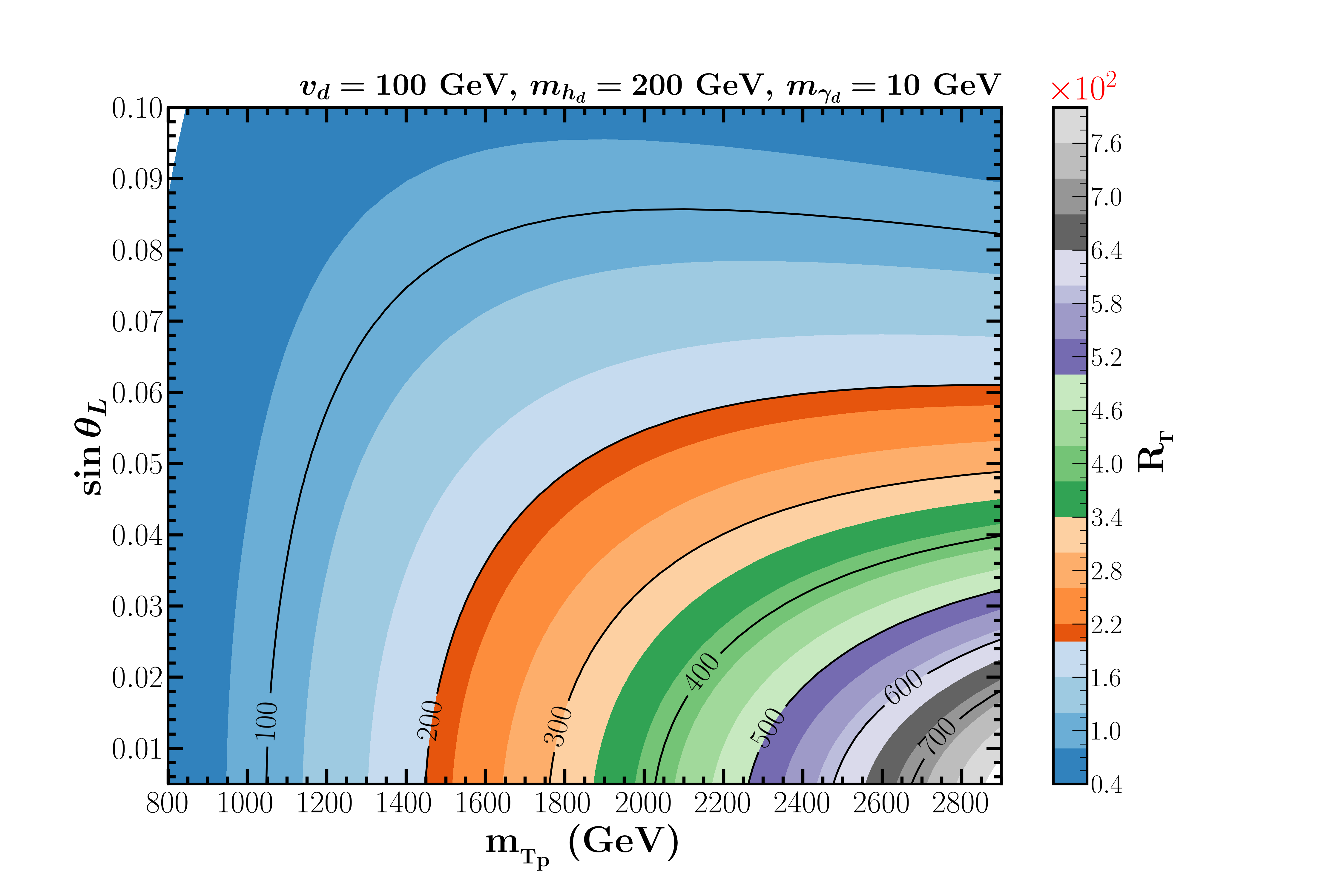}}

\subfloat[\label{fig:RG_vd_200_gev}]{\includegraphics[width=0.5\columnwidth]{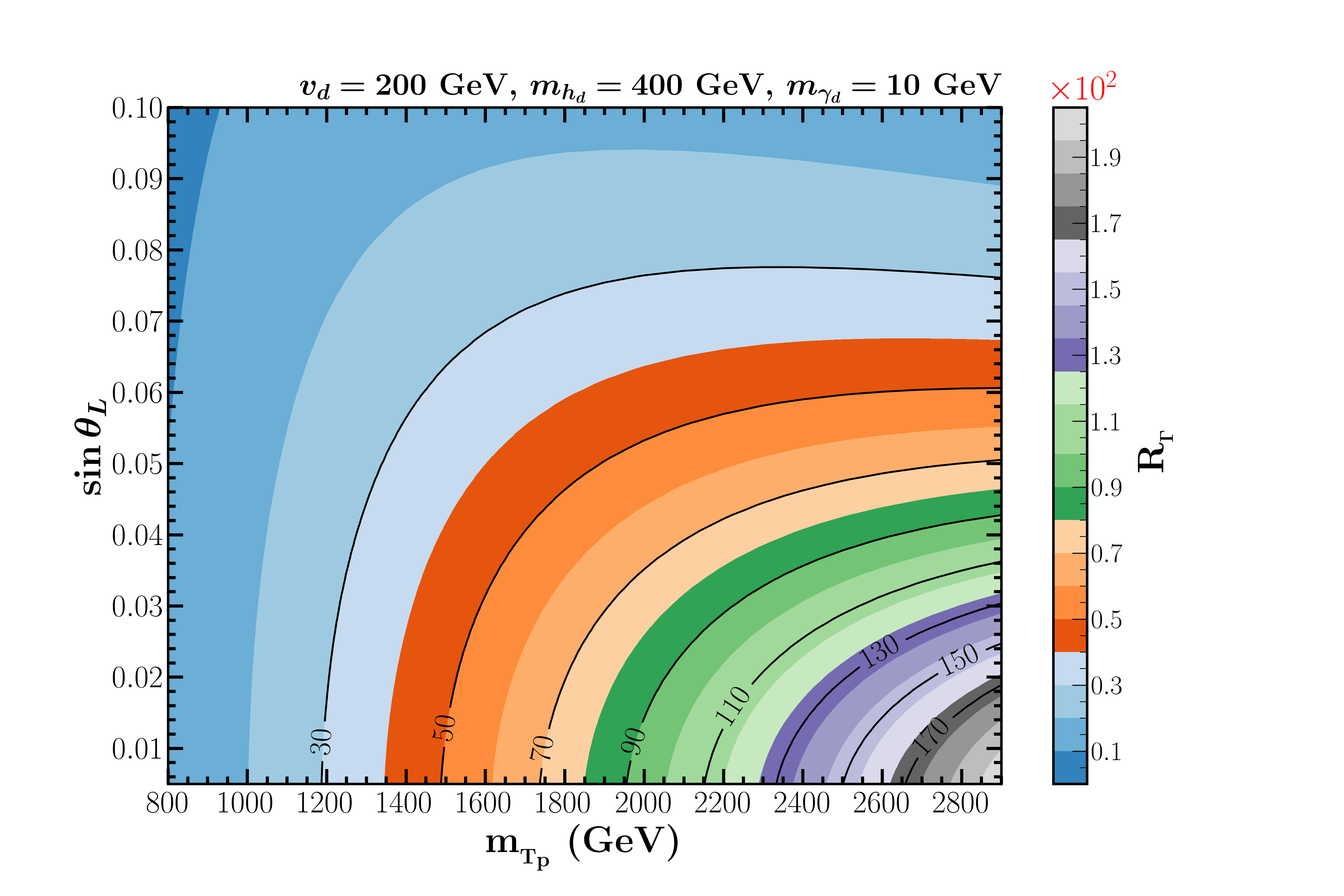}}
\subfloat[\label{fig:RG_vd_246_gev}]{\includegraphics[width=0.5\columnwidth]{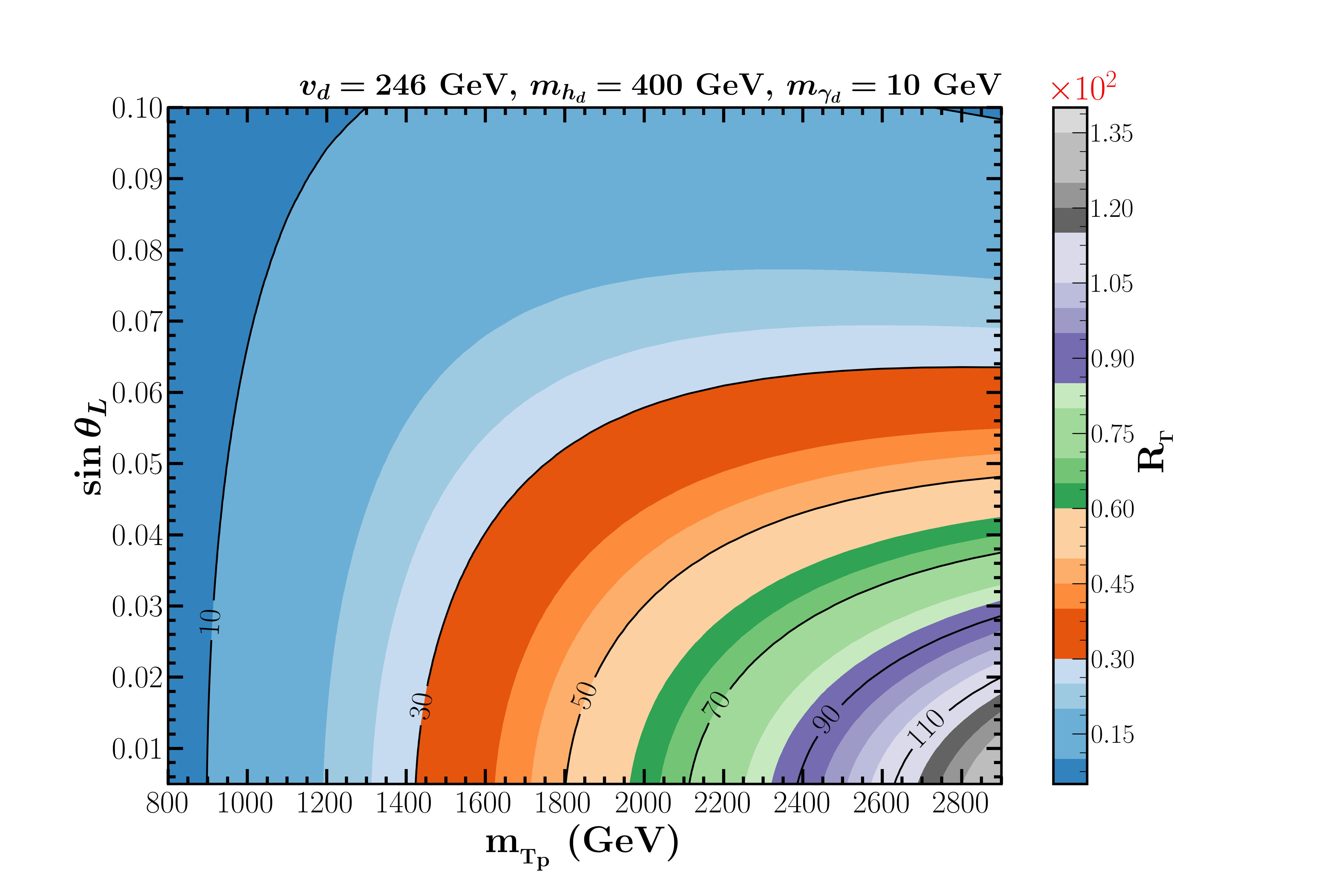}}

\caption{{ \scriptsize Ratio of the decay widths in the nonstandard ($t \gamma_d, ~t h_d$) and traditional ($b W, ~t Z, ~t h$) modes of the maverick top partner in the $\sin\theta_{L} - m_{_{T_{p}}}$ plane for different choices of $v_{d}$ consistent with perturbative unitarity bound.}}
\label{fig:width_ratio}
\end{figure}

Fig.~\ref{fig:width_ratio} shows that the ratio of the decay widths (${\rm R}_{_{\Gamma}}$) of the top partner in the nonstandard and the traditional modes without making any approximation mentioned in Eq.~(\ref{eq:RG_2}). The ${\rm R}_{_{\Gamma}}$ is calculated using Eq.~(\ref{eq:RG_1}) and supplying various decay widths provided in Appendix \ref{App:B}.

{To calculate various decay widths, we use the following values for the relevant SM parameters \cite{PDG:2020ssz},

$$
    m_h = 125.5 {\rm ~ GeV}, ~ 
   	m_t = 173.2 {\rm ~ GeV}, ~
    m_b = 4.18 {\rm ~ GeV} 
$$
$$
    m_W = 80.377 {\rm ~ GeV}, ~
	m_Z = 91.187 {\rm ~ GeV},~
    v_{_{\rm EW}} = 246 {\rm ~ GeV}$$
  
  \noindent  
Throughout our analyses we set, $
|\sin\theta_S| = 10^{-6}
$.
}

\begin{figure}[H]
\centering
\subfloat[\label{fig:BR_vd_100_50_gev}]{\includegraphics[width=0.5\columnwidth]{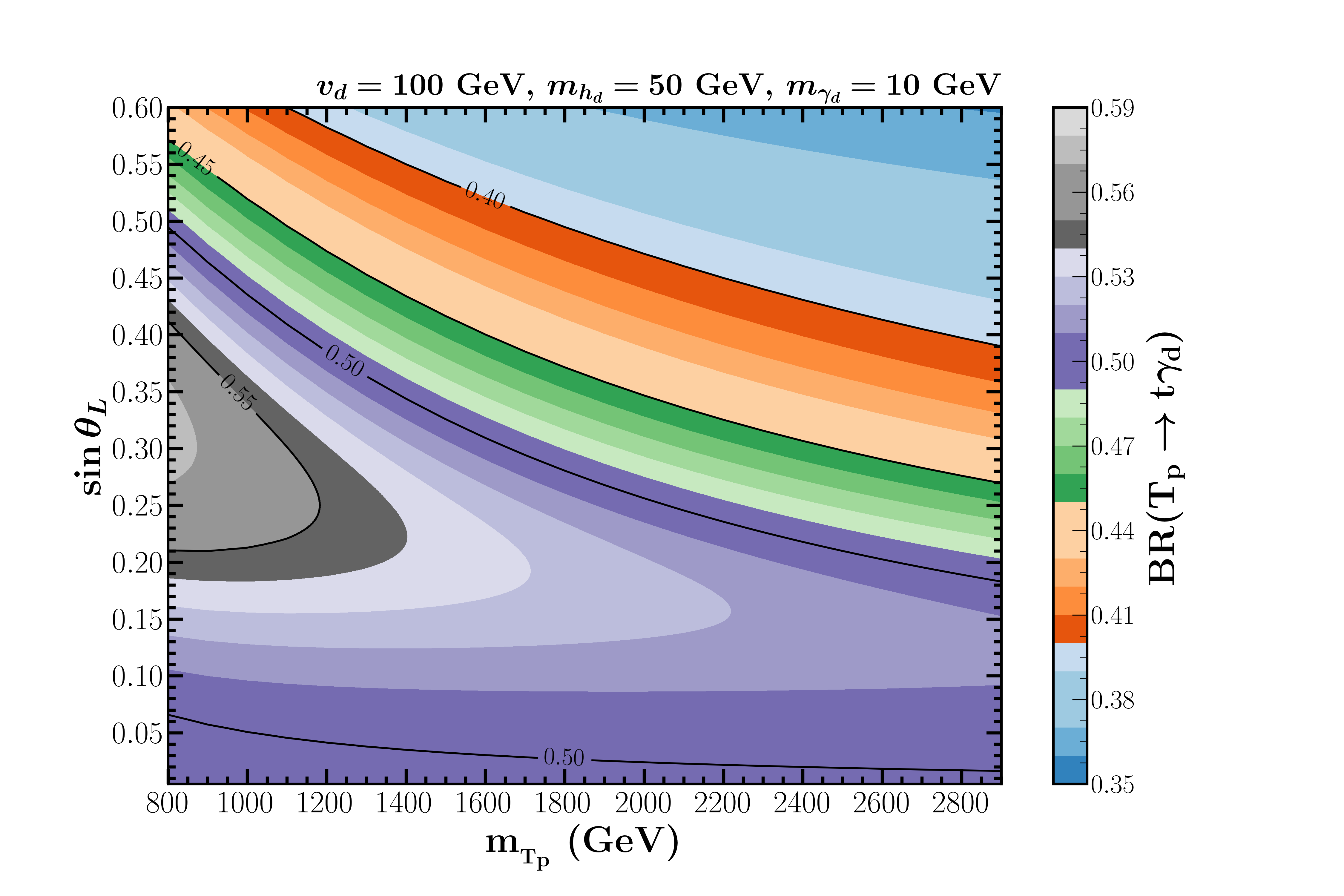}}
\subfloat[\label{fig:BR_vd_100_gev}]{\includegraphics[width=0.5\columnwidth]{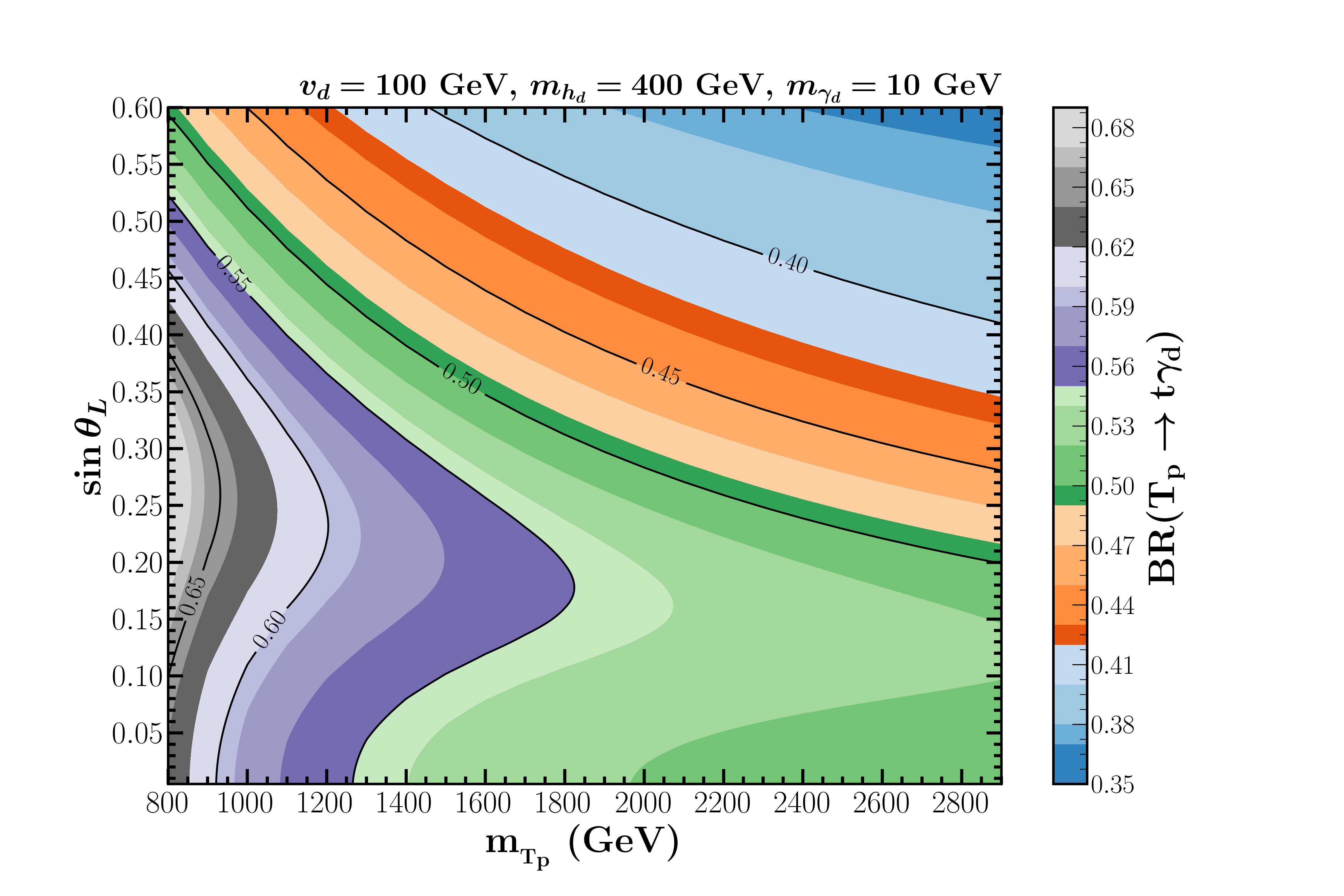}}

\subfloat[\label{fig:BR_vd_200_gev}]{\includegraphics[width=0.5\columnwidth]{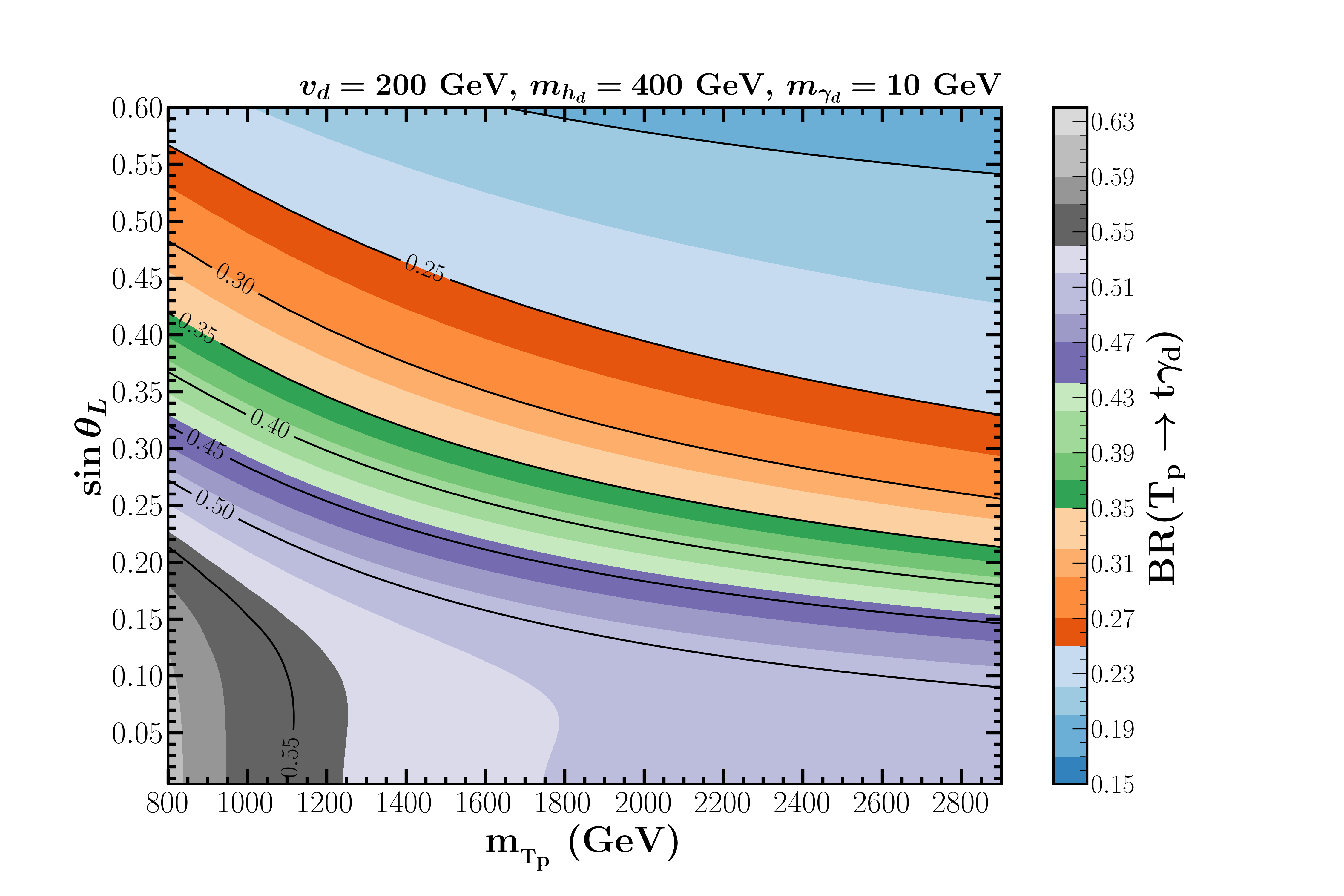}}
\subfloat[\label{fig:BR_vd_246_gev}]{\includegraphics[width=0.5\columnwidth]{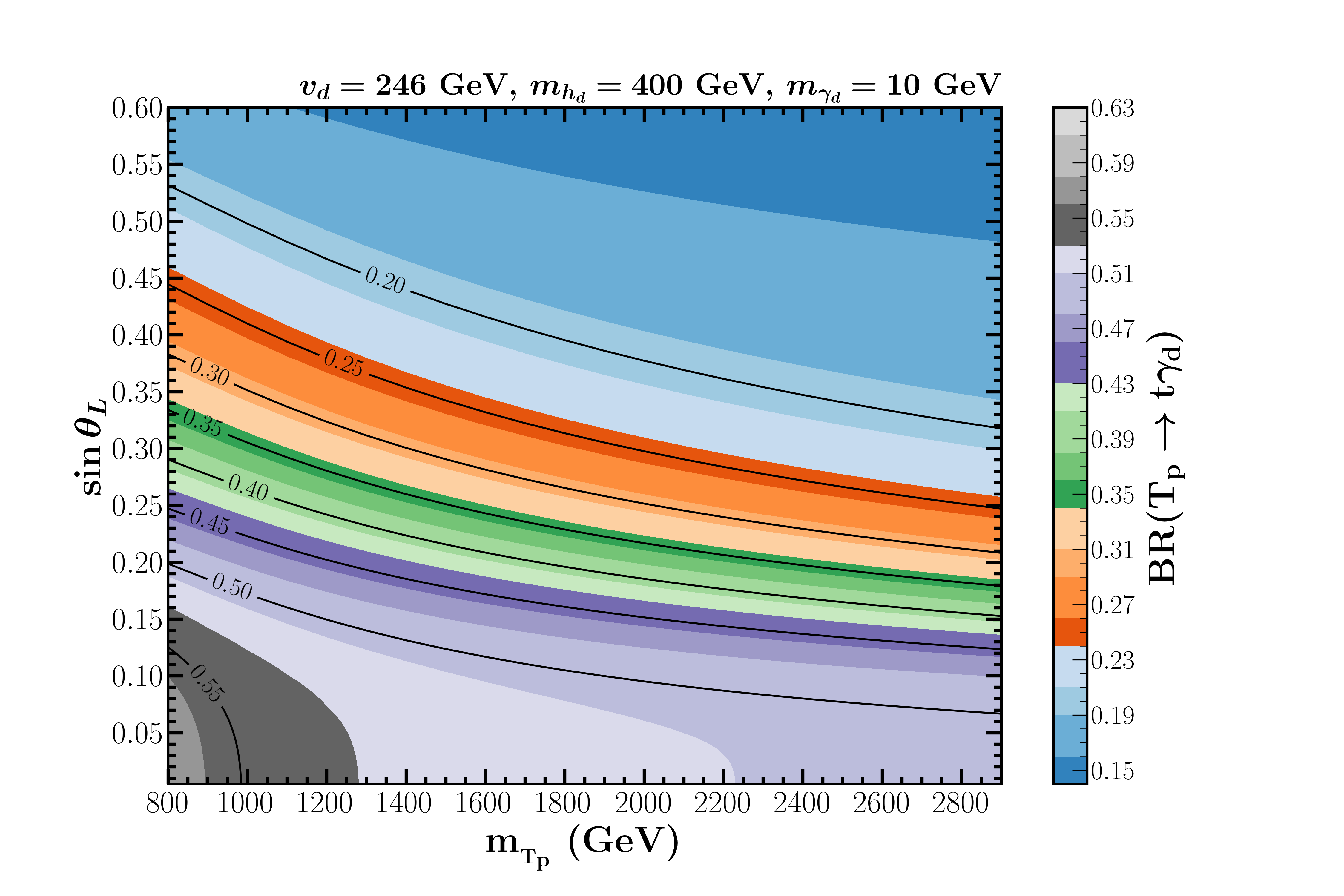}}

\caption{{ \scriptsize Variation of the branching ratio of $T_p \to t \gamma_d$ in the $\sin\theta_{L} - m_{_{T_{p}}}$ plane for several choices of $v_{d}$, $m_{{\gamma}_d}$ and $m_{h_d}$.}}
\label{fig:br_tdph}
\end{figure}

The branching ratio in the $\sin\theta_{L}$ - $m_{_{T_{p}}}$ plane is displayed in Fig.~\ref{fig:br_tdph} for various choices 
of $v_d$, $m_{\gamma_d}$, and $m_{h_d}$. {In Table \ref{table:branchingratios} we list the branching ratios of the top partner decaying into a top quark and a dark photon 
for various benchmark points}. We have mostly restricted ourselves to the case of light dark photon in the mass range
$1-10$ GeV. In this mass range the corresponding bound on the kinetic mixing parameter ($\varepsilon$) is found to be $\varepsilon \lesssim 10^{-3}$
\cite{Jaeckel:2010ni, PhysRevD.43.2314, Brockway:1996yr}. For small kinetic mixing the $U(1)_Y$ and $U(1)_d$ gauge sectors remain practically decoupled and it hardly plays any role in the remaining discussions.

For the present analysis, we assume that the dark photon 
decays to an invisible mode (for example a pair of dark
matter particles) so that it gives rise to $\cancel{E}_{T}$ signature in a collider experiment. To achieve this, one needs to augment the present model with an additional dark sector particle charged under $U(1)_d$ (a possible dark matter candidate\footnote{We do not present explicit details of the DM content of the DS here. For more details we refer \cite{PhysRevD.99.115024}.}$^{,}$\footnote{ Alternatively, the dark photon can be made stable at the length scale of the detector provided the kinetic mixing parameter is very small ($\lesssim 10^{-7}$).}).
 Throughout this article we, therefore, simply assume that the dark photon
is invisible and evade any detection by the detector.

\begin{table}[H]
\centering
\resizebox{\columnwidth}{!}{
\begin{tabu}{lcccccccccccc}
\hline 
\hline
\bm{$m_{_{T_p}}$}& 800 & 1000 & 1200 & 1400 & 1600 & 1800 & 2000 & 2200 & 2400 & 2500 & 2600 & 2800 \\ 

\textbf{BR} & 0.593 & 0.560 & 0.542 & 0.531  & 0.522 & 0.516 & 0.511 & 0.507 & 0.503 & 0.502  & 0.500
 & 0.497\\
\hline  
\hline
\end{tabu} 
}
\caption{{ \scriptsize Branching ratios of $T_p \to t \gamma_d$ for various benchmark points, assuming $\sin\theta_L=0.1$, $v_d = 200$ GeV, $m_{\gamma_d}=10$ GeV and $m_{h_d}= 400$ GeV.}}
\label{table:branchingratios}
\end{table}

\section{Top partner at the LHC}\label{Tp@LHC}

In this section we describe in detail our collider analysis in the context of the LHC. 
We have considered both the single ($pp \to T_p j $) and the pair ($pp \to  T_p \overline{T}_p$) production  of top partner(s). 
We have used M{\scriptsize AD}G{\scriptsize RAPH}5\_aMC@NLO (version 3.2.0) \cite{Alwall:2014hca} event generator to simulate both the single and pair production of top partner in the context of 13 and 14 TeV LHC energies. In order to 
achieve this, we have used F{\scriptsize EYN}R{\scriptsize ULES} \cite{Alloul:2013bka} where  we have implemented the model described in Sec. \ref{sec:model} at the Lagrangian level. The Universal F{\scriptsize EYN}R{\scriptsize ULES} Output \cite{Degrande:2011ua} is
then interfaced with M{\scriptsize AD}G{\scriptsize RAPH}5\_aMC@NLO for event generation. We have used parton distributions provided by NNPDF (version 2.3) \cite{Ball:2012cx} to simulate the parton-parton hard scattering in a $pp$ collision. Events generated by M{\scriptsize AD}G{\scriptsize RAPH}5 are then interfaced with P{\scriptsize YTHIA} (version 6.4)\footnote{For $pp \to $ QCD multijet simulation we have used P{\scriptsize YTHIA} (version 8.3)\cite{Bierlich:2022pfr} in order to implement the jet parton matching.} \cite{Sjostrand:2006za} for parton shower, hadronization, and further analysis.

{The energy and momenta of all the final state objects are smeared with appropriate Gaussian smearing function \cite{Park:2642494, CMS:2016lmd} to take into account the finite detector resolution effects and the resolution function considered is,

\bea
\frac{\sigma(x)}{x} &=& \frac{N}{x}  \oplus  \frac{S}{\sqrt{x}}\oplus C
\eea

where, $x = p_{_T}$ (for electron and muon) or $E$ (for jet),  $N$ encapsulates the effect of electronic and pile-up noise, $S$ characterizes stochastic effects arising from the sampling nature of calorimeters and C describes the  $x$ independent offset term.  

\begin{table}[H]
\centering
\resizebox{0.6\columnwidth}{!}{
\begin{tabu}{lcccc}
\hline 
\hline
 & & $N \left({\rm GeV}\right)$ & $S \left({\rm GeV}\right)^{1/2}$ & $C$ \\ 
\hline 
\hline
\multirow{3}{*}{Jet} & $ \eta \leq 1.7$ & 1.59 & 0.521 & 0.030 \\ 
 
& $1.7<\eta \leq 3.2$ & 0 & 0.706 & 0.05 \\ 

&$ 3.2 < \eta \leq 5.0$& 0& 1& 0.0942 \\
\hline
\multirow{2}{*}{Electron} &$\eta \leq 1.5$ & 0.05 & 0 & 1.7 $\times 10^{-3}$ \\

& $ 1.5 < \eta \leq 2.5$ & 0.15 & 0 & 3.1 $\times 10^{-3}$ \\
\hline
\multirow{2}{*}{Muon} &$\eta \leq 1.5$ & 0.015 & 0 & 1.5 $\times 10^{-4}$ \\

& $ 1.5 < \eta \leq 2.5$ & 0.025 & 0 & 3.5 $\times 10^{-4}$ \\
\hline  
\hline
\end{tabu} 
}
\caption{{ \scriptsize N, S and C parameters for resolution function.}}
\label{table:nsc_resfunc}
\end{table}

We follow the functional form of resolution function as well as the fitted values of N, S, and C parameters as given in Table \ref{table:nsc_resfunc} from default ATLAS card in D{\scriptsize ELPHES} (version 3.5.0) \cite{deFavereau:2013fsa}. }
 
Since the top partner dominantly decays to top-quark and a dark photon $T_p \to t \gamma_d$ in our model, presence of top-quark is ubiquitous in both the  single and pair production channels. We will mostly confine ourselves to the fully hadronic decay of the top-quark ($t\to bW \to b q\bar{q'}$). In the following we consider the top partner in the mass range 1 TeV (800 GeV) $\leq m_{_{T_p}} \leq$ 2.6 (1.6) TeV for the single (pair) production of the top partner at the LHC. The top-quarks thus produced in the decay of the top partners are highly boosted for most of the ranges of top partner mass. Hence, the top decay products are highly collimated. Since the identification of top-quark and reconstruction of its four momenta will play an important  role in our analysis, we make use of the Johns Hopkins (J-H) Top Tagger \cite{Kaplan:2008ie} to identify and reconstruct boosted top-quark initiated jet(s) implemented within the framework of FastJet (version 3.4.0) \cite{Cacciari:2011ma} jet finding and analysis package. 

In the boosted top tagging analysis, we consider only those events that have total transverse momentum ($E_{\rm tot}$) greater than 400 GeV. 
We use Cambridge-Achen (C-A) clustering algorithm \cite{Bentvelsen:1998ug} to define a fat jet with jet radius ($R$) having values in accordance 
with the $E_{\rm tot}$ of these events and these are tabulated in Table.~\ref{table:delrvalues}. The jets thus obtained are then sorted in decreasing order of their transverse momentum. The J-H top tagger iteratively declusters a C-A clustered jet to find the substructure inside a fat jet of radius $R$. 

The J-H top tagging algorithm requires the specification of the following additional parameters: the fraction of the jet $p_{_T}$ carried by a subjet and the Manhattan distance (defined as $|\Delta \eta| + |\Delta \phi|$) between two subjets
to satisfy minimum values, $\delta p$  and $\delta r$, respectively to be considered as hard and resolved. We tabulate the values of $R$, $\delta p$  and $\delta r$ for different total transverse energy of an event, $E_{\rm tot}$ in Table.~\ref{table:delrvalues}.

\begin{table}[H]
\centering
\resizebox{0.7\columnwidth}{!}{
\begin{tabu}{lcccccc}
\hline 
\hline
$E_{\rm tot}$ & 400 & 600 & 800 & 1000 & 1600 & 2600 \\ 
\hline 
\hline
R & 1.4 & 1.2 & 1.0 & 0.8 & 0.6 & 0.4 \\

$\delta p$ & 0.10 & 0.10 & 0.05 & 0.05 & 0.05 & 0.05 \\
  
$\delta r$ & 0.19 & 0.19 & 0.19 & 0.19 & 0.19 & 0.19\\
\hline
\hline
\end{tabu} 
}
\caption{{ \scriptsize Choices of R, $\delta p,~ \delta r$ for various total transverse momentum range.}}
\label{table:delrvalues}
\end{table}

The efficiency of tagging a true top quark initiated jet in the  $p_{_{T}} $ range 400-1000 GeV is found to be 25 \%-50 $\%$ and corresponding light quark/gluon initiated jet being tagged as a top is found to be in the range 0.5 \%-1.5 $\%$ in the same range.

\subsection{Pair production of top partner}\label{subsec:tp@lhc}

\begin{figure}[H]
\centering
\subfloat[\label{fig:feynmandiag_tp_a}]{\includegraphics[width=0.4\columnwidth]{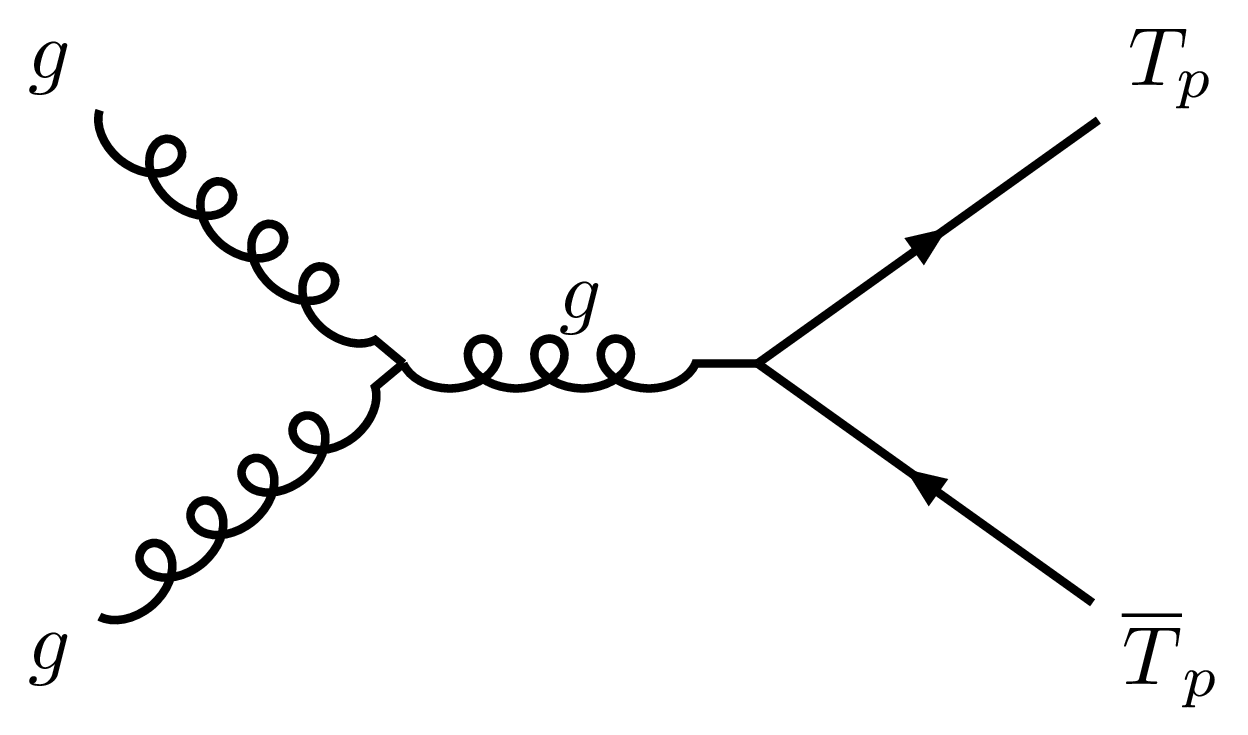}}~
\subfloat[\label{fig:feynmandiag_tp_b}]{\includegraphics[width=0.4\columnwidth]{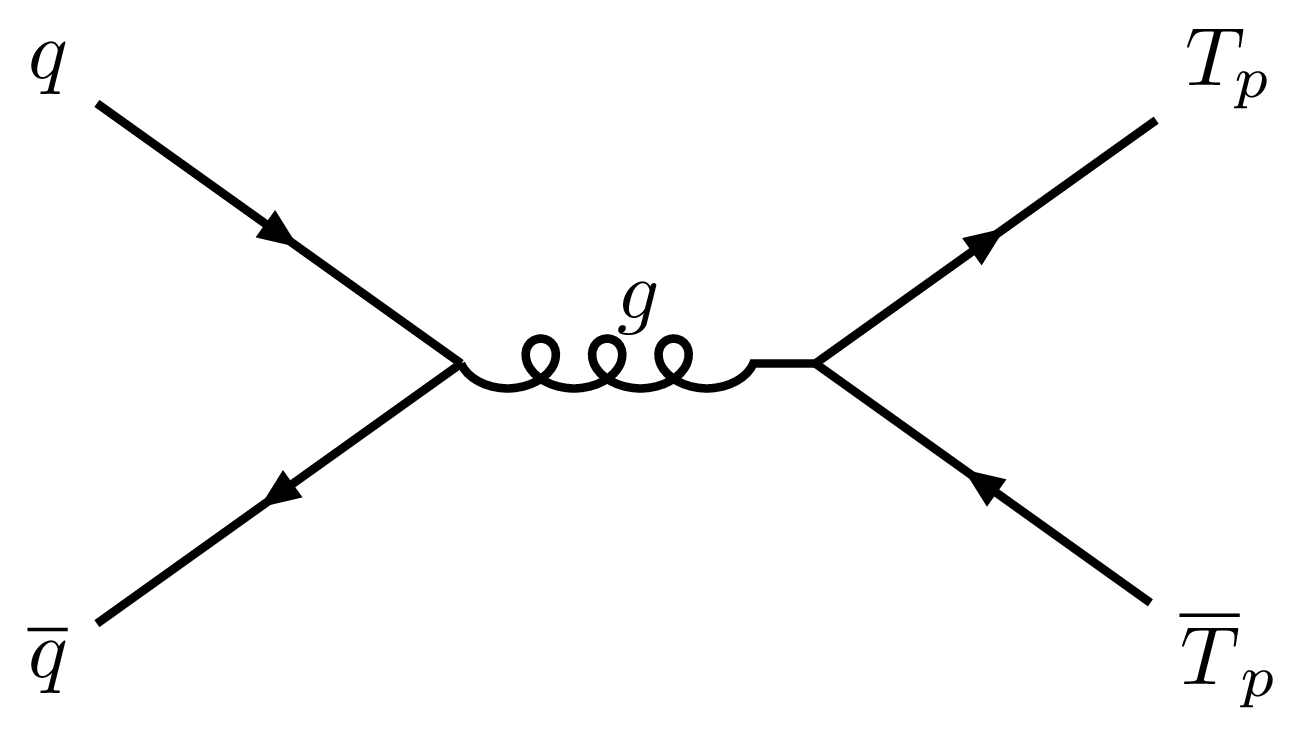}}

\subfloat[\label{fig:feynmandiag_tp_c}]{\includegraphics[width=0.33\columnwidth]{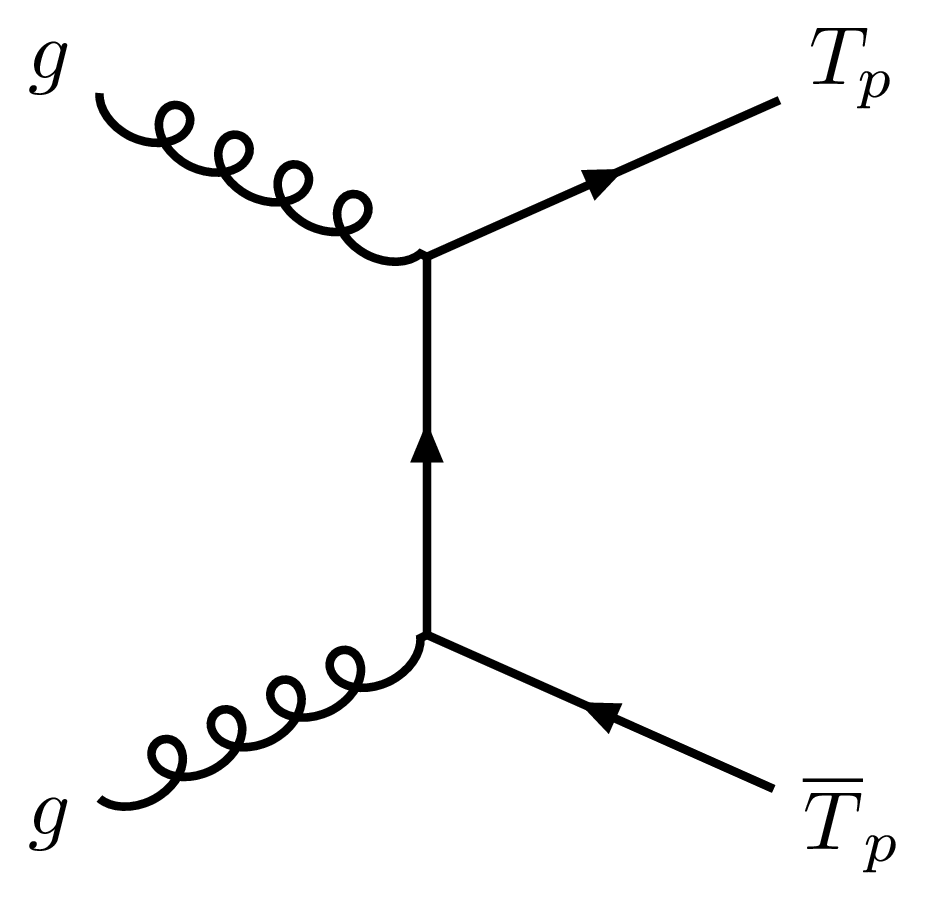}}
\caption{{\scriptsize Feynman diagrams for the pair production of top partner at the LHC.}}\label{fig:fd_tp}
\end{figure}

\begin{figure}[H]
    \centering
    \includegraphics[width=0.6\columnwidth]{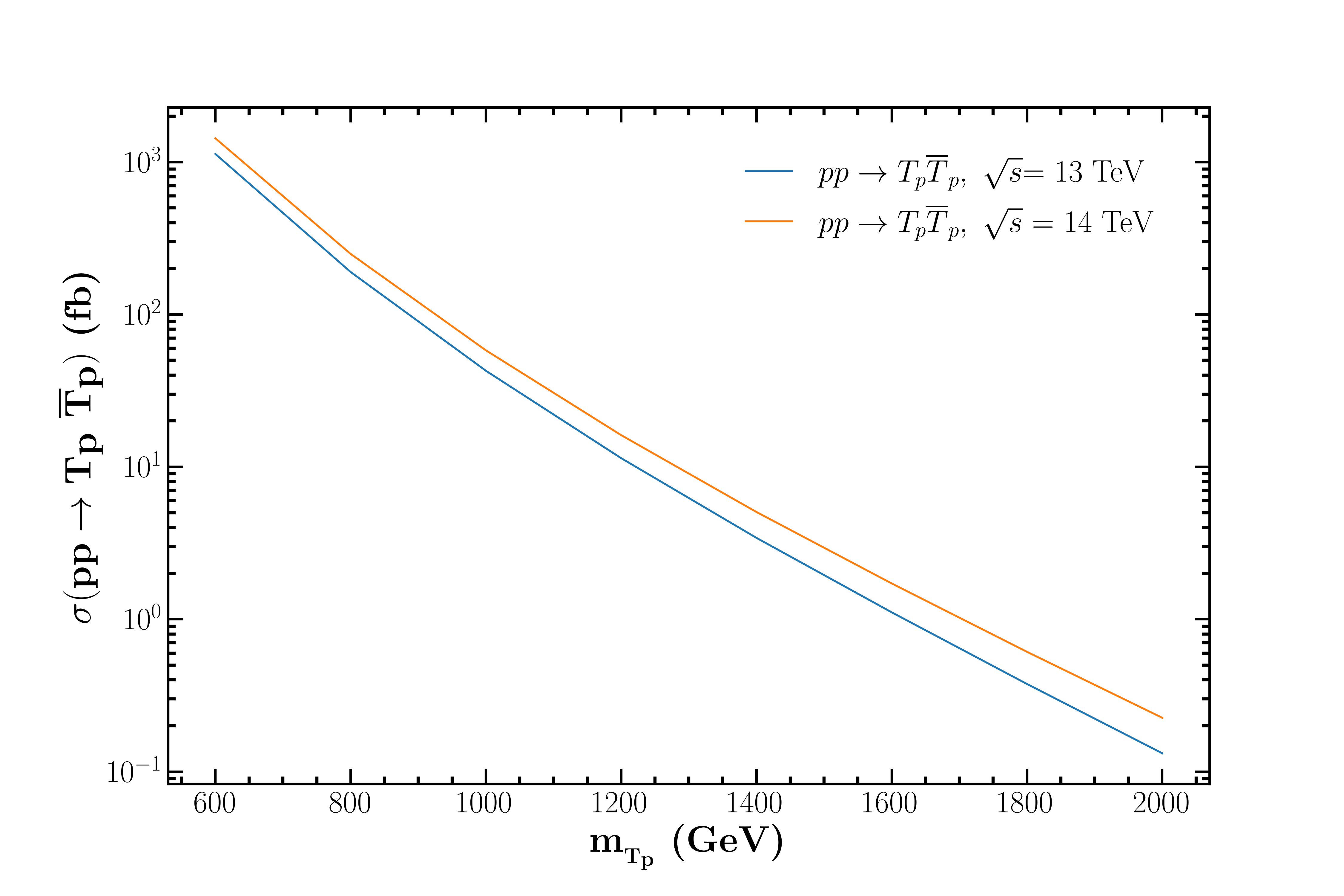}
    \caption{{\scriptsize $pp\to T_{p}\overline{T}_{p}$ cross  section as a function of the top partner mass at $\sqrt{s} = 13 ~{\rm and}~ 14$ TeV. {[$\alpha_{_S}(m_{_Z}) = 0.1179$ \cite{PDG:2020ssz} is used to estimate above cross section.]} }}
    \label{fig:tp_xsec_13_14}
\end{figure}

The pair production of top partner at the LHC is dominated by strong production and it depends on the strong coupling constant ($\alpha_{_S}$), the mass of the top partner ($m_{_{T_p}}$) {and it's spin}. This channel is less model dependent, {\it i.e.}, the production cross section is independent of the details
of the model parameters other than the top partner mass. The relevant Feynman diagrams for this process are depicted in Fig.~\ref{fig:fd_tp}. The production 
cross section as a function of $m_{_{T_p}}$ is presented in Fig.~\ref{fig:tp_xsec_13_14} for 13 and 14 TeV LHC centre of mass energies. {The NNLO corrected cross sections at $\sqrt{s}=13$ TeV are quoted from \cite{Matsedonskyi:2014mna}. We also use these NNLO corrected cross sections at $\sqrt{s}=13$ TeV to extract the $m_{_{T_{p}}}$ dependent $k$-factors by comparing them to the LO cross sections obtained from M{\scriptsize AD}G{\scriptsize RAPH}5\_aMC@NLO at the same center of mass energy. These $k$-factors are then used to get the NNLO corrected cross section at $\sqrt{s}=14$ TeV from the LO cross section provided by M{\scriptsize AD}G{\scriptsize RAPH}5\_aMC@NLO at this center of mass energy. }

\begin{figure}[H]
\begin{center}
\includegraphics[width=0.5\columnwidth]{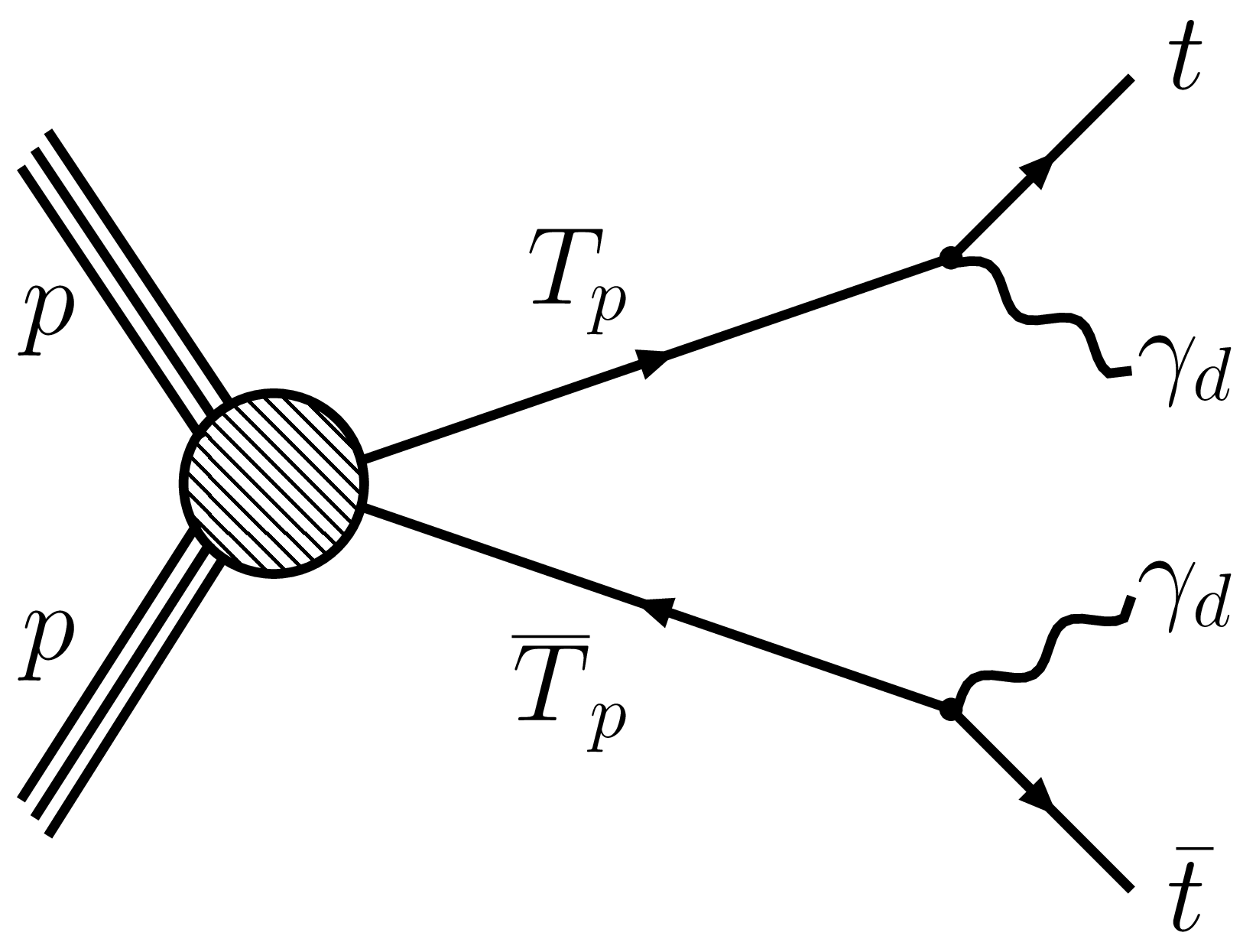}
\end{center}
\caption{{\scriptsize Schematic diagram for the signal process $pp \to T_{p}\overline{T}_{p}$ and further decay of $T_{p}\to t \gamma_{d}$.}}
\label{fig:schematic_tp}
\end{figure}

The top partner thus produced further decays substantially to a top quark and a dark photon in this model. So the final state consists of 
a $t\bar{t}$ pair and a pair of invisible dark photons ($pp \to T_{p}\overline{T}_{p} \to  t \bar{t} + 2  \gamma_d$). The schematic diagram for 
the same is shown in Fig.~\ref{fig:schematic_tp}.

 The presence of dark photons in the final state
gives rise to missing transverse energy signature as it goes undetected at the detector for the reasons already mentioned 
in previous sections.

 Since the top quark produced in the decay of the top partner are highly boosted we have considered fully hadronic decays 
of the top quark  to implement the boosted top tagging algorithm efficiently. 
The final state thus consists of {\it jets + missing transverse energy} 
with additional top structure present in it.
We require {\it at least two }central jets with $p_{_T}>20$ GeV, $|\eta|<2.5$ and no isolated leptons. {The SM backgrounds that contribute to such a final state are $t\bar{t}$, $t\bar{t}V$ (where $V=Z/W$), $tW$, and QCD multijet processes. Since our final state requirement is \textit{at least one} boosted top quark jet, SM $tj$ process in principle constitutes a potential background. However, the contribution from this background can be eliminated after further event selection criteria that we discuss later.}
We have listed the SM backgrounds relevant for the pair production analysis in Table.~\ref{table:tpbckcross} along with their cross sections.


\begin{table}[H]
\centering
\resizebox{0.6\columnwidth}{!}{\begin{tabu}{lcc}
\hline
\hline
\multirow{2}{*}{Process}& \multicolumn{2}{c}{Cross section (fb)} \\
\cline{2-3}
&13 TeV& 14 TeV\\
\hline
\hline
$pp \to t\bar{t}$&  $8.39\times 10^{5}$ ~ \cite{Kidonakis:2022hfa} &$9.9\times 10^{5}$  ~\cite{Kidonakis:2022hfa}\\

$pp \to t\bar{t}Z$ & $8.63\times 10^{2}$ ~\cite{Kulesza:2018tqz} & $1.045\times 10^{3}$ ~\cite{Kulesza:2018tqz}\\

$pp \to t\bar{t}W$& $5.66\times 10^{2}$ ~\cite{Kulesza:2018tqz}& $6.53\times 10^{2}$ ~\cite{Kulesza:2018tqz}\\

$pp \to tW$ & $7.95\times 10^{4}$ ~\cite{Kidonakis:2021vob}&$9.4\times 10^{4}$  ~\cite{Kidonakis:2021vob}\\

$pp \to $ QCD multijet& $1.96\times 10^{11}$&$2.16\times 10^{11}$\\
\hline
\hline
\end{tabu}
}
\caption{{ \scriptsize cross sections of the background processes considered for the top partner pair production analysis at $\sqrt{s} = 13 ~{\rm and}~ 14$ TeV.}}
\label{table:tpbckcross}
\end{table}
The cross sections quoted in Table.~\ref{table:tpbckcross} are beyond leading order\footnote{In this analysis the $t\bar{t}$ and $tW$ is considered at aN$^{3}$LO and $t\bar{t}Z$, $t\bar{t}W$ are considered at NLO+NNLL in QCD.} except for the QCD multijet background which is at leading order.
{For the simulation of QCD multijet background we have generated $pp\to 2j,~3j,~4j$ samples using M{\scriptsize AD}G{\scriptsize RAPH}5\_aMC@NLO and interfaced with P{\scriptsize YTHIA} $8.3$ to implement jet parton matching using the MLM matching scheme \cite{Mangano:2001xp}. }

\subsection*{Event selection criteria}

{We require the final state to have {\it at least one} boosted top quark jet. If the event contains {\it exactly one} tagged top quark jet, we take the untagged one of the two hardest jets as a top candidate.} 
{
We have compared the significances for the two different categories: {\it i.e.}, requiring {\it exactly two} boosted top quarks vs {\it at least one} boosted top quark requirement. It is the later one which gives better significance because of higher signal acceptance ratio. In addition, the kinematic variables we have used later in our analysis (for example, stransverse mass ($M_{_{T_2}}$) variable)  capture the topology of an event where the top partners are pair produced and decay to semi-invisible mode.}

We propose several kinematic observables including $M_{_{T_2}}$ which can be useful to efficiently discriminate the signal from the SM backgrounds. Below, we define
these variables and present their corresponding distributions. 

\begin{itemize}

\item Transverse momentum ($\bsym{p_{_{T}}}$): The transverse momentum of a particle is defined as, 
\bea
p_{_{T}} = \sqrt{p_{x}^{2}+p_{y}^{2}}
\eea
and corresponding transverse momentum distribution of the tagged top quark jet for both the signal benchmark points and SM backgrounds are shown in 
Fig.~\ref{fig:tp_pt}.

\begin{figure}[H]
\centering
\subfloat[\label{fig:tp_pt_dis_13tev}]{\includegraphics[width=0.5\columnwidth]{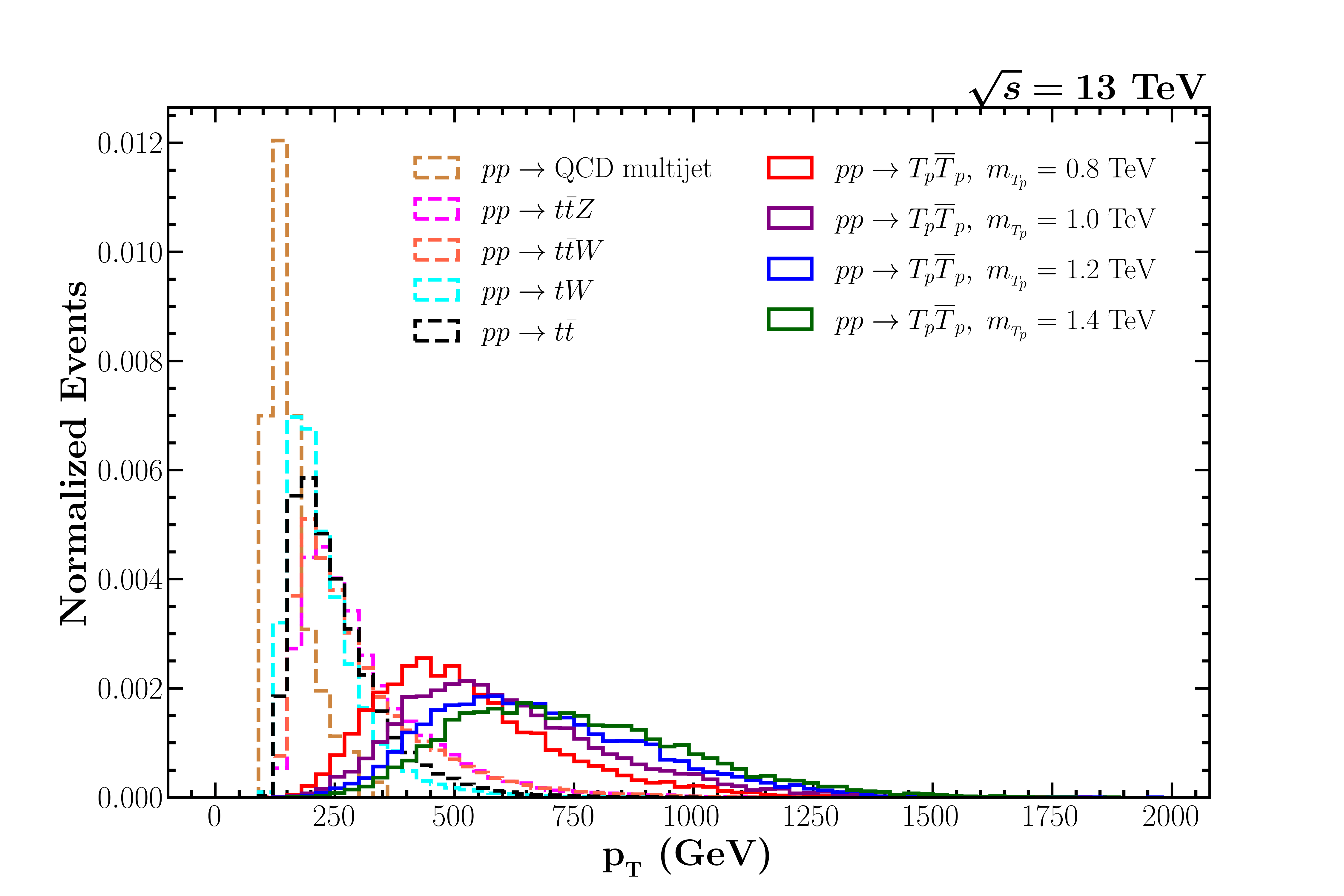}}
\subfloat[\label{fig:tp_pt_dis_14tev}]{\includegraphics[width=0.5\columnwidth]{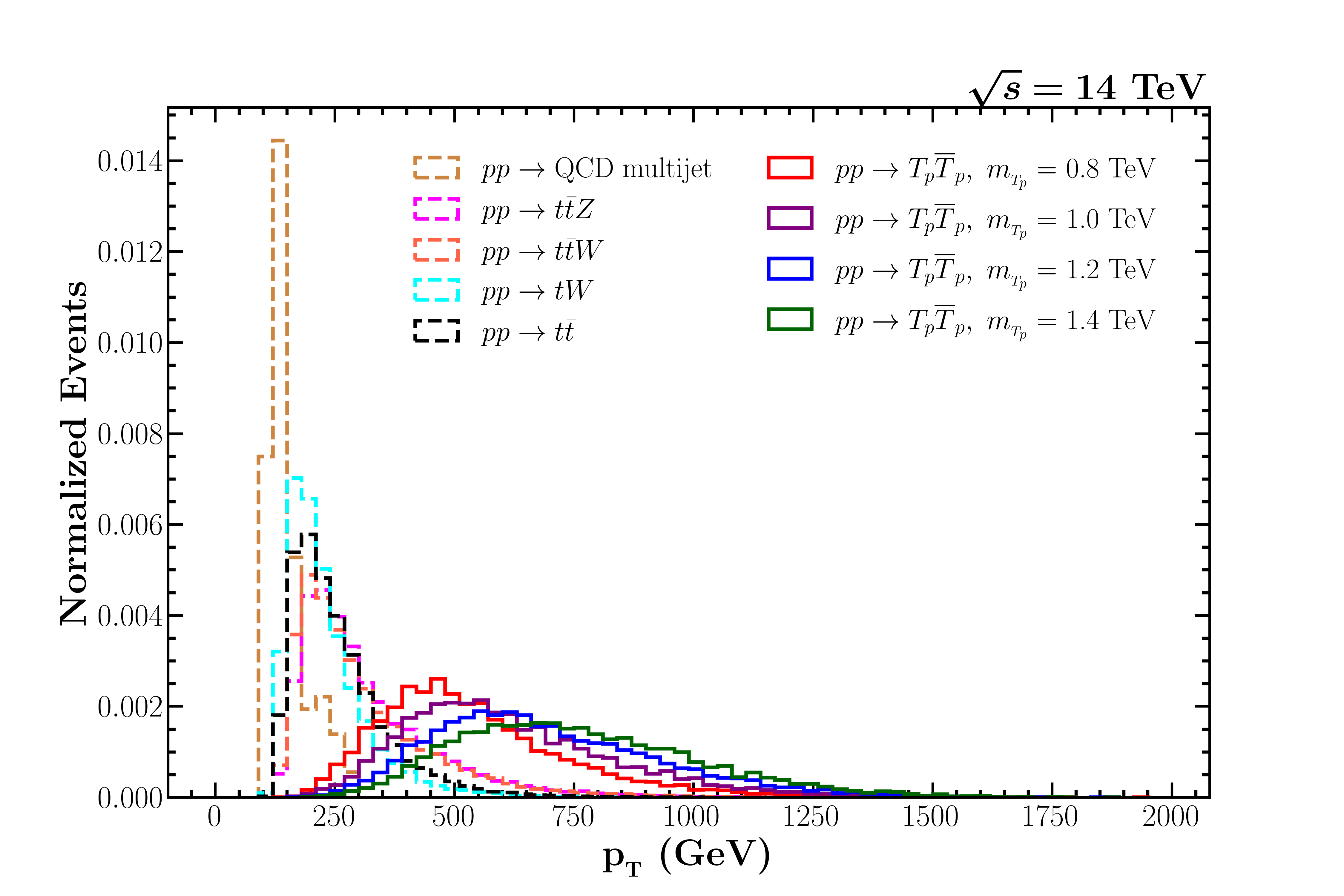}}
\caption{{\scriptsize Transverse momentum ($p_{_{T}}$) distributions of leading (or only) tagged top quark jet for various signal benchmark points 
and background processes at 13 and 14 TeV center of mass energies.}}\label{fig:tp_pt}
\end{figure}

\item Missing transverse energy($\bsym{\cancel{E}_T}$): The missing transverse energy is associated with the particles which go undetected at the
detector. In a hadronic collider such as the LHC one can use the momentum conservation in the transverse plane to find the transverse components of 
the total missing momentum associated with all the invisible particles 

\bea
\vec{\cancel{p}}_{_T} = -\sum_{ i \in \rm visible} \vec{p}^{~i}_{_T}
\eea

The missing transverse energy is then simply, $\cancel{E}_{T} = |\vec{\cancel{p}}_{_T}|$. Corresponding distributions for both the signal benchmark points and SM backgrounds are shown in Fig.~\ref{fig:tp_met}.

\begin{figure}[H]
\centering
\subfloat[\label{fig:tp_met_dis_13tev}]{\includegraphics[width=0.5\columnwidth]{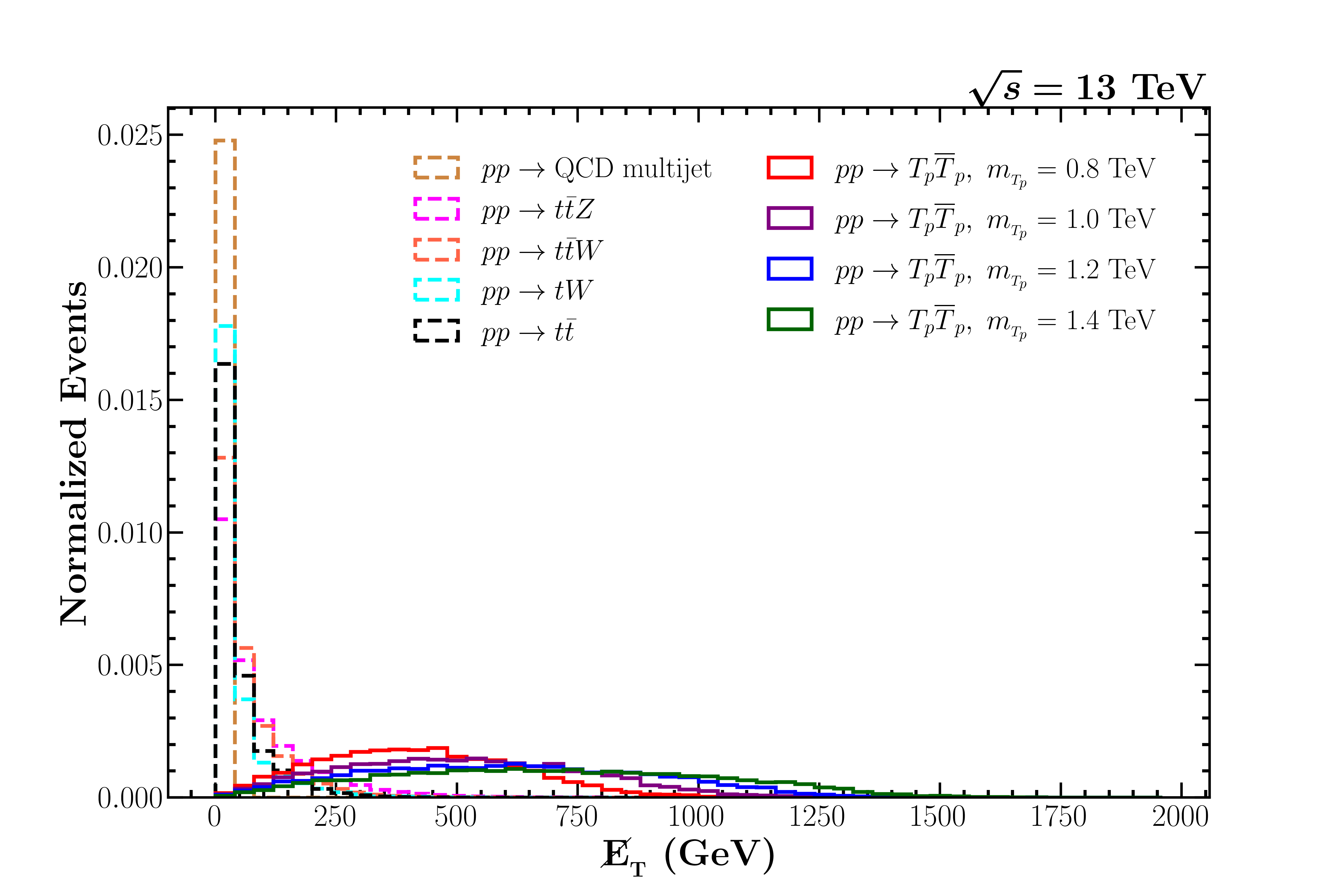}}
\subfloat[\label{fig:tp_met_dis_14tev}]{\includegraphics[width=0.5\columnwidth]{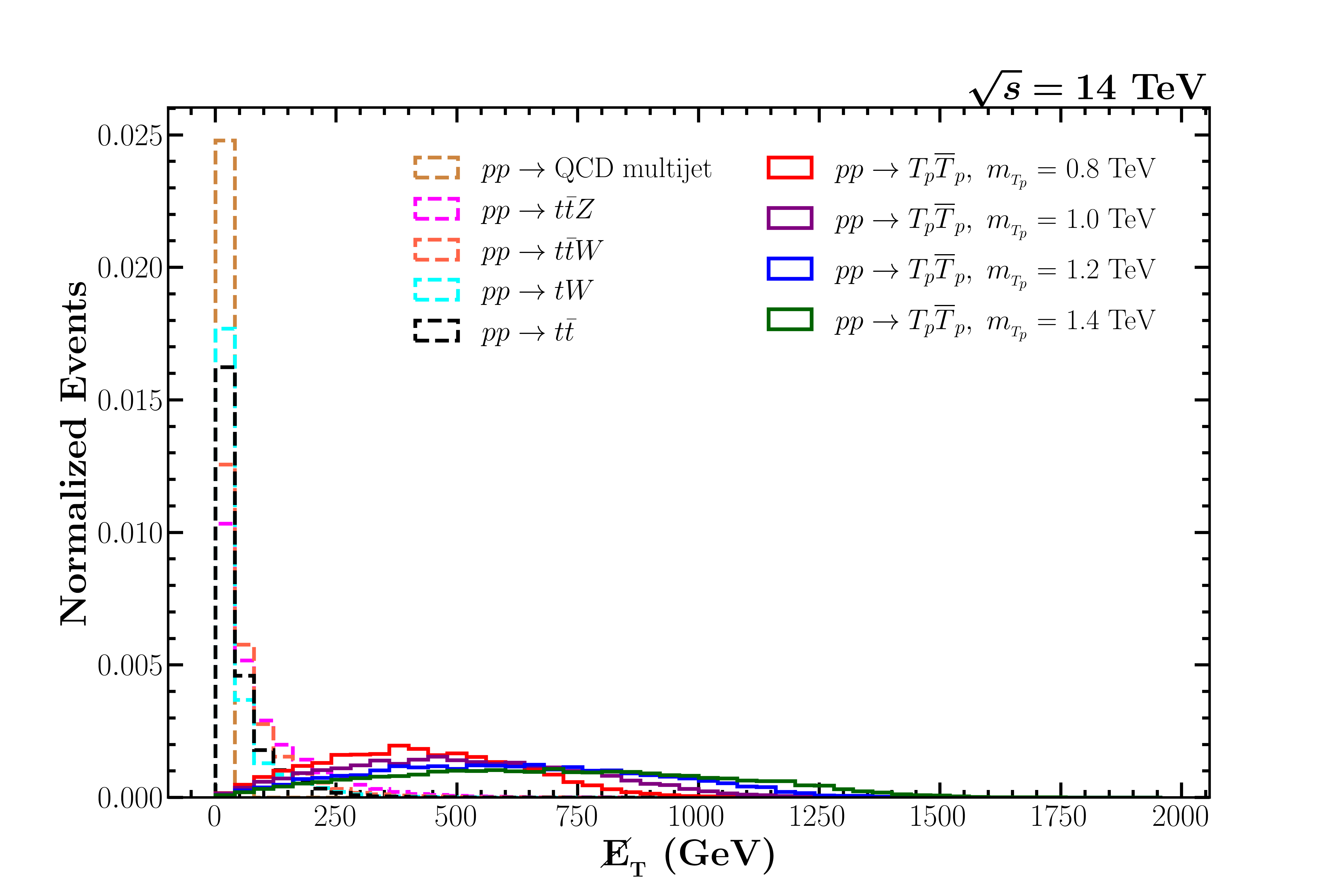}}
\caption{ {\scriptsize Distribution of missing transverse energy ($\cancel{E}_{T}$) variable for signal (top partner pair production) and background processes at 13 and 14 TeV center of mass energies.}}\label{fig:tp_met}
\end{figure}

\item Stransverse mass ($\mathbf{M_{T2}}$): When a heavy particle is pair produced and subsequently decays to a mode which contains both
visible as well as a invisible particles in such a situation the final state contains at least two invisible particles coming from both end of the 
two decay chains (see Fig.~\ref{fig:schematic_tp}). The transverse mass variable ($M_{T}$) defined as   

\bea\label{eq:mt}
M_{T} &=& \sqrt{\left(E_{T, {\rm vis}}+\cancel{E}_{T}\right)^{2}-\left(\vec{p}_{_{T, {\rm vis}}}+\vec{\cancel{p}}_{_T}\right)^{2}}
\eea

is not useful since it will be difficult to reconstruct the missing momentum carried by the individual invisible particle.
In this case, one can then make use of the stransverse mass variable \cite{mt2} defined as

\bea
M_{T2}^{2}&=& \min_{\vec{\cancel{p}}_{_1}+\vec{\cancel{p}}_{_2} = \vec{\cancel{p}}_{_T}} \{ \max\left[M_{T}^{2}(\vec{p}_{_{1,{\rm vis}}}, \vec{\cancel{p}}_{_1})+M_{T}^{2}(\vec{p}_{_{2,{\rm vis}}}, \vec{\cancel{p}}_{_2})\right]\} \leq m_{_{T_{p}}}^{2}
\eea 
where $\vec{p}_{_{1,{\rm vis}}}$ and $\vec{p}_{_{2,{\rm vis}}}$ represent the momenta of the reconstructed top-jets originated
in the two decay chains as shown in Fig.~\ref{fig:schematic_tp}. The corresponding $M_{T2}$ distributions
for signal benchmark points and backgrounds are shown in Fig.~\ref{fig:tp_mt2}.

\begin{figure}[H]
\centering
\subfloat[\label{fig:tp_mt2_dis_13tev}]{\includegraphics[width=0.5\columnwidth]{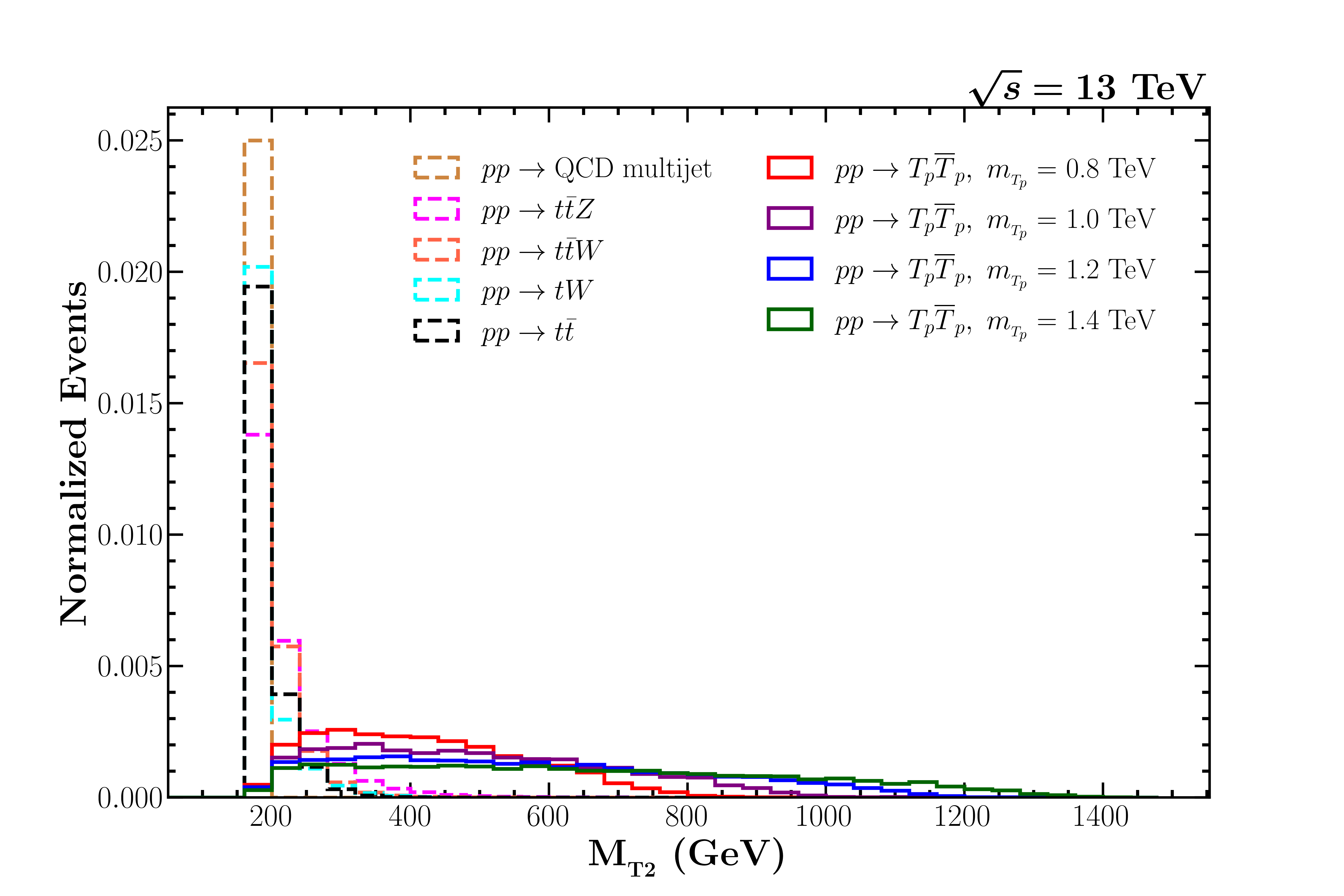}}
\subfloat[\label{fig:tp_mt2_dis_14tev}]{\includegraphics[width=0.5\columnwidth]{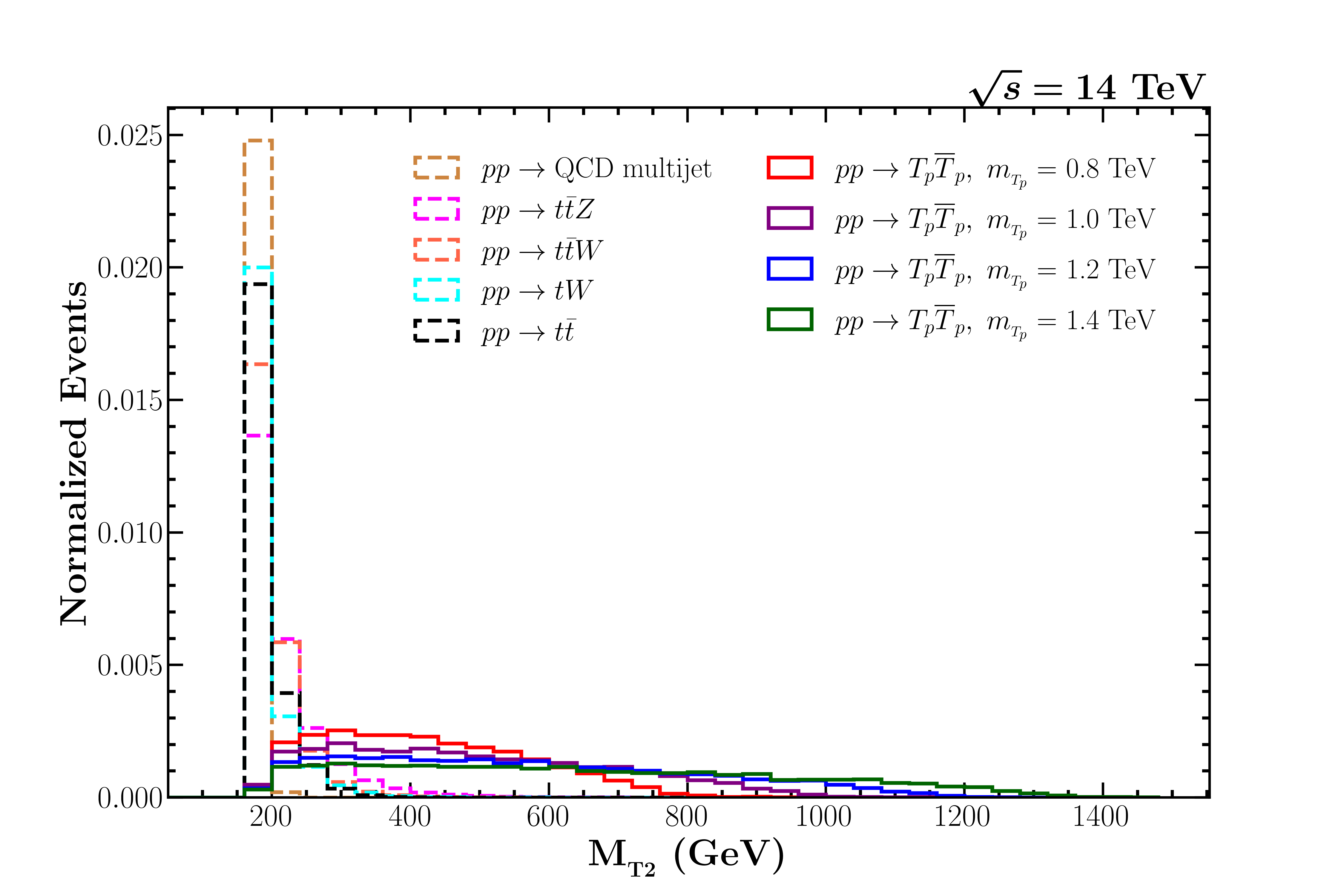}}
\caption{ {\scriptsize Distribution of stransverse mass ($M_{T2}$) variable for signal (top partner pair production) and background processes at 13 and 14 TeV center of mass energies.}}\label{fig:tp_mt2}
\end{figure}

\end{itemize}

All these variables play a crucial role in separating the signal from the SM backgrounds quite efficiently
as evident from the respective distributions. The optimized event selection criteria\footnote{The choice of benchmark point dependent cuts can be justified by the fact that the endpoint of stransverse mass distribution contains the information about the mass of the top partner.} are summarized in Table.~\ref{table:tpcutsval} 
after comparing the kinematic distributions for various signal benchmark points and the corresponding SM backgrounds. 

\begin{table}[H]
\centering
\resizebox{0.8\columnwidth}{!}{
\begin{tabu}{ccc}
\hline
\hline
Minimum values&13 TeV& 14 TeV\\
\hline 
\hline
\multirow{2}{*}{$M_{T2_{0}}$}&$467\left(m_{T_{p}} < 1000\right)$&\multirow{2}{*}{$472$}\\
&$523\left(m_{T_{p}} \geq 1000\right)$&\\

$p_{_{T_{0}}}$&$300 $& $300$ \\
 
\multirow{2}{*}{$\cancel{E}_{T_{0}}$}&$ 463 ~\left(m_{_{T_{p}}} < 1000\right)$& $ 0.674\times m_{_{T_{p}}}  ~\left(m_{_{T_{p}}} < 1200\right)$ \\

& $663 \left(m_{_{T_{p}}}\geq 1000\right) $& $708~ \left(m_{_{T_{p}}} \geq 1200\right)$\\
\hline
\hline
\end{tabu}
}
\caption{{ \scriptsize Minimum values of the observables $M_{T2},~ p_{_T}$ and $\cancel{E}_{T}$ in the context of $pp\to T_{p}\overline{T}_{p}$ analysis at $\sqrt{s}=13
 ~{\rm and}~ 14$ TeV.}}
\label{table:tpcutsval}
\end{table}

\subsection{Single production of top partner}\label{subsec:st@lhc}

\begin{figure}[H]
\begin{center}

\subfloat[\label{fig:feynmandiag_st_a}]{\includegraphics[width=0.3\columnwidth]{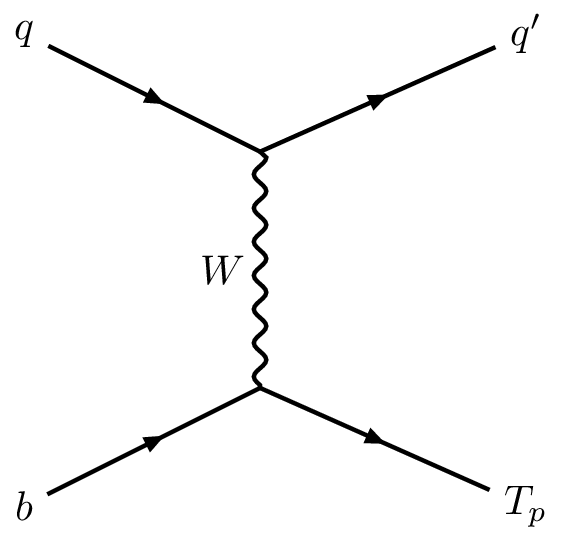}}~
\subfloat[\label{fig:feynmandiag_st_b}]{\includegraphics[width=0.3\columnwidth]{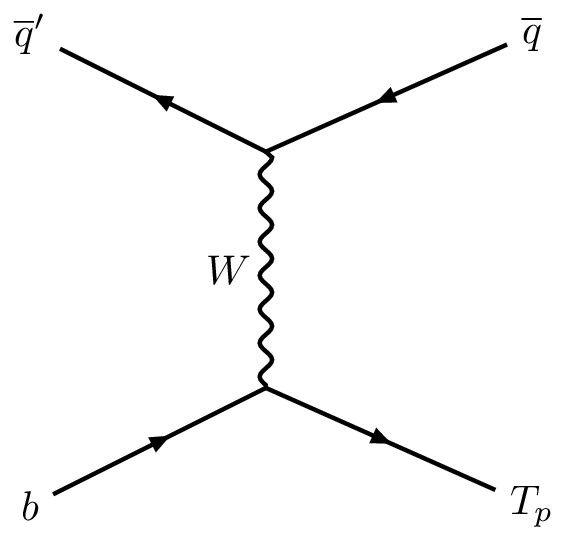}}

\end{center}
\caption{{ \scriptsize Feynman Diagrams for the single production of top partner at the LHC.}}\label{fig:fd_st}
\end{figure}

\begin{figure}[H]
    \centering
    \includegraphics[width=0.6\columnwidth]{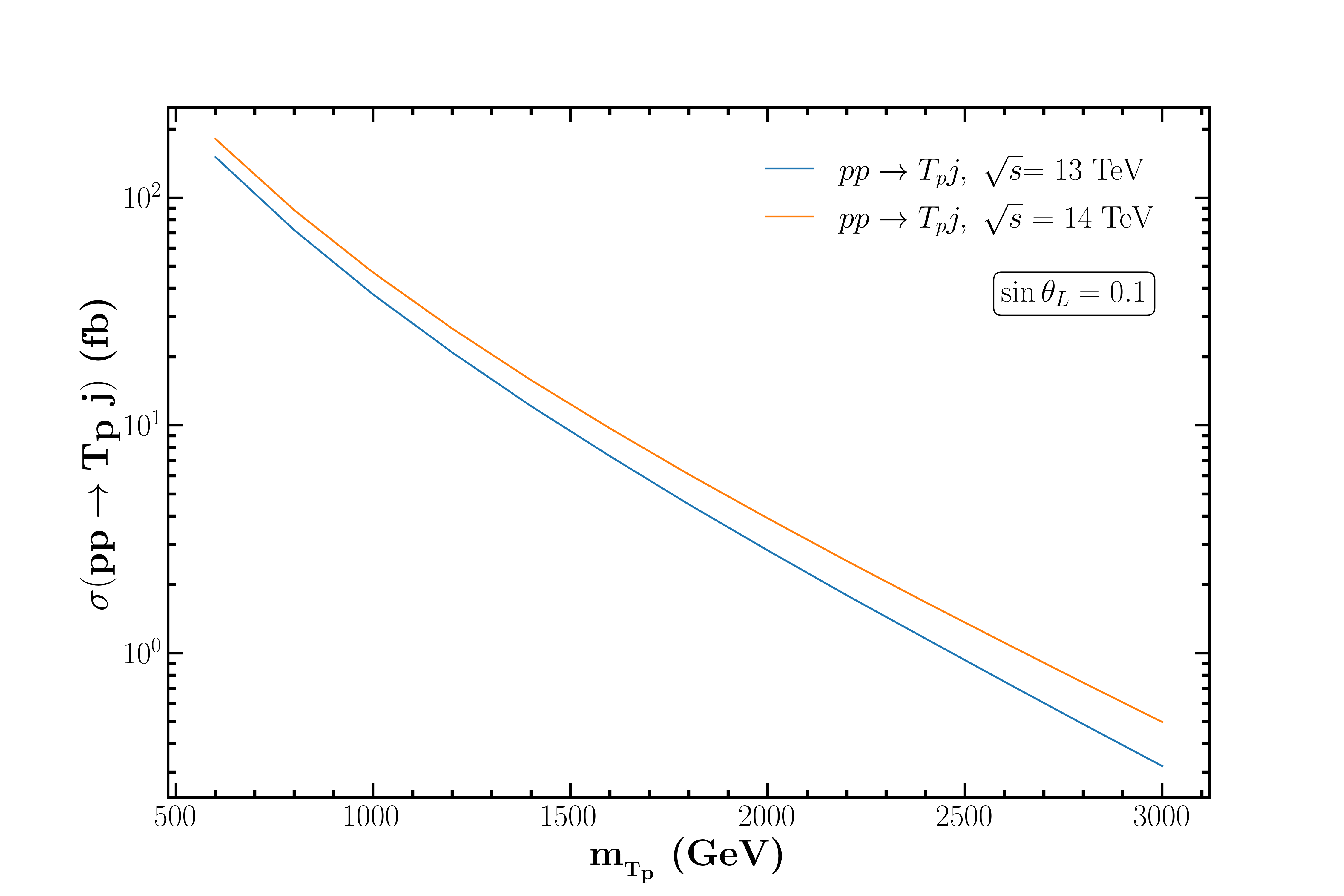}
    \caption{{\scriptsize The cross sections for $pp\to T_{p}j$ process at $\sqrt{s} = 13~  {\rm and} ~14$ TeV respectively, as a function of the top partner mass. {[$\alpha_{_{EM}}^{\rm -1} = 137.036$ \cite{PDG:2020ssz} is used to estimate the above cross section.]} }}
    \label{fig:st_cross_13_14}
\end{figure}

The single production of top partner in association with a light-quark jet occurs via $t$-channel $W$
boson exchange and requires the presence of $b$-quark Parton Distribution Function (PDF) in the 
five flavor PDF scheme \cite{Maltoni:2012pa}. The relevant Feynman diagrams for this process are shown in Fig.~\ref{fig:fd_st}. Since the top partner is $SU(2)_L$ singlet, the single production
 cross section depends on the mixing of the top-quark with the top partner, parametrized by the mixing angle $\sin \theta_L$. 
 The $T_{p}-b-W$ coupling being proportional to the mixing angle $\sin \theta_L$, the single top partner production cross section is proportional to $\sin^2 \theta_L$. 
 Fig.~\ref{fig:st_cross_13_14} depicts the $\sigma(pp \to T_p j )$ as a function of the top partner mass ($m_{_{T_P}}$) at 13 and 14 TeV LHC center of mass energies assuming $\sin \theta_L = 0.1$. In this case the cross sections are quoted at leading order (LO) using M{\scriptsize AD}G{\scriptsize RAPH}5\_aMC@NLO.

\begin{figure}[H]
\centering
{\includegraphics[width=0.5\columnwidth]{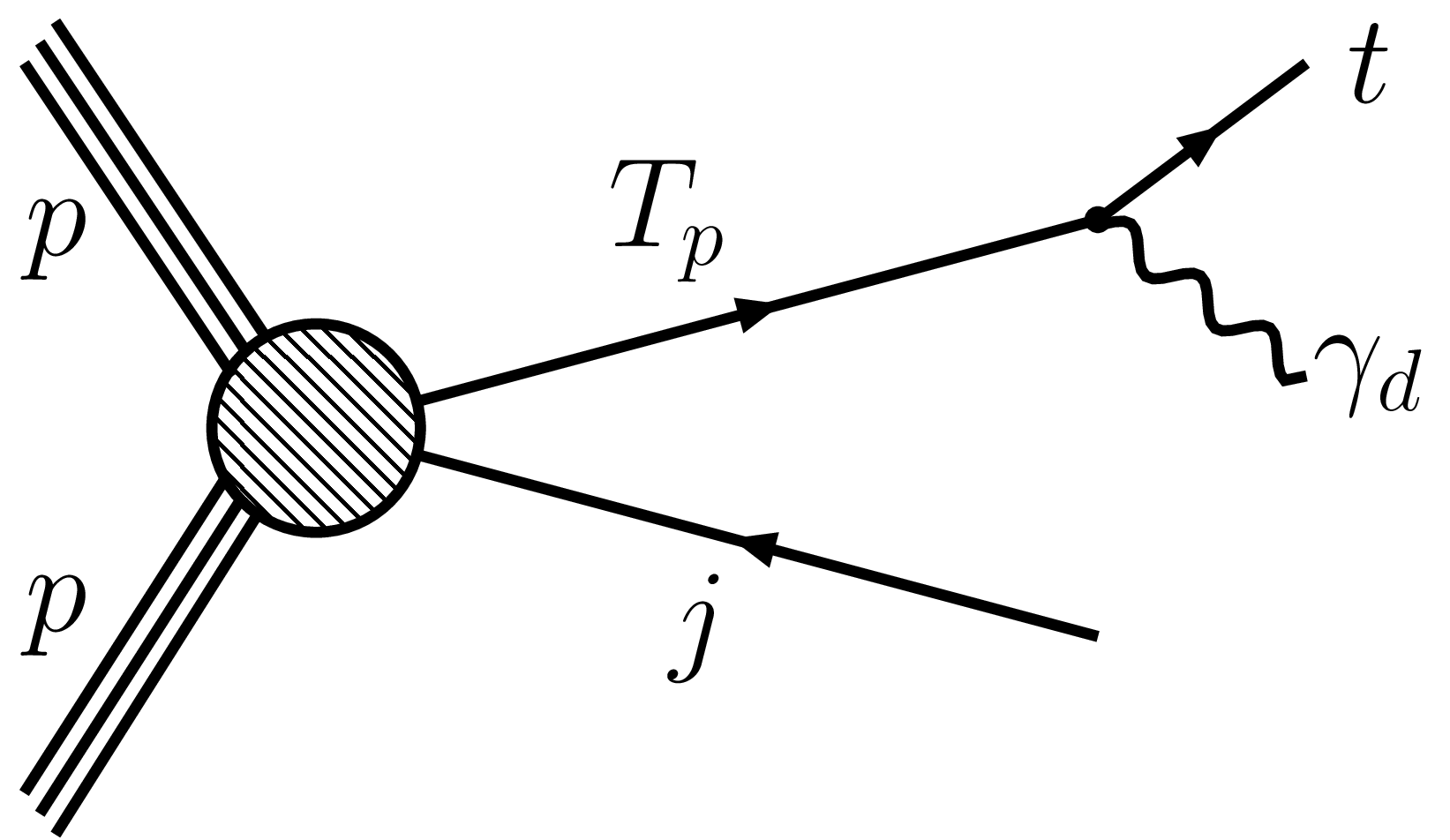}}

\caption{{\scriptsize Schematic diagram for the process $pp \to T_{p} j \to (T_{p} \to t\gamma_d) j $.}}
\label{fig:schematic_st}
\end{figure}

The top partner produced in $pp$ collision further decays into a top-quark and a dark photon ($T_p \to t \gamma_d$). Thus giving rise to $t+\cancel{E}_T+jet$ final state. 
Fig.~\ref{fig:schematic_st} represents a schematic diagram for $pp \to T_p j \to t \gamma_d j$ process. In this analysis the fully hadronic decay of the top-quark 
has been considered for the same reason as that in the previous section. We also require the final state to have \textit{at least two} central jets with $p_T>20$ GeV, $|\eta|<2.5$ 
and no isolated leptons.

{The relevant SM backgrounds in this case are $t\bar{t}$, $t\bar{t}V$ (where $V=Z/W$), $tW$, $t j$ and QCD multijet processes.} The dominant contribution comes from the $t\bar{t}$ process and to some extent QCD multijet process given their overwhelming production rate at the LHC.
The process $pp \to t Z j$ with $Z\to \bar{\nu}\nu$ constitutes a irreducible background, however, we have estimated its contribution to be negligible and hence, not considered in the following discussion. In Table.~\ref{table:stbckcross} we list these background  processes along with their cross sections.

\begin{table}[H]
\centering
\resizebox{0.6\columnwidth}{!}{\begin{tabu}{lcc}
\hline
\hline
\multirow{2}{*}{Process}& \multicolumn{2}{c}{Cross section (fb)} \\
\cline{2-3}
&13 TeV& 14 TeV\\
\hline
\hline
$pp \to t\bar{t}$&  $8.39\times 10^{5}$ ~ \cite{Kidonakis:2022hfa} &$9.9\times 10^{5}$  ~\cite{Kidonakis:2022hfa}\\

$pp \to t\bar{t}Z$ & $8.63\times 10^{2}$ ~\cite{Kulesza:2018tqz} & $1.045\times 10^{3}$ ~\cite{Kulesza:2018tqz}\\

$pp \to t\bar{t}W$& $5.66\times 10^{2}$ ~\cite{Kulesza:2018tqz}& $6.53\times 10^{2}$ ~\cite{Kulesza:2018tqz}\\

$pp \to tW$ & $7.95\times 10^{4}$ ~\cite{Kidonakis:2021vob}&$9.4\times 10^{4}$  ~\cite{Kidonakis:2021vob}\\

$pp \to tj$ & $1.579\times 10^{5}$ ~\cite{Kidonakis:2021vob}&$1.797\times 10^{5}$  ~\cite{Kidonakis:2021vob}\\

$pp \to $ QCD multijet& $1.96\times 10^{11}$& $2.16\times 10^{11}$	\\
\hline
\hline
\end{tabu}
}
\caption{{\scriptsize cross sections of the background processes considered for the analysis of single top partner production at $\sqrt{s}=13  ~{\rm and}~ 14$ TeV.}}
\label{table:stbckcross}
\end{table}

The cross sections quoted in Table.~\ref{table:stbckcross} are beyond leading order\footnote{In this analysis the $t\bar{t}$, $tW$ and $tj$ are considered at aN$^{3}$LO in QCD.} except for the QCD multijet background which is at leading order. 

\subsection*{Event selection criteria}

In the context of collider simulation of the single top partner production in $pp$ collision {we require the final state to have {\it exactly one} boosted top quark jet.}
Furthermore, we again propose certain kinematical variables that can discriminate between 
the signal and the SM backgrounds. Some of these variables are already defined in the context of pair production of top partner in $pp$ collision (see Sec. \ref{subsec:tp@lhc})
such as transverse momentum ($p_{_{T}}$) of the reconstructed top and missing transverse energy ($\cancel{E}_T$) will be useful in this analysis as well. 

However, instead of 
the stranverse mass variable used in the previous analysis we will use the transverse mass variable ($M_T$) itself as the final state in this case contain only one invisible particle barring the neutrinos from the decay of the top-quarks.

 We also use an additional forward jet rapidity cut ($|\eta| > 2.1$) where the forward jet is defined as the maximum rapidity jet {which is not tagged as a top quark initiated jet}. Fig.~\ref{fig:st_pt}, \ref{fig:st_met}, and \ref{fig:st_mt} display the distributions of $p_{_{T}}$ of the
top quark, $\cancel{E}_T$, and $M_T$ of the top quark and the invisible system.

\vfill
\begin{itemize}

\item Transverse momentum ($\bsym{p_{_{T}}}$):
\begin{figure}[H]
\centering
\subfloat[\label{fig:st_pt_dis_13tev}]{\includegraphics[width=0.5\columnwidth]{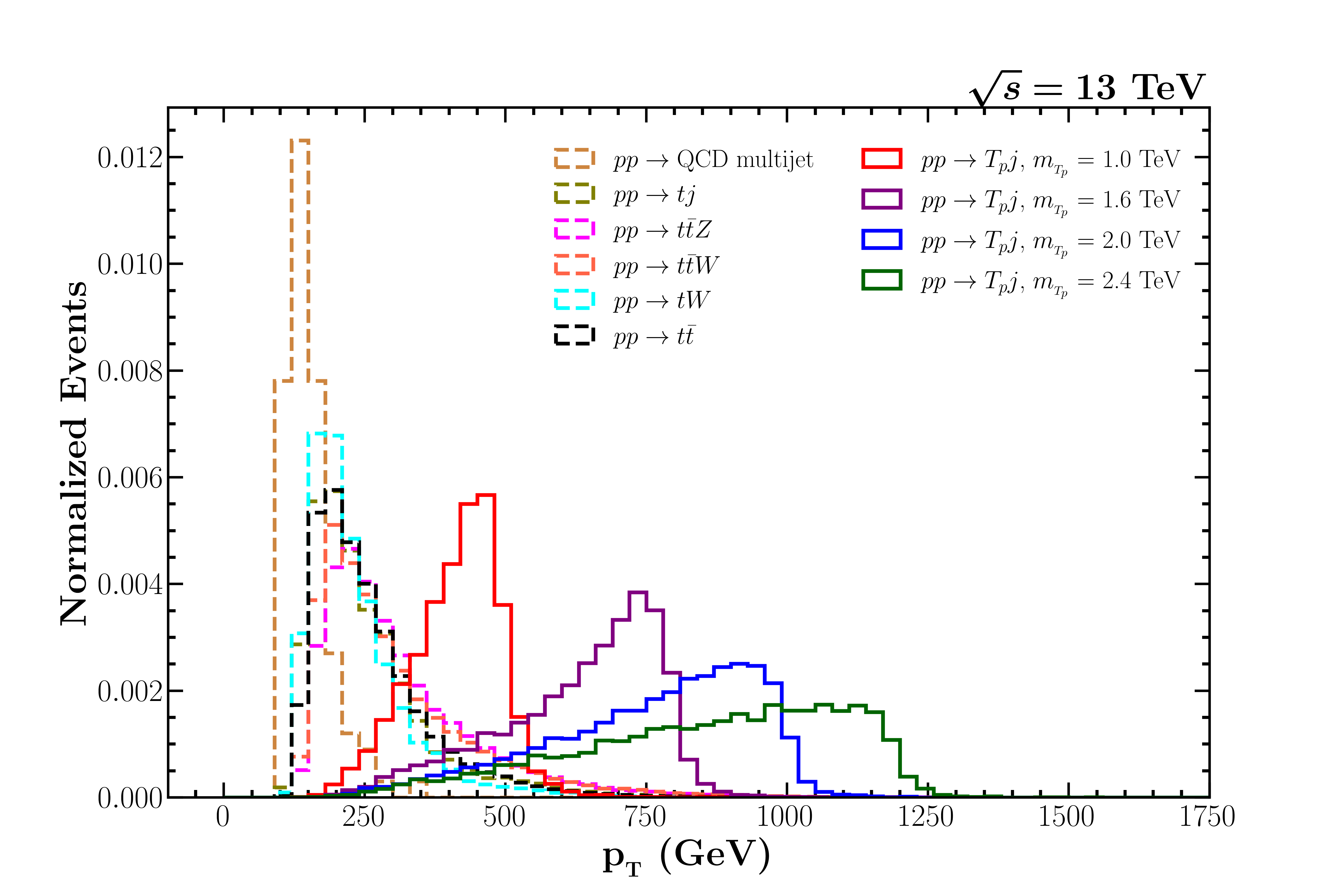}}
\subfloat[\label{fig:st_pt_dis_14tev}]{\includegraphics[width=0.5\columnwidth]{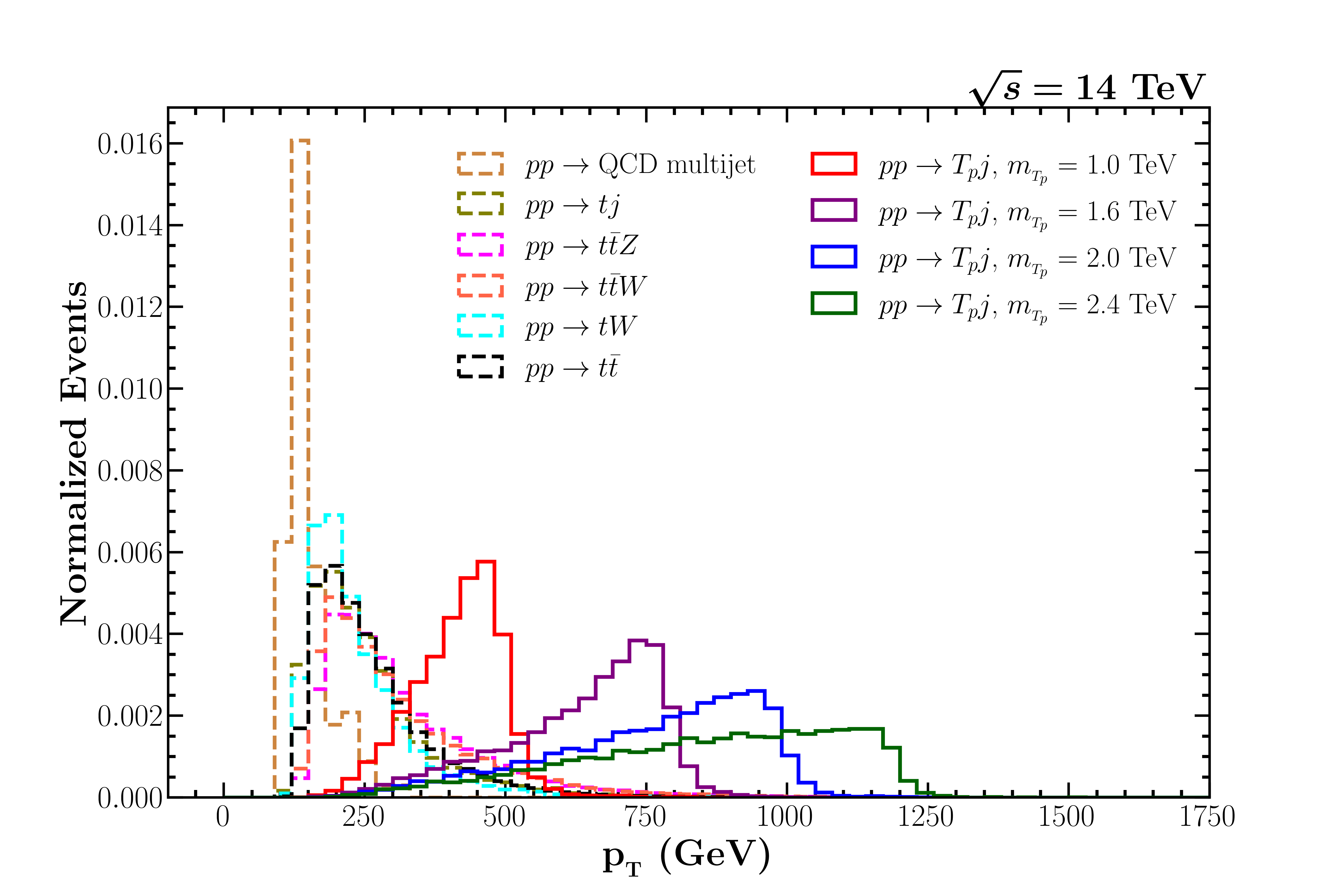}}
\caption{{ \scriptsize Transverse momentum ($p_{_{T}}$) distribution of tagged top quark for signal and background processes at 13 and 14 TeV center of mass energies.}}\label{fig:st_pt}
\end{figure}

\item Missing transverse energy ($\bsym{\cancel{E}_T}$):

\begin{figure}[H]
\centering
\subfloat[\label{fig:st_met_dis_13tev}]{\includegraphics[width=0.5\columnwidth]{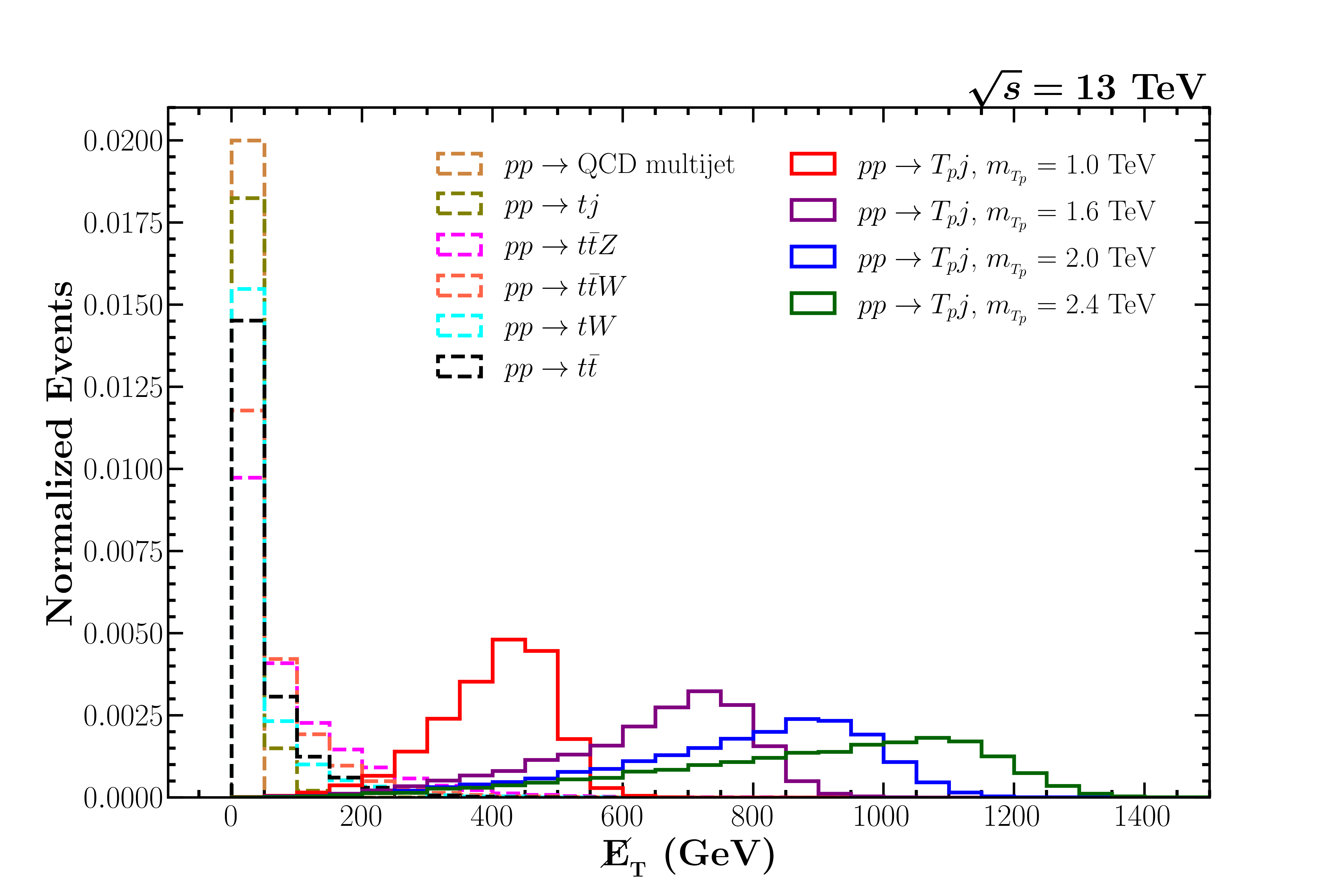}}
\subfloat[\label{fig:st_met_dis_14tev}]{\includegraphics[width=0.5\columnwidth]{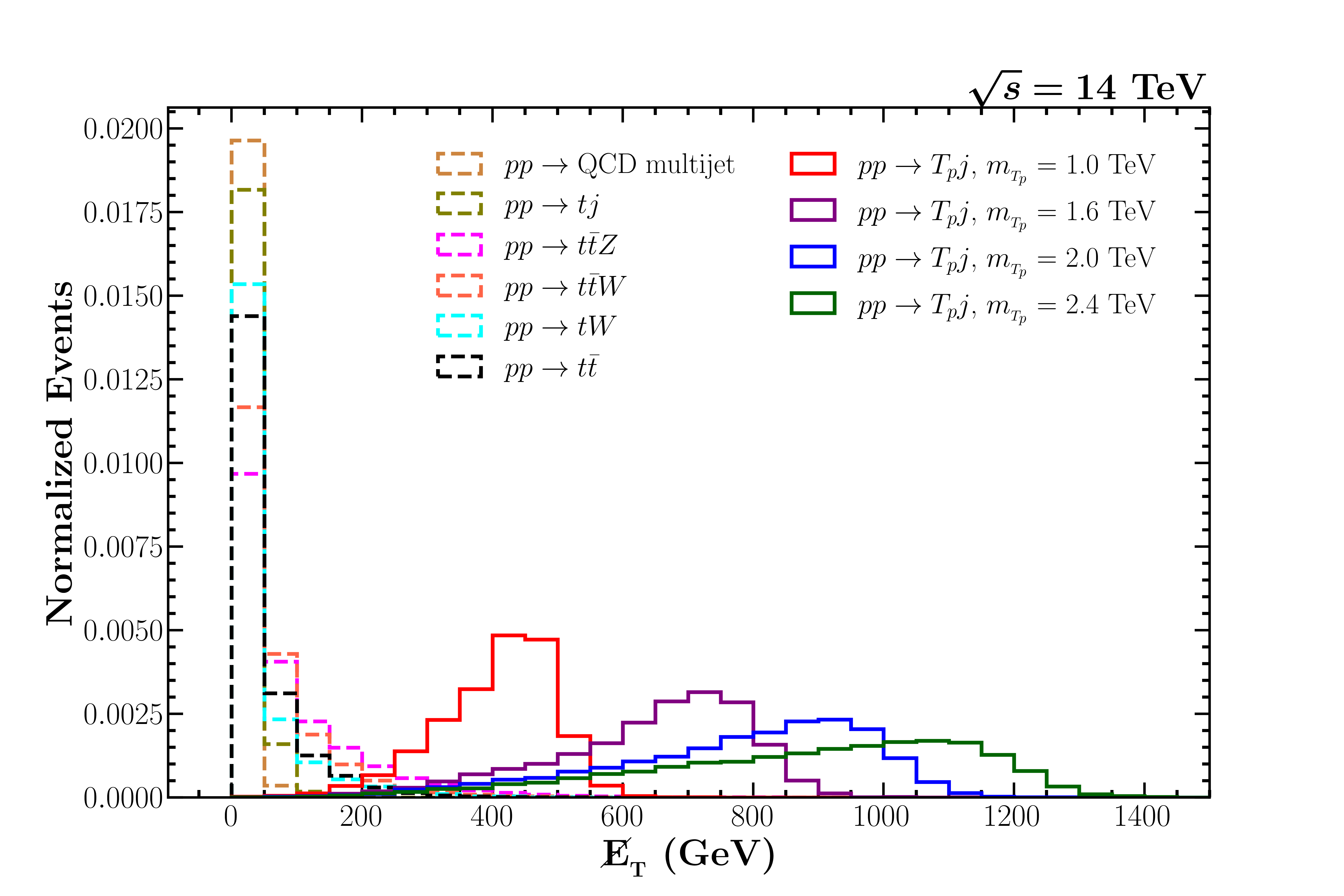}}
\caption{{ \scriptsize Distribution of missing transverse energy ($\cancel{E}_T$) variable for signal (single top partner production) and background processes at 13 and 14 TeV center of mass energies.}}\label{fig:st_met}
\end{figure}

\item Transverse mass ($\mathbf{M_{T}}$):

The transverse mass $M_{T}$ of the reconstructed top quark jet and the missing momentum system following the definition in Eq.~(\ref{eq:mt}) is given by,

\bea
M_{T} &=& \sqrt{\left(E_{T,t}+\cancel{E}_{T}\right)^{2}-\left(\vec{p}_{_{T},t}+\vec{\cancel{p}}_{_T}\right)^{2}} \leq m_{_{T_{p}}}
\eea
where $E_{T,t}$ and $\vec{p}_{_{T},t}$ represent the reconstructed energy and momentum of the top-jet, respectively.

\begin{figure}[H]
\centering
\subfloat[\label{fig:st_mt_dis_13tev}]{\includegraphics[width=0.5\columnwidth]{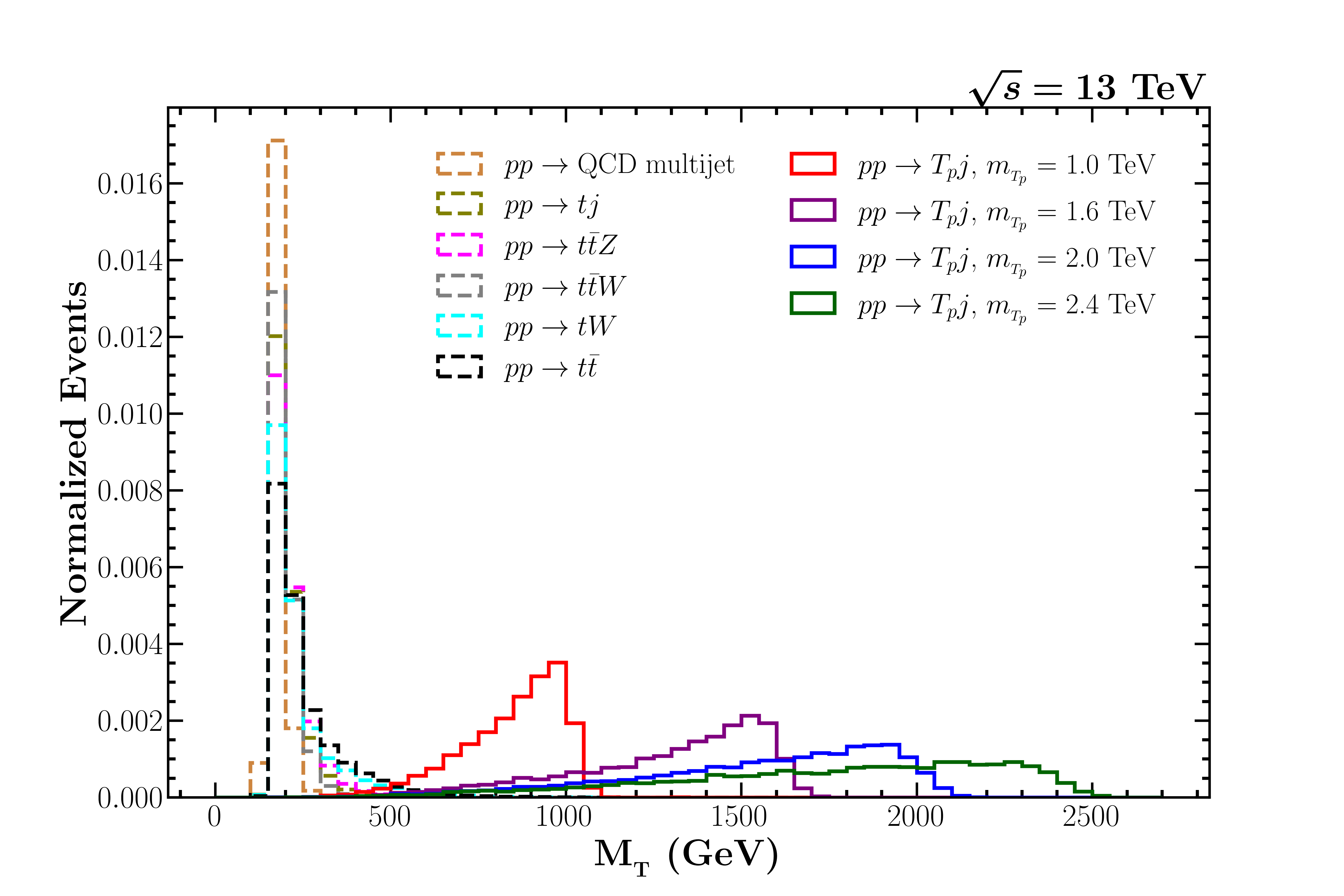}}
\subfloat[\label{fig:st_mt_dis_14tev}]{\includegraphics[width=0.5\columnwidth]{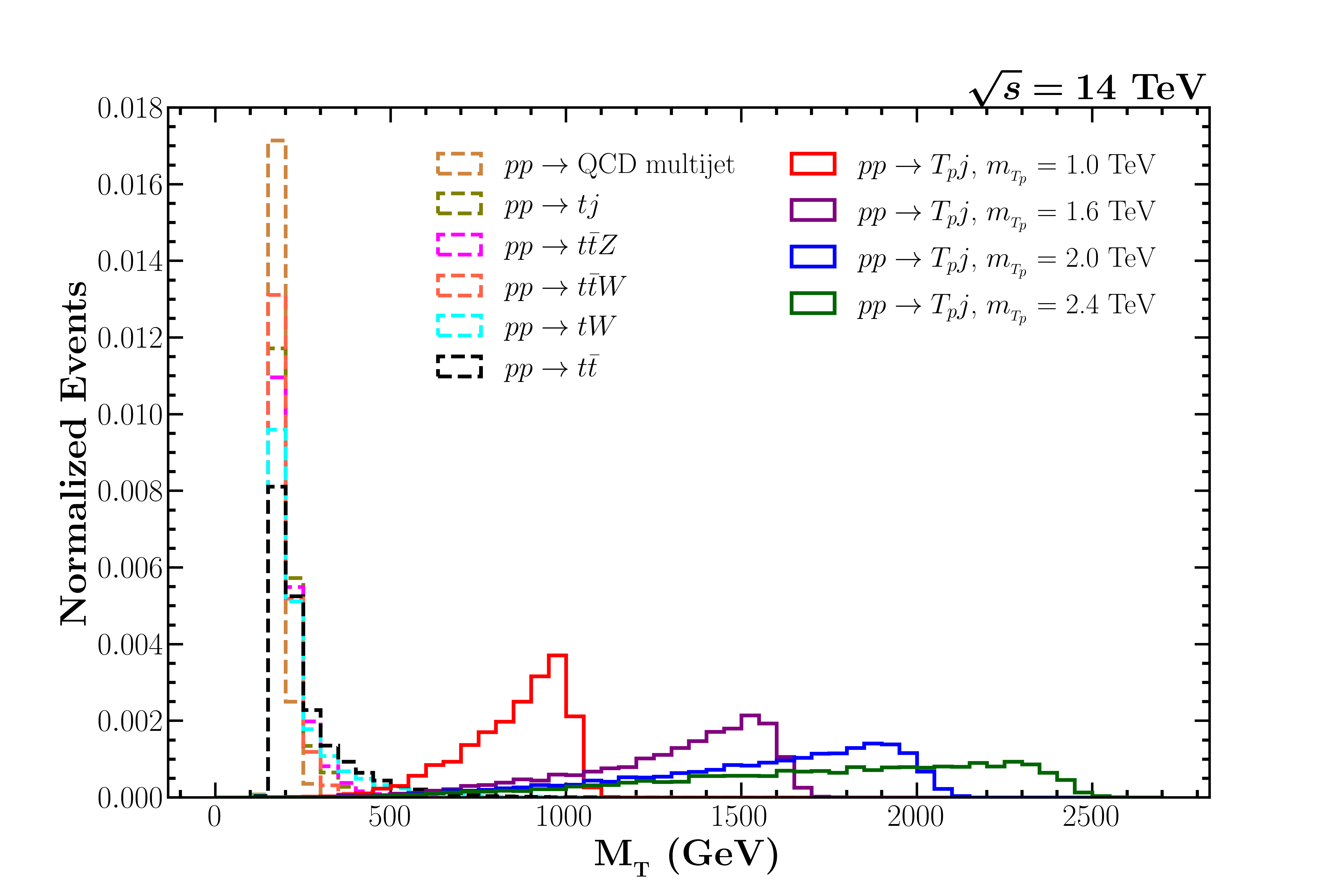}}
\caption{{ \scriptsize Distribution of transverse mass ($M_{T}$) variable for signal (single top partner production) and background processes for 13 and 14 TeV center of mass energies.}}\label{fig:st_mt}
\end{figure}

\end{itemize}

Depending on above distributions, we have imposed the following set of cuts presented in Table.~\ref{table:stcutsval}, in order to optimize the signal significance.

\begin{table}[H]
\centering
\resizebox{0.8\columnwidth}{!}{\begin{tabu}{ccc}
\hline
\hline
Minimum values&13 TeV& 14 TeV\\
\hline 
\hline
$M_{T_{0}}$&$0.5\times m_{_{T_{p}}} $&$0.5\times m_{_{T_{p}}} $\\

$p_{_{T_{0}}}$&$300 $& $300$ \\
 
\multirow{2}{*}{$\cancel{E}_{T_{0}}$}&$ 0.40\times m_{_{T_{p}}}  ~\left(m_{_{T_{p}}} < 1600\right)$& $ 0.42\times m_{_{T_{p}}}  ~\left(m_{_{T_{p}}} < 1800\right)$ \\

& $708~\left(m_{_{T_{p}}} \geq 1600\right) $& $774~ \left(m_{_{T_{p}}} \geq 1800\right)$\\
\hline
\hline
\end{tabu}
}
\caption{{ \scriptsize Minimum values of the observables $M_{T},~ p_{_T}$ and $\cancel{E}_{T}$ in the context of $pp\to T_{p}j$ analysis at $\sqrt{s}=13  ~{\rm and}~ 14$ TeV.}}
\label{table:stcutsval}
\end{table}

\section{Results and discussions}\label{results}

In this section, we present the results of our analysis for both the single and pair production of top partner in $pp$ collisions
at two different LHC center of mass energies (13 and 14 TeV) after implementing the event selection criteria detailed 
in the previous section for both of these production modes.

\subsection{Pair production case}\label{results_tp}

For the top partner pair production process, we consider $5\times 10^{4}$ simulated  $pp\to T_{p}\overline{T}_{p}$ signal events for various choices of the top partner's mass in the range $\{0.8,1,1.2,1.4,1.6\}$ TeV against the $5\times 10^{6}$ simulated $t\bar{t}$ , {and $10^{6}$ simulated $tW,~t\bar{t}Z,~ t\bar{t}W,$ and $1.21\times 10^{6}$ QCD~multijet background events.}
cross sections for each of the background processes are quoted in Table.~\ref{table:tpbckcross} and the values for $M_{T2_{0}},~p_{_{T_{0}}},~ \cancel{E}_{T_{0}} $ are given 
in Table.~\ref{table:tpcutsval}. 

We illustrate the effects of the cut flow on cross sections for both the signal benchmark points and the estimated total SM background in Table.~\ref{table:tpp13tev} and \ref{table:tpp14tev}. The estimated signal significance (defined as ${S}/{\sqrt{S+B}}$, where $S$ and $B$ are the number of signal and background events after applying the hard cuts) at 13  (14) TeV LHC collision energy assuming an integrated luminosity of 139 (300) fb$^{-1}$ is also mentioned in Table.~\ref{table:tpp13tev} (\ref{table:tpp14tev}).

\begin{table}[H]
\centering
\resizebox{0.9\columnwidth}{!}{
\begin{tabu}{lccccccc}
\hline
\hline
\rowfont \normalfont
$m_{_{T_{p}}}$ &Production & Basic &  Boosted top  & $M_{T2}$ cut &$p_{_{T}}$ cut & $\cancel{E}_T$ cut  & Significance \\
\rowfont \normalfont
(TeV)&cross section& cuts& requirement& $M_{T2}>M_{T2_{0}}$&$p_{_{T}}>p_{_{T_{0}}}$&$\cancel{E}_T>\cancel{E}_{T_{0}}$& $\left( \frac{S}{\sqrt{S+B}} \right)$ \\
\rowfont \normalfont
&(fb)& (fb)& (fb)&(fb)&(fb)&(fb)& \\

\hline
\hline

\multirow{2}{*}{0.8}&  {190.0} &  {124.24} & { 31.71} &  {11.35 }&  {11.03 }&   {9.86} &\multirow{2}{*}{16.6 ({\color{blue}32.9})}\\ 
\rowfont{\color{red}}
& $1.96\times 10^{11}$ &  $7.42\times 10^{10}$ &  $1.93\times 10^{7}$ &  $6.14$ &  $5.87$ &  $2.61$ &  \\

\multirow{2}{*}{1.0}& {42.70} &  {28.69} &   {8.45} &  {3.45} &  {3.42 }&   {2.15} &\multirow{2}{*}{8.4 ({\color{blue}16.5})}\\ 
\rowfont{\color{red}}
& $1.96\times 10^{11}$ &  $7.42\times 10^{10}$ &  $1.93\times 10^{7}$ &  $2.48$ &  $2.47$ &  $0.22$ &  \\

\multirow{2}{*}{1.2}& {11.40} &  {7.93} &  {2.47} &  {1.28} &  {1.27} &  {0.96} &\multirow{2}{*}{4.7 ({\color{blue}10.4})}\\ 
\rowfont{\color{red}}
& $1.96\times 10^{11}$ &  $7.42\times 10^{10}$ &  $1.93\times 10^{7}$ &  $2.48$ &  $2.47$ &  $0.22$ &  \\

\multirow{2}{*}{1.4}& {3.42 }&  {2.44} &  {0.82} &  {0.49} &  {0.49} & { 0.40} &\multirow{2}{*}{2.3 ({\color{blue}6.0})}\\ 
\rowfont{\color{red}}
& $1.96\times 10^{11}$ &  $7.42\times 10^{10}$ &  $1.93\times 10^{7}$ &  $2.48$ &  $2.47$ &  $0.22$ &  \\

\multirow{2}{*}{1.6}& {1.11} &  {0.82} &  {0.32} &  {0.20} &  {0.20} &  {0.17} &\multirow{2}{*}{1.1 ({\color{blue}3.3})}\\ 
\rowfont{\color{red}}
& $1.96\times 10^{10}$ &  $7.42\times 10^{10}$ &  $1.93\times 10^{7}$ &  $2.48$ &  $2.47$ &  $0.22$ &  \\

\hline
\hline
\end{tabu}
}
\caption{{ \scriptsize Columns [2-7] represent the effects of cut flow on cross sections for both the signal ({\bf black}) and the total SM background ({\color{red}red}) at $\sqrt{s}=13$ TeV. $\sin \theta_{L} = 0.1$ is assumed to estimate the branching ratios for the various benchmark choices of $m_{_{ T_{P}}}$. {The cross sections presented in column [2-7] assume  BR$(T_{p}\to t \gamma_{d}) = 100\%$. The estimated signal significance is quoted in the last column assuming \textit{actual} BR ({\color{blue}100\% BR})  with an integrated luminosity of 139 fb$^{-1}$. [$v_d = 200$ GeV, $m_{\gamma_d}=10$ GeV and $m_{h_d}= 400$ GeV are assumed to obtain the above results.]} } }
\label{table:tpp13tev}
\end{table}


\begin{table}[H]
\centering
\resizebox{0.9\columnwidth}{!}{
\begin{tabu}{lccccccc}
\hline
\hline
$m_{_{T_{p}}}$ &Production & Basic &  Boosted top  & $M_{T2}$ cut &$p_{_{T}}$ cut & $\cancel{E}_T$ cut  & Significance \\
(TeV)&cross section& cuts& requirement& $M_{T2}>M_{T2_{0}}$&$p_{_{T}}>p_{_{T_{0}}}$&$\cancel{E}_T>\cancel{E}_{T_{0}}$& $\left( \frac{S}{\sqrt{S+B}} \right)$ \\
\rowfont \normalfont
&(fb)& (fb)& (fb)&(fb)&(fb)&(fb)& \\

\hline
\hline

\multirow{2}{*}{0.8}& {249.61} &   {162.60} &   {41.52} &  {14.71} &  {14.31} &  {10.06} &\multirow{2}{*}{26.2 ({\color{blue}50.3})}\\ 
\rowfont{\color{red}}
& $2.16\times 10^{11}$ &  $8.15\times 10^{10}$ &  $2.14\times 10^{7}$ & $8.55$&$8.04$ &  $1.94$ &  \\

\multirow{2}{*}{1.0}& {58.16} &  {38.91} &  {11.28 }& { 5.50} &  {5.44} &  {2.90} &\multirow{2}{*}{13.0 ({\color{blue}27.0})}\\ 
\rowfont{\color{red}}
& $2.16\times 10^{11}$ &  $8.15\times 10^{10}$ &  $2.14\times 10^{7}$ &   $8.55$ &  $8.04$ &  $0.57$ &  \\

\multirow{2}{*}{1.2}& {16.15} &  {11.15} &  {3.51} &  {2.07} &  {2.06} &  {1.25} &\multirow{2}{*}{7.5 ({\color{blue}17.1})}\\ 
\rowfont{\color{red}}
& $2.16\times 10^{11}$ &  $8.15\times 10^{10}$ &  $2.14\times 10^{7}$ &   $8.55$ &  $8.04$ &  $0.36$ &  \\

\multirow{2}{*}{1.4}&  {5.06} & { 3.62} &  {1.21} & { 0.80} & { 0.80} & { 0.57} &\multirow{2}{*}{3.8 ({\color{blue}10.2})}\\ 
\rowfont{\color{red}}
& $2.16\times 10^{11}$ &  $8.15\times 10^{10}$ &  $2.14\times 10^{7}$ &   $8.55$ &  $8.04$ &  $0.36$ &  \\

\multirow{2}{*}{1.6}&  {1.72} &  {1.25} &  {0.49} &  {0.33} &  {0.33} &  {0.26} &\multirow{2}{*}{1.9 ({\color{blue}5.7})}\\ 
\rowfont{\color{red}}
& $2.16\times 10^{11}$ &  $8.15\times 10^{10}$ &  $2.14\times 10^{7}$ &   $8.55$ &  $8.04$ &  $0.36$ &  \\

\hline
\hline
\end{tabu}
}
\caption{{ \scriptsize Columns [2-7] represent the effects of cut flow on cross sections for both the signal ({\bf black}) and the total SM background ({\color{red}red}) at $\sqrt{s}=14$ TeV.  $\sin \theta_{L} = 0.1$ is assumed to estimate the branching ratios for various benchmark choices of $m_{_{ T_{P}}}$.  {The cross sections presented in column [2-7] assume  BR$(T_{p}\to t \gamma_{d}) = 100\%$. The estimated signal significance is quoted in the last column assuming \textit{actual} BR ({\color{blue}100\% BR})  with an integrated luminosity of 300 fb$^{-1}$. [$v_d = 200$ GeV, $m_{\gamma_d}=10$ GeV and $m_{h_d}= 400$ GeV are assumed to obtain the above results.]} }}
\label{table:tpp14tev}
\end{table}

\begin{figure}[H]
\centering
\subfloat[\label{fig:tp_brxsigma_mtp_13tev}]{\includegraphics[width=0.5\columnwidth]{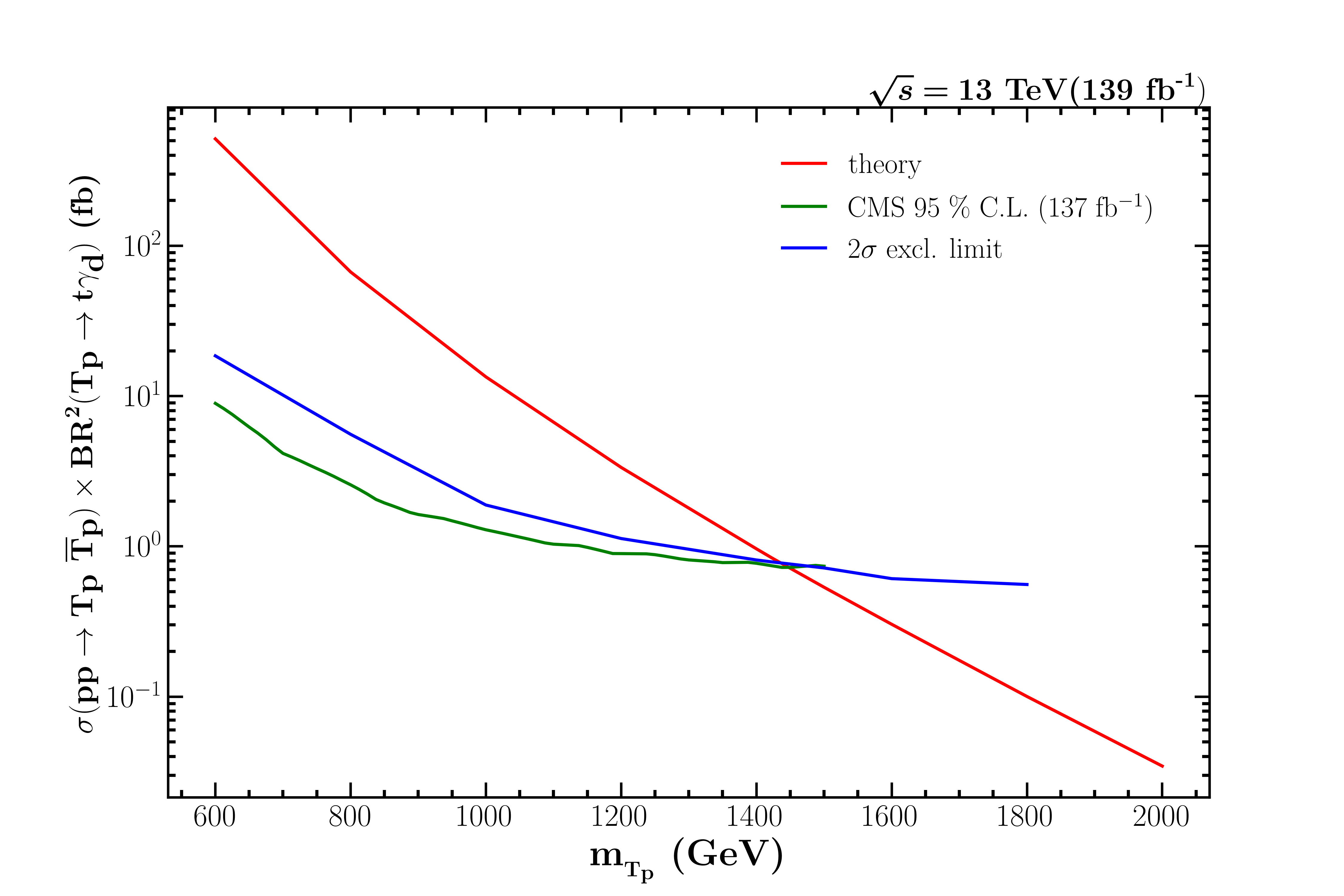}}
\subfloat[\label{fig:tp_brxsigma_mtp_14tev}]{\includegraphics[width=0.5\columnwidth]{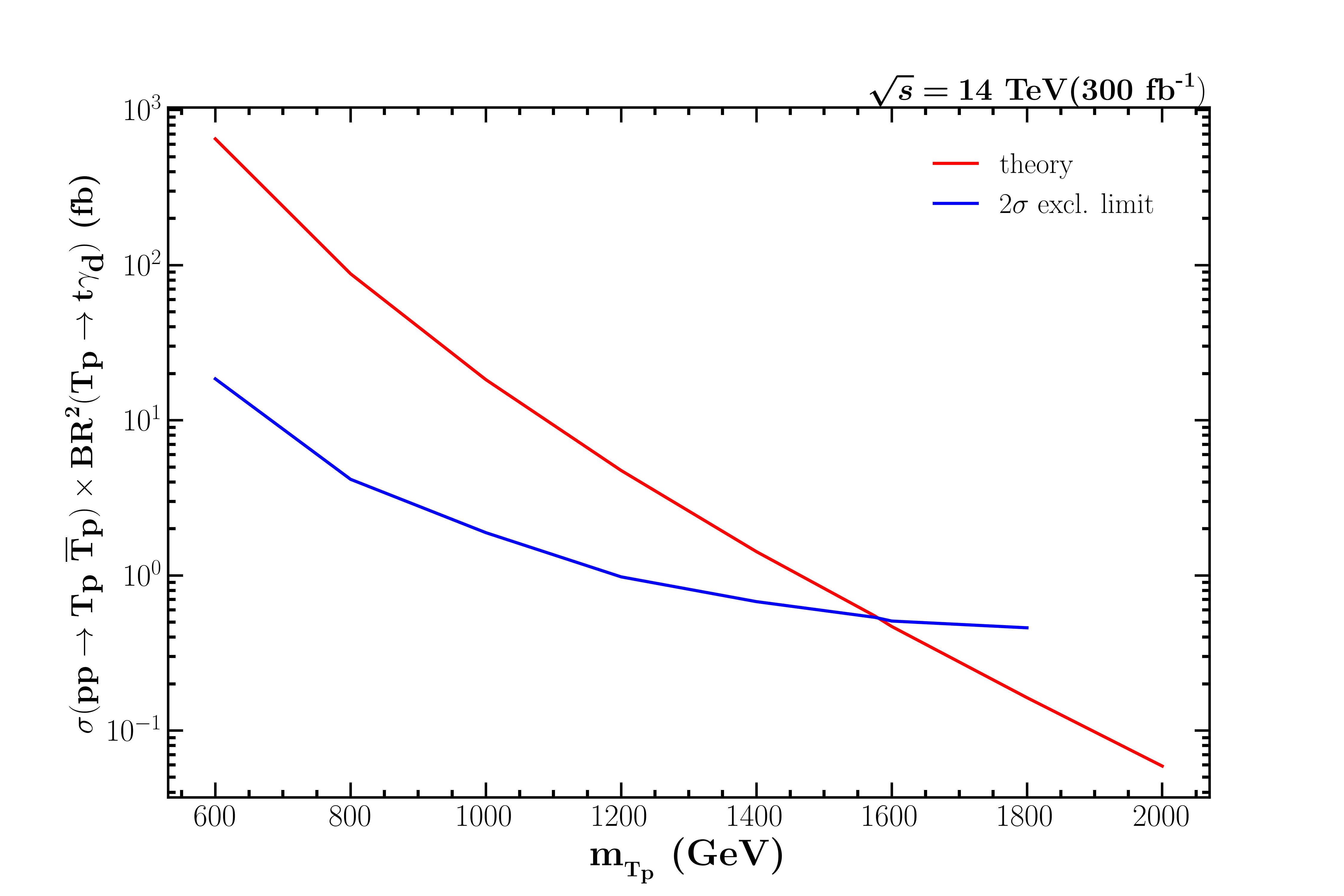}}
\caption{{\scriptsize $2\sigma$ exclusion limit on the $\sigma(pp\to T_{p}\overline{T}_{p}) \times {\rm BR}^2(T_{p}\to t\gamma_{d})$ as a function of $m_{_{T_{p}}}$ at (a) 13 TeV (139 fb$^{-1}$) and (b) 14 TeV (300 fb$^{-1}$). {[$\sin\theta_L = 0.1$, $v_d = 200$ GeV, $m_{\gamma_d}=10$ GeV and $m_{h_d}= 400$ GeV are assumed to obtain the above results.]} }}\label{fig:tp_excl}
\end{figure}

One can see from Table.~\ref{table:tpp14tev} that at 14 TeV LHC center of mass energy with 300 fb$^{-1}$ integrated luminosity it is possible 
to exclude $m_{_{T_{p}}} \leq 1.6$ TeV at $2\sigma$ significance level. The expected $2\sigma$ exclusion limits on the 
$\sigma(pp\to T_{p}\overline{T}_{p})\times {\rm BR}^2(T_p \to t \gamma_d)$ for both the $\sqrt{s}=$13 and 14 TeV are depicted in Fig.~\ref{fig:tp_excl}. We also depict the constraint coming from the CMS stop searches ($pp\to \tilde{t}\tilde{t}^{*},~ \tilde{t}\to t \tilde{\chi}_{_1}^{0}$) \cite{CMS:2021beq} in the {\it $t\bar{t}$+missing transverse energy} channel in Fig.~\subref*{fig:tp_brxsigma_mtp_13tev}. It is evident that the CMS analysis in this channel gives better limit compared to our analysis in the low $m_{_{T_{p}}}$ region and converges near $m_{_{T_{p}}} = 1.4$ TeV.

\subsection{Single top partner production case}\label{results_st}

To analyze the $pp\to T_{p} j$ process, we have simulated $5\times 10^{4}$ signal events for top partner mass in the range 1-2.6 TeV
against $5\times 10^{6}$ simulated $t\bar{t}$ \footnote{For $m_{_{T_p}} \geq $ 1.6 (1.8) TeV, we have simulated $10^{7}$ $t\bar{t}$ background events at center of mass energy 13 (14) TeV.}, $10^{6}$ simulated $tW,~t\bar{t}Z,~t\bar{t}W,~tj$ and $1.21\times 10^{6}$ QCD multijet background events.

The signal cross sections for various choices of the top partner masses and the effects of various kinematic cuts (listed in Tab \ref{table:stcutsval}) 
for $\sqrt{s}=13$ TeV are depicted in Table.~\ref{table:singletop13tev}. It also contains the estimated signal significance at 13 TeV LHC
collision energy assuming an integrated luminosity of 139 fb$^{-1}$. The results of the corresponding 14 TeV analysis are depicted in Table.~\ref{table:singletop14tev} 
which contains the signal cross sections, effects of the cut flow on the signal as well as  SM background cross sections and the estimated signal significance 
assuming an integrated luminosity of 300 fb$^{-1}$.

\begin{table}[H]
\centering
\resizebox{0.9\columnwidth}{!}{
\begin{tabu}{lcccccccc}
\hline
\hline
\rowfont \normalfont
$m_{_{ T_{p}}}$ &Production & Basic &  Boosted top  & & $M_{T}$ cut &$p_{_{T}}$ cut & $\cancel{E}_{T}$ cut &  Significance  \\
\rowfont \normalfont
(TeV)&cross section& cuts& requirement& $|\eta| > 2.1$& $M_{T}>M_{T_{0}}$ & $p_{_{T}}>p_{_{T_{0}}}$ & $\cancel{E}_{T}>\cancel{E}_{T_{0}}$ &  $\left( \frac{S}{\sqrt{S+B}} \right)$ \\
\rowfont \normalfont
&(fb)& (fb)& (fb)&&(fb)&(fb)&(fb)& \\

\hline
\hline

\multirow{2}{*}{1}&  37.57 &  28.38 &  7.26 &  6.45 &  6.30 &  5.82 &  3.61 & \multirow{2}{*}{5.2 ({\color{blue}9.0})}\\
\rowfont{\color{red}}
& $1.96\times 10^{11}$ &  $7.42\times 10^{10}$ &  $1.81\times 10^{7}$ &  $3.91\times 10^{6}$ &  $567.88$ &  $427.88$ &  $19.04$ &\\

\multirow{2}{*}{1.2}& 20.92 &  16.05 &  4.58 &  4.12 &  3.94 &  3.85 &  2.35 &   \multirow{2}{*}{5.1 ({\color{blue}8.8})}\\ 
\rowfont{\color{red}}
& $1.96\times 10^{11}$ &  $7.42\times 10^{10}$ &  $1.81\times 10^{7}$ &  $3.91\times 10^{6}$ &  $254.31$ &  $240.55$ &  $7.43$ &\\

\multirow{2}{*}{1.4}& 12.16 &  9.37 &  2.89 &  2.62 &  2.47 &  2.46 &  1.50 &   \multirow{2}{*}{5.3 ({\color{blue}9.0})}\\
\rowfont{\color{red}}
& $1.96\times 10^{11}$ &  $7.42\times 10^{10}$ &  $1.81\times 10^{7}$ &  $3.91\times 10^{6}$ &  $113.26$ &  $113.26$ &  $2.28$ & \\

\multirow{2}{*}{1.6}&  7.33 &  5.65 &  1.85 &  1.69 &  1.56 &  1.56 &  0.66 &  \multirow{2}{*}{5.6 ({\color{blue}8.4})}\\ 
\rowfont{\color{red}}
& $1.96\times 10^{11}$ &  $7.42\times 10^{10}$ &  $1.81\times 10^{7}$ &  $3.91\times 10^{6}$ &  $54.5$ &  $54.5$ &  $0.19$ & \\

\multirow{2}{*}{1.8}& 4.50 &  3.46 &  1.16 &  1.06 &  0.97 &  0.97 &  0.62 & \multirow{2}{*}{5.3 ({\color{blue}8.1})}\\ 
\rowfont{\color{red}}
& $1.96\times 10^{11}$ &  $7.42\times 10^{10}$ &  $1.81\times 10^{7}$ &  $3.91\times 10^{6}$ &  $27.2$ &  $27.2$ &  $0.19$ & \\

\multirow{2}{*}{2.0}& 2.83 &  2.14 &  0.71 &  0.64 &  0.58 &  0.58 &  0.43 &\multirow{2}{*}{4.1 ({\color{blue}6.5})}\\ 
\rowfont{\color{red}}
& $1.96\times 10^{11}$ &  $7.42\times 10^{10}$ &  $1.81\times 10^{7}$ &  $3.91\times 10^{6}$ &  $14.6$ &  $14.6$ &  $0.19$ & \\

\multirow{2}{*}{2.2}& 1.80 &  1.34 &  0.42 &  0.39 &  0.34 &  0.34 &  0.28 &\multirow{2}{*}{3.0 ({\color{blue}4.9})}\\ 
\rowfont{\color{red}}
& $1.96\times 10^{11}$ &  $7.42\times 10^{10}$ &  $1.81\times 10^{7}$ &  $3.91\times 10^{6}$ &  $9.23$ &  $9.23$ &  $0.19$ & \\

\multirow{2}{*}{2.4}& 1.16 &  0.85 &  0.25 &  0.23 &  0.19 &  0.19 &  0.17 &\multirow{2}{*}{2.0 ({\color{blue}3.4})}\\ 
\rowfont{\color{red}}
& $1.96\times 10^{11}$ &  $7.42\times 10^{10}$ &  $1.81\times 10^{7}$ &  $3.91\times 10^{6}$ &  $5.60$ &  $5.60$ &  $0.19$ & \\

\hline
\hline
\end{tabu}
}
\caption{{ \scriptsize Columns [2-8] depict the effects of cut flow on cross sections for both the signal ({\bf black}) and total background ({\color{red}red}) at 
$\sqrt{s}=13$ TeV. $\sin \theta_{L} = 0.1$ is assumed to estimate the signal cross sections and branching ratios at various benchmark choices of $m_{_{ T_{P}}}$. {The cross sections presented in column [2-8] assume BR$(T_{p}\to t \gamma_{d}) = 100\%$. The estimated signal significance is quoted in the last column assuming \textit{actual} BR ({\color{blue}100\% BR})  with an integrated luminosity of 139 fb$^{-1}$. [$v_d = 200$ GeV, $m_{\gamma_d}=10$ GeV and $m_{h_d}= 400$ GeV are assumed to obtain the above results.]} }}
\label{table:singletop13tev}
\end{table}

\begin{table}[H]
\centering
\resizebox{0.9\columnwidth}{!}{
\begin{tabu}{lcccccccc}
\hline
\hline
\rowfont{\normalfont}
$m_{_{T_{p}}}$ & Production & Basic &  Boosted top  & &  $M_{T}$ cut &$p_{_{T}}$ cut& $\cancel{E}_{T}$ cut & Significance  \\
\rowfont{\normalfont}
(TeV)& cross section& cuts& requirement& $|\eta| > 2.1$& $M_{T}>M_{T_{0}}$ & $p_{_{T}}>p_{_{T_{0}}}$ & $\cancel{E}_{T}>\cancel{E}_{T_{0}}$ &  $\left( \frac{S}{\sqrt{S+B}} \right)$ \\
\rowfont \normalfont
&(fb)& (fb)& (fb)&&(fb)&(fb)&(fb)& \\

\hline
\hline

\multirow{2}{*}{1}&  46.89 &  35.63 &   9.01 &  8.07 &  7.88 &  7.35 &  3.91 &\multirow{2}{*}{8.2 ({\color{blue}14.1})}\\
\rowfont{\color{red}}
& $2.16\times 10^{11}$ &  $8.17\times 10^{10}$ &  $2.00\times 10^{7}$ &  $5.73\times 10^{6}$ &  $713.3$ &  $533.1$ &  $19.2$ &\\

\multirow{2}{*}{1.2}& 26.65 &  20.46 &  5.85 &  5.29 &  5.07 &  4.96 &  2.56 &\multirow{2}{*}{8.3 ({\color{blue}14.4})}\\ 
\rowfont{\color{red}}
& $2.16\times 10^{11}$ &  $8.17\times 10^{10}$ &  $2.00\times 10^{7}$ &  $5.73\times 10^{6}$ &  $320.5$ &  $303.2$ &  $7.0$ &\\

\multirow{2}{*}{1.4}&  15.81 &  12.14 &  3.79 &  3.46 &  3.25 &  3.24 &  1.69 &\multirow{2}{*}{7.8 ({\color{blue}13.4})}\\ 
\rowfont{\color{red}}
& $2.16\times 10^{11}$ &  $8.17\times 10^{10}$ &  $2.00\times 10^{7}$ &  $5.73\times 10^{6}$ &  $147.8$ &  $147.7$ &  $3.1$ & \\

\multirow{2}{*}{1.6}&  9.70 &  7.48 &  2.48 &  2.26 &  2.10 &  2.10 &  1.09 &\multirow{2}{*}{8.0 ({\color{blue}13.2})}\\ 
\rowfont{\color{red}}
& $2.16\times 10^{11}$ &  $8.17\times 10^{10}$ &  $2.00\times 10^{7}$ &  $5.73\times 10^{6}$ &  $73.4$ &  $73.4$ &  $0.95$ & \\

\multirow{2}{*}{1.8}& 6.09 &  4.67 &  1.56 &  1.43 &  1.31 &  1.31 &  0.60 &\multirow{2}{*}{7.3({\color{blue}11.5})}\\
\rowfont{\color{red}}
& $2.16\times 10^{11}$ &  $8.17\times 10^{10}$ &  $2.00\times 10^{7}$ &  $5.73\times 10^{6}$ &  $36.2$ &  $36.2$ &  $0.23$ & \\

\multirow{2}{*}{2.0}& 3.91 &  2.96 &  0.97 &  0.90 &  0.80 &  0.80 &  0.50 &\multirow{2}{*}{6.4 ({\color{blue}10.2})}\\ 
\rowfont{\color{red}}
& $2.16\times 10^{11}$ &  $8.17\times 10^{10}$ &  $2.00\times 10^{7}$ &  $5.73\times 10^{6}$ &  $18.0$ &  $18.0$ &  $0.23$ & \\

\multirow{2}{*}{2.2}& 2.54 &  1.89 &  0.59 &  0.54 &  0.48 &  0.48 &  0.35 &\multirow{2}{*}{4.8 ({\color{blue}7.9})}\\ 
\rowfont{\color{red}}
& $2.16\times 10^{11}$ &  $8.17\times 10^{10}$ &  $2.00\times 10^{7}$ &  $5.73\times 10^{6}$ &  $10.8$ &  $10.8$ &  $0.23$ & \\

\multirow{2}{*}{2.4}& 1.67 &  1.22 &  0.36 &  0.33 &  0.28 &  0.28 &  0.22 &\multirow{2}{*}{3.3 ({\color{blue}5.7})}\\ 
\rowfont{\color{red}}
& $2.16\times 10^{11}$ &  $8.17\times 10^{10}$ &  $2.00\times 10^{7}$ &  $5.73\times 10^{6}$ &  $6.53$ &  $6.53$ &  $0.23$ & \\

\multirow{2}{*}{2.5}& 1.36 &  0.98 &  0.28 &  0.26 &  0.22 &  0.22 &  0.18 &\multirow{2}{*}{2.7 ({\color{blue}4.8})}\\ 
\rowfont{\color{red}}
& $2.16\times 10^{11}$ &  $8.17\times 10^{10}$ &  $2.00\times 10^{7}$ &  $5.73\times 10^{6}$ &  $4.54$ &  $4.54$ &  $0.23$ & \\

\multirow{2}{*}{2.6}& 1.11 &  0.79 &  0.22 &  0.20 &  0.17 &  0.17 &  0.14 &\multirow{2}{*}{2.3 ({\color{blue}4.1})}\\ 
\rowfont{\color{red}}
& $2.16\times 10^{11}$ &  $8.17\times 10^{10}$ &  $2.00\times 10^{7}$ &  $5.73\times 10^{6}$ &  $3.14$ &  $3.14$ &  $0.23$ & \\

\hline
\hline
\end{tabu}
}
\caption{{ \scriptsize Columns [2-8] depict the effects of cut flow on cross sections for both the signal ({\bf black}) and total background ({\color{red}red}) at $\sqrt{s}=14$ TeV. $\sin \theta_{L} = 0.1$ is assumed to estimate the signal cross sections and branching ratios at various benchmark choices of $m_{_{ T_{P}}}$.  {The cross sections presented in column [2-8] assume BR$(T_{p}\to t \gamma_{d}) = 100\%$. The estimated signal significance is quoted in the last column assuming \textit{actual} BR ({\color{blue}100\% BR})  with an integrated luminosity of 300 fb$^{-1}$.  [$v_d = 200$ GeV, $m_{\gamma_d}=10$ GeV and $m_{h_d}= 400$ GeV are assumed to obtain the above results.]} }}
\label{table:singletop14tev}
\end{table}


{In Table. \ref{table:singletop13tev}  (\ref{table:singletop14tev}), the final background cross sections for $m_{_{T_P}}> 1.6~(1.8)$ TeV at $\sqrt{s}= 13~{\rm TeV}~(14~{\rm TeV})$ are all same even though different cross sections are expected as the cuts depend on benchmark point choices. This is visible after the $M_{T}$ cut level for $m_{_{T_P}}> 1.6~(1.8)$ TeV at $\sqrt{s}= 13~{\rm TeV}~(14~{\rm TeV})$ and also after the $p_{_T}$ cut level one has the effects of mass dependence in the cross section even though the $p_{_T}$ cut is same for these benchmark point. However, after the $\cancel{E}_{T}$ cut we have {only \textit{two} simulated $t\bar{t}$ background event left out of $10^7$} for these benchmark points due to high $\cancel{E}_{T}$ cut and a small but nonzero $t\bar{t}V$ background contribution. This is the reason why the final cross section are same for all these benchmark points.
}

\begin{figure}[H]
\centering
\subfloat[\label{fig:st_sin_mtp_cross_13tev}]{\includegraphics[width=0.5\columnwidth]{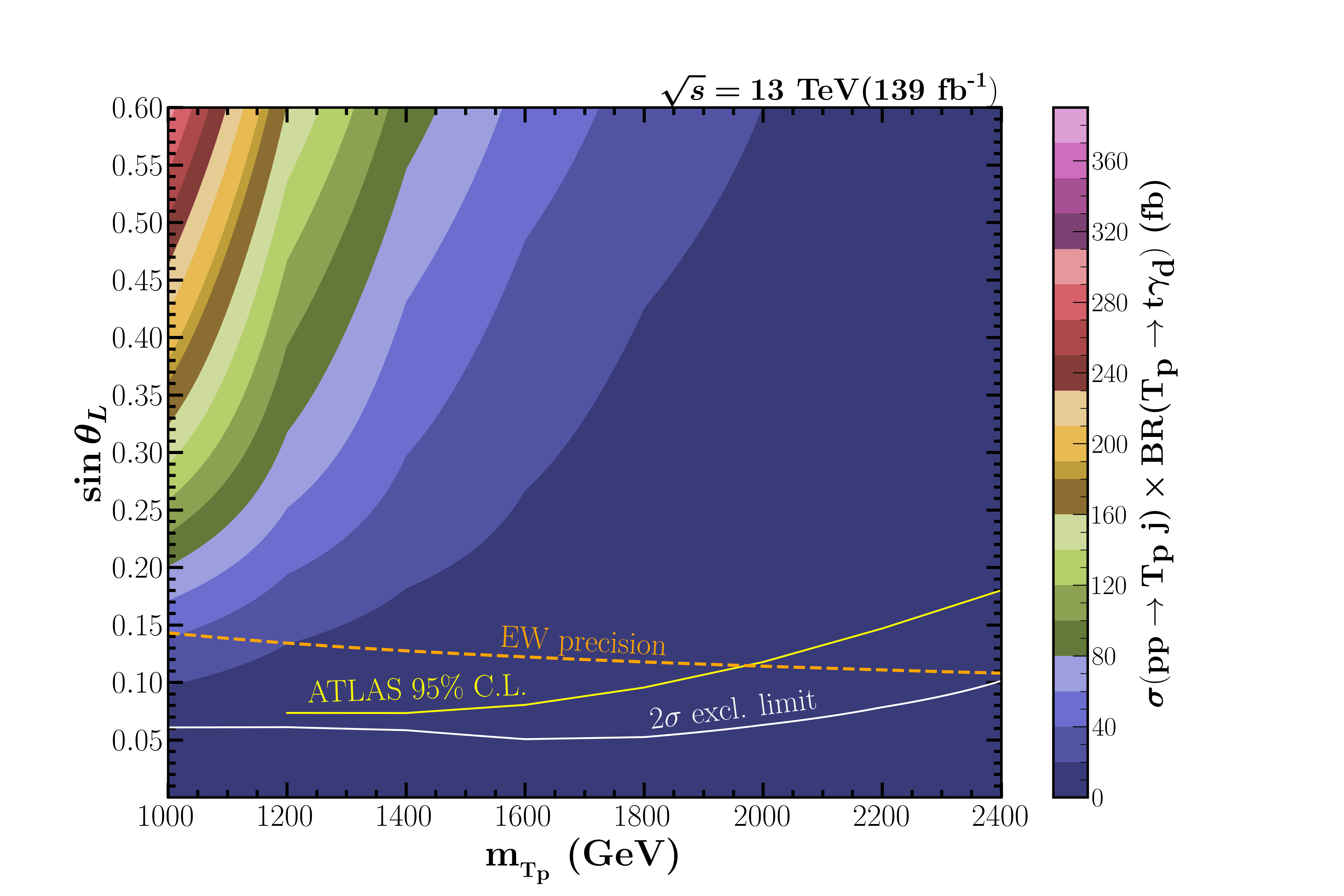}}
\subfloat[\label{fig:st_sin_mtp_cross_14tev}]{\includegraphics[width=0.5\columnwidth]{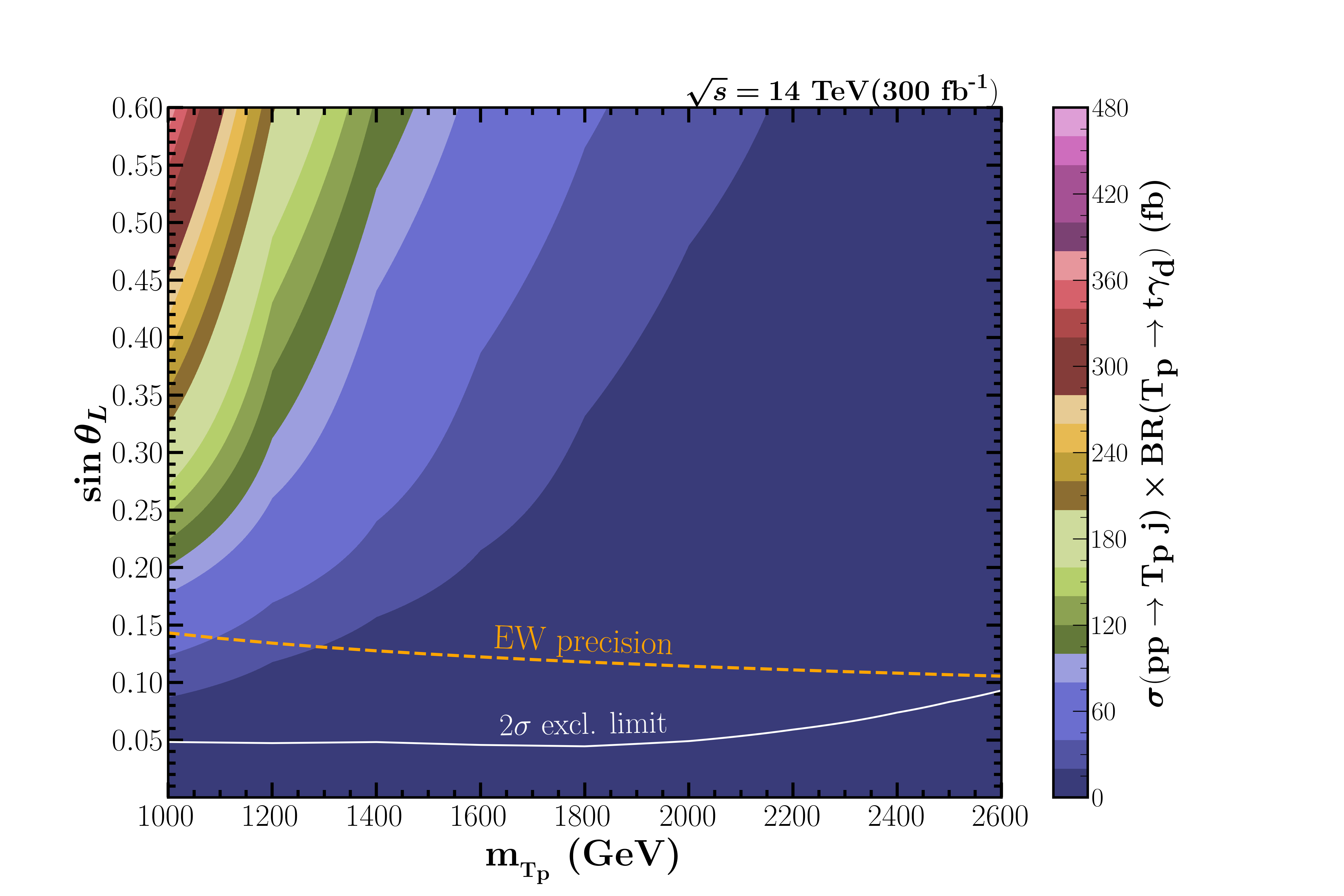}}

\subfloat[\label{fig:st_sin_mtp_cross_14tev_3000}]{\includegraphics[width=0.5\columnwidth]{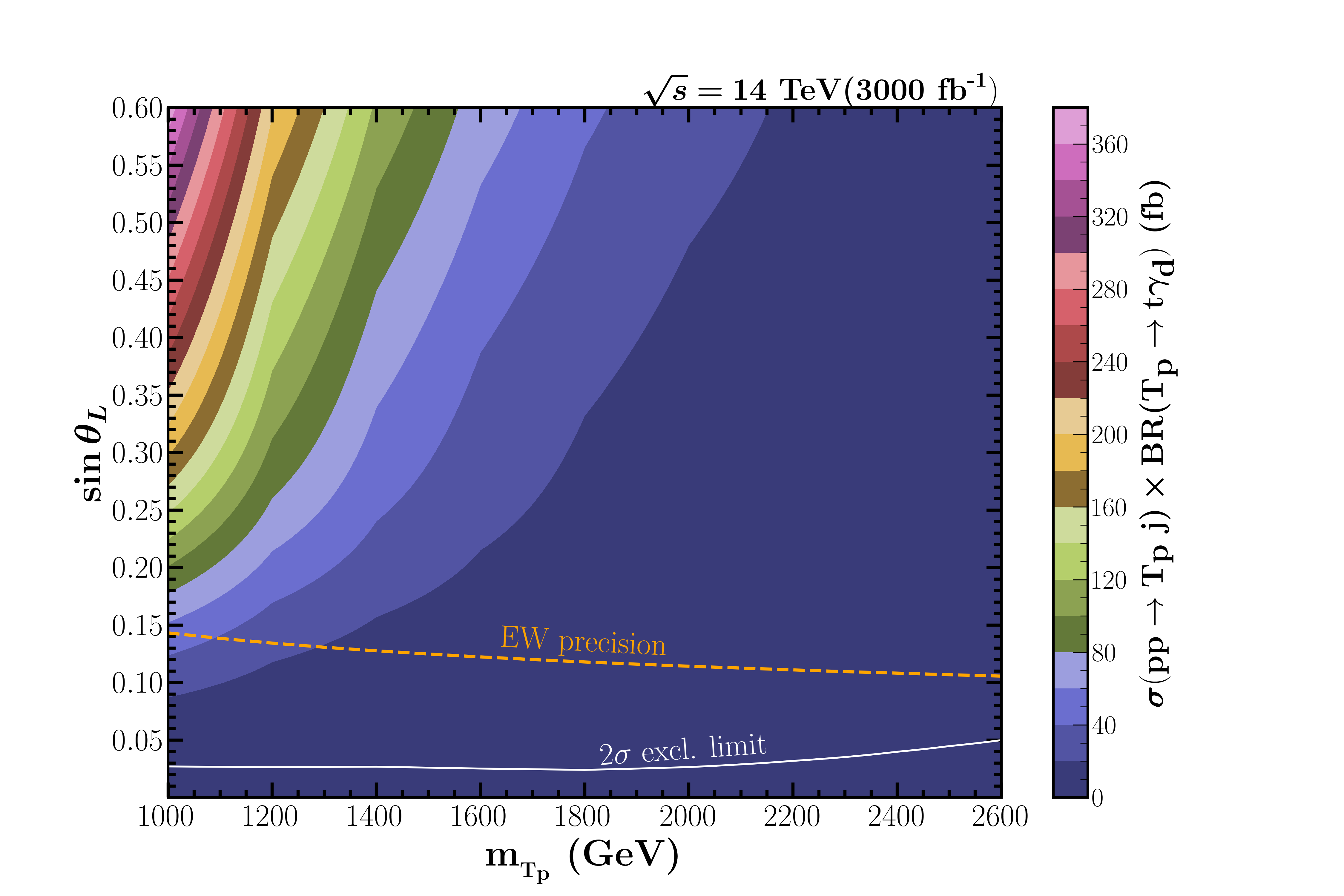}}
\caption{{\scriptsize $2\sigma$ exclusion limit on $\sigma(pp\to T_{p}j) \times {\rm BR}(T_{p}\to t\gamma_{d})$ in the  $\sin\theta_L - m_{T_{p}}$ plane at (a) 13 TeV(139 fb$^{-1}$), (b) 14 TeV(300 fb$^{-1}$) and (c) 14 TeV(3000 fb$^{-1}$). [$v_d = 200$ GeV, $m_{\gamma_d}=10$ GeV and $m_{h_d}= 400$ GeV are assumed to obtain the above results.]}}
\end{figure}

The $2\sigma$ exclusion limits on $\sigma(pp\to T_{p} j) \times {\rm BR}(T_{p}\to t\gamma_{d})$ in the $\sin\theta_L - m_{_{T_p}}$ plane for 13 TeV (139 fb$^{-1}$) and 
14 TeV (300 fb$^{-1}$ and 3000 fb$^{-1}$) LHC centre of mass energies have been presented in Figs.~\subref*{fig:st_sin_mtp_cross_13tev}, \subref*{fig:st_sin_mtp_cross_14tev} and
\subref*{fig:st_sin_mtp_cross_14tev_3000}, respectively, assuming $v_d = 200$ GeV, $m_{\gamma_d}=10$ GeV and $m_{h_d}= 400$ GeV.

In these figures, we also depict the constraints coming  from various other observations in the $\sin\theta_L - m_{_{T_p}}$ plane. The most stringent of these
constraints is the one coming from the latest LHC data which rules out $\sin\theta_L>0.11 $ for $m_{_{T_p}} \leq 1.95$ TeV \cite{ATLAS:2022vff}. However, for $m_{_{T_p}} \geq 
1.95$ TeV the EWP data sets more stringent limit on $\sin\theta_L$, which rules out $\sin\theta_L>0.11$ in this mass range.

 Fig.~\subref*{fig:st_sin_mtp_cross_13tev} shows that the region of parameter space that can be ruled out using ATLAS data \cite{ATLAS:2022vff}, namely, $\sin\theta_L \gtrsim  0.1$ for $m_{_{T_p}}\gtrsim 2$ TeV is already excluded by the EWP data for the singlet top partner case irrespective of its decay branching ratio. Our analysis at $\sqrt{s}=13$ TeV with integrated luminosity of 139 fb$^{-1}$ in the single top partner mode gives comparatively better limit in the $\sin\theta_L$ - $m_{_{T_p}}$ plane.
 
  More importantly, in the {high-luminosity} phase of the LHC run which will be able to probe even a smaller production cross section corresponding to $\sin\theta_L \sim  0.03-0.05$ in the mass range 1.0 TeV $\lesssim m_{_{T_p}} \lesssim$ 2.6  TeV, the analysis we presented here for the single top partner channel will be of much relevance. This is because it is exactly the range of parameter space where the decay width of the top partner in the traditional modes are highly suppressed compared to the nonstandard modes (see Fig.~\ref{fig:width_ratio}).

In fact, the branching ratio of $T_p \to t \gamma_d$ could be as large as 60\%-70\% (see Fig.~\ref{fig:br_tdph}) depending on the masses of the dark photon and dark Higgs and the choice of $v_d$ \footnote{For low, $v_d \sim 1$ GeV the allowed ranges of $\sin\theta_L$ consistent with the perturbative unitarity bound are so small ($|\sin\theta_L|\lesssim 0.002$, see Fig.~\ref{fig:lamt_mtp_st_vd_1_gev}) that it is not sensitive to LHC analyses.} .
  

The study of the single production of the top partner at the LHC in the ${\rm \it single ~top} + \cancel{E}_T+{\rm \it ~jet}$ final state and the event selection criteria listed in Sec. \ref{Tp@LHC} shows if the top partner dominantly decays to a top quark and an invisible dark photon, it will be possible to set potentially stringent limit in the $\sin\theta_L - m_{_{T_p}}$ plane. We obtained a $2\sigma$ exclusion limit on $\sin\theta_L$ as low as $\sim 0.025$ up to $m_{_{T_p}} \sim 2.0$ TeV {at $\sqrt{s} = 14$ TeV with $3$ ab$^{-1}$ integrated luminosity.} For $m_{_{T_p}}$ up to 2.6 TeV $\sin\theta_L \ge 0.05$ can be excluded at $2\sigma$ significance with the same specification.

\begin{figure}[H]
\centering
\subfloat[\label{fig:st_sig_lum_sintheta0pt1}]{\includegraphics[width=0.5\columnwidth]{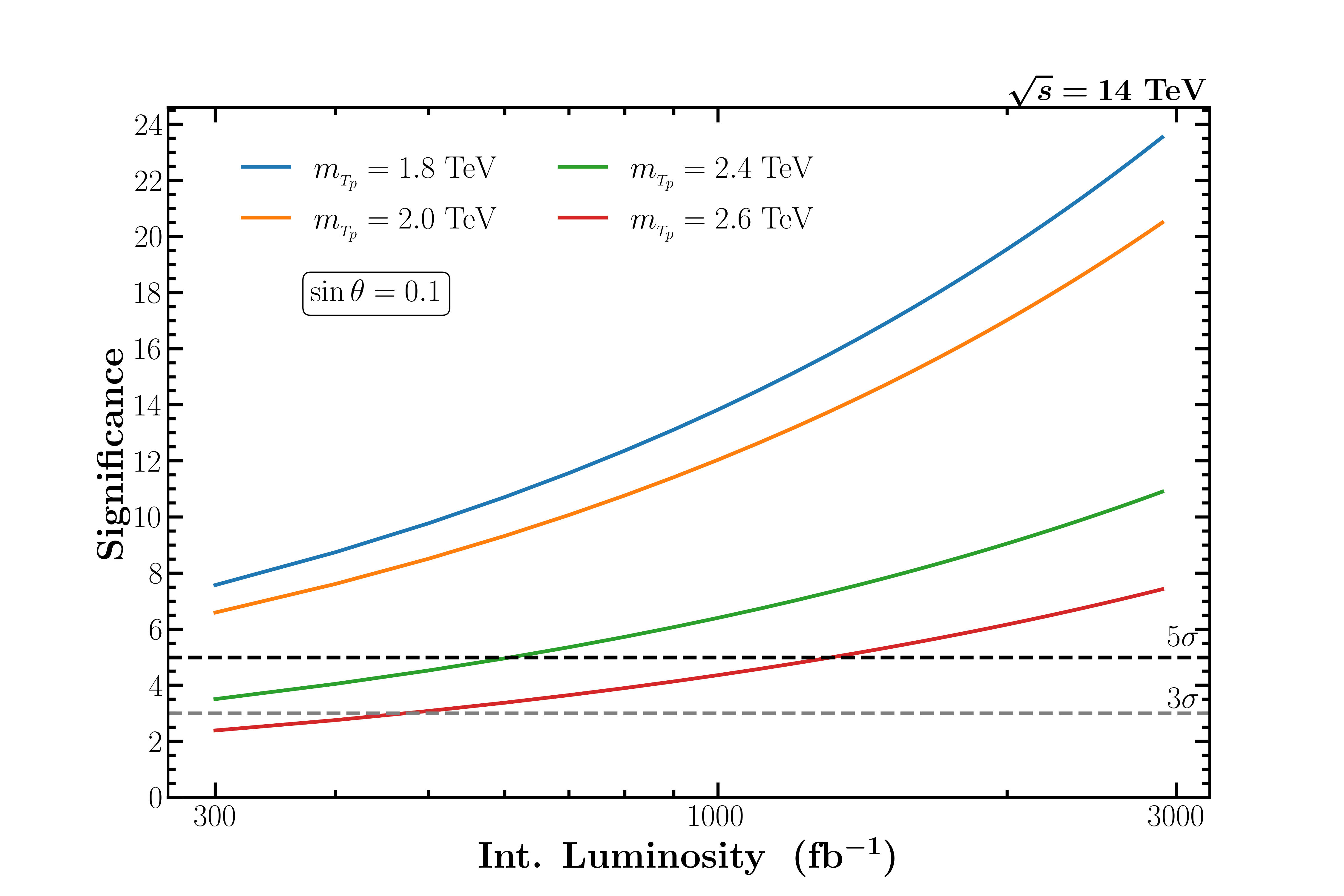}}
\subfloat[\label{fig:st_sig_lum_sintheta0pt05}]{\includegraphics[width=0.5\columnwidth]{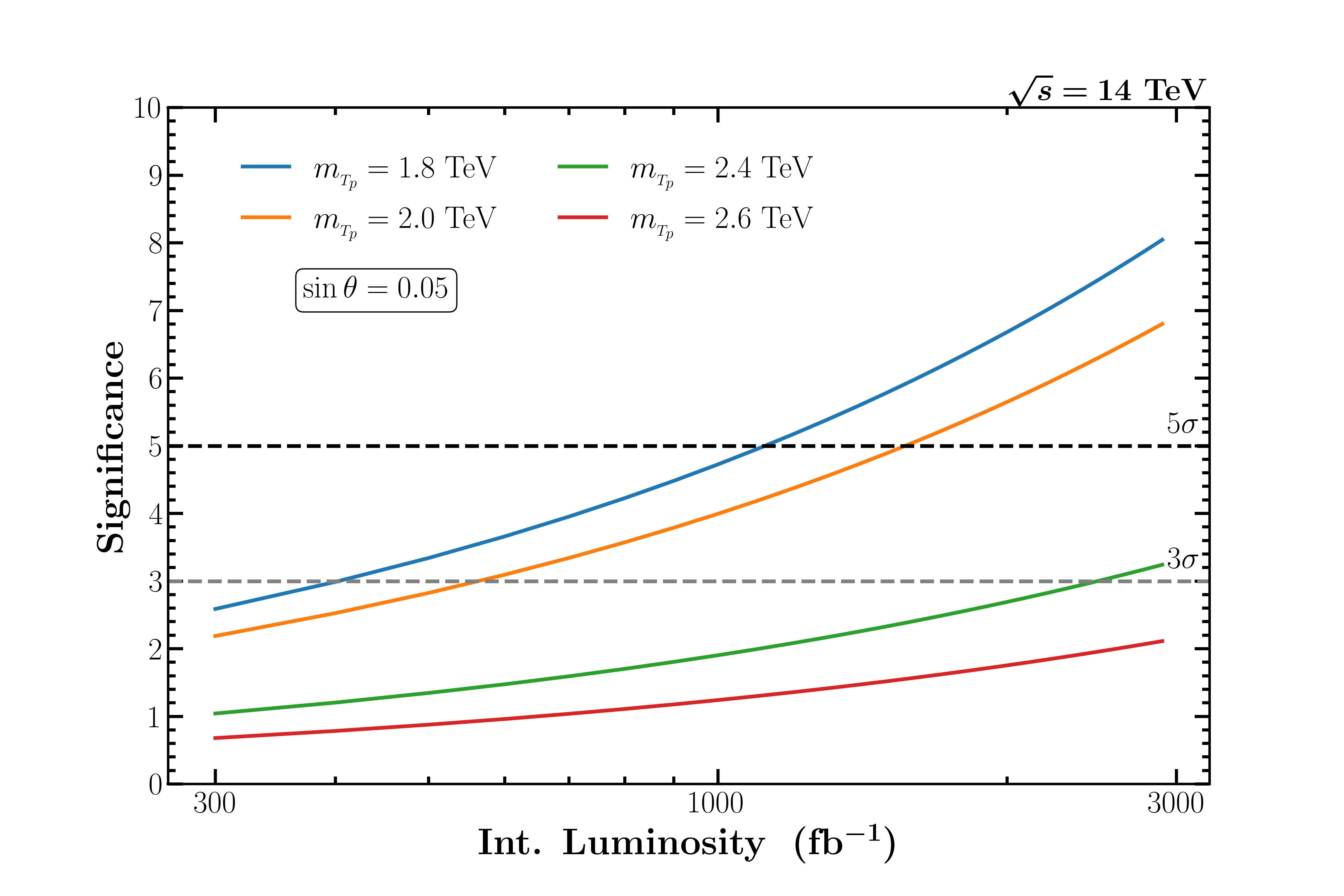}}

\caption{{\scriptsize The LHC reach for $m_{_{T_p}}$ at $3\sigma$ and $5\sigma$ significance level as a function of the integrated luminosity at $\sqrt{s}=$14 TeV for two different choices of (a). $\sin\theta_L = 0.1$ and (b). $\sin\theta_L = 0.05$. }}\label{fig:lhcreach}
\end{figure}

We also present a $5\sigma$ ($3\sigma$) discovery (exclusion) limit on $m_{_{T_p}}$ using the single production of top partner analysis as a function of the integrated luminosity at $\sqrt{s}=$14 TeV, assuming $\sin\theta_L = 0.1, 0.05$.~Fig.~\ref{fig:lhcreach} shows that at $\sqrt{s}=$14 TeV assuming $3 ~{\rm ab}^{-1}$ integrated luminosity one can exclude $m_{_{T_p}} \leq 2.4$ TeV for $\sin\theta_L = 0.05$ at $3\sigma$ significance level.

\section{Conclusions}\label{conclusions}

We have discussed the phenomenology of a VLQ which is charged under both the SM gauge interaction and also carries an additional dark $U(1)_d$ charge. This gives rise to some new consequences in terms of its decay.  We have shown that in a large region of the parameter space, the top partner predominantly decays to a top and a dark photon/dark Higgs ($T_p\to t+\gamma_d/h_d$) with it's
traditional decay modes suppressed. This not only helps one to evade the strong bounds coming from LHC searches  for a top partner  
in the traditional channels such as $T_p \to b W, t Z, {\rm ~or~} t h$  it opens up a new possibility to search for them at the LHC. We have 
focused on the $T_p \to t \gamma_d$ decay mode and analyzed the pair and single production of top partner at the LHC in the $t\bar{t}+\cancel{E}_T$
and $t +\cancel{E}_T$ final states, respectively, assuming that the dark photon decays to an invisible system or is stable at the length scale of the detector. 

A striking feature of our signal is that the top quark produced in the decay of the top partner is highly boosted and one can use the boosted top tagging technique to reconstruct the momentum of 
the top quark initiated jet in the hadronic channel. We have proposed several kinematic variables such as transverse momentum of the reconstructed top quark, missing transverse momentum and stransverse mass or transverse mass to suppress the dominant SM backgrounds in these channels. We have shown that top partner mass up to 1.6 TeV
can be exclude with more than $2\sigma$ significance using the top-pair production channel. We have also incorporated the limit coming from the LHC analysis for {\it stop pair searches} ($pp\to \tilde{t}\tilde{t}^{*}\to ~(t \tilde{\chi}_{_1}^{0}) (\bar{t} \tilde{\chi}_{_1}^{0})$) in the same final state.  

The single production of the top partner is sensitive to the mass of the top partner, $m_{_{T_p}}$ and the mixing angle, $\sin\theta_L$. 
The constraints coming from the EWPO and perturbative unitarity mostly prefer a very small mixing angle ($|\sin\theta_L| \lesssim 0.1$).
Such a small mixing angle will imply small cross section in the single top partner production case. The future run of LHC at 14 TeV center of mass energy
with 3 ab$^{-1}$ integrated luminosity will be sensitive to such small mixing angle, {\it i.e.}, $\sin\theta_L \sim 0.05 - 0.1$.  For $m_{_{T_p}} \sim 2.5$ TeV, this is 
exactly the range consistent with the hierarchy  $|\sin\theta_L| \ll m_t /m_{_{T_p}} \ll 1$. This is also the case where one has the branching ratio enhancement
in the $t \gamma_d$ and $t h_d$ channels. Using simple cut based analysis we have set an exclusion limit in the $\sin\theta_L - m_{_{T_p}}$ plane. We have shown 
that $\sin\theta_L \sim 0.025$ (0.05) can be ruled out for $m_{_{T_p}} \leq 2.0 $ (2.6) TeV at a 95\% confidence level using the future LHC data corresponding to 3 ab$^{-1}$ integrated luminosity and $\sqrt{s}=14$ TeV. We have also considered bounds coming from the latest LHC data in the single top partner channel with $T \to t Z \to t (Z\to \nu\bar{\nu})$. We have also presented LHC reach for top partner mass as a function of integrated luminosity at 14 TeV center of mass energy. Our collider analysis will also be applicable whenever the top partner decays to a top and an invisible system other than dark photon assuming $m_{_{T_p}},~ \sin\theta_L~{\rm and} ~{\rm BR}(T_p \to t+invisible)$ as independent parameters.


\section*{Acknowledgement}

SB is thankful to Emidio Gabrielli and Subhadeep Mondal for useful discussions. 
SV is thankful for support provided by Council of Scientific $\&$ Industrial Research (CSIR, India) under CSIR-UGC NET Fellowship (File No.: 09/934(0017)/2020-EMR-I). 
AC would like to thank Indian Institute of Technology, Kanpur for supporting this work by means of Institute Post-Doctoral Fellowship (Ref.No. DF/PDF197/2020-IITK/970). 

\appendix

\section{EWPO}

\setcounter{equation}{0}
\numberwithin{equation}{section}
  
For a singlet top partner model, we follow \cite{Chen:2017hak} to get the general expression for corrections to oblique parameters and $Zb\bar{b}$ couplings,
\bea
\Delta T &=& \frac{N_{c} m_{t}^{2}}{16\pi \sin^{2}_{W} m_{W}^{2}}\sin^{2}\theta_{L}\left[-(1+\cos^{2}\theta_{L})+2\cos^{2}\theta_{L}\frac{r_{T}}{r_{T}-1}\log(r_{T})+r_{T} \sin^{2}\theta_{L}\right]
\label{eq:deltat}\\
\nonumber
\Delta S &=& \frac{N_{c}}{18\pi}\sin^{2}\theta_{L}\left[\log(r_{T})+\cos^{2}\theta_{L}\left(\frac{5(r^{2}_{T}+1)-22 r_{T}}{\left(1-r_{T}\right)^{2}}\right. \right.\\
&& \left. \left. \hspace{2cm} +\frac{3(r_{T}+1)\left(r^{2}_{T}-4r_{T}+1\right)}{\left(1-r_{T}\right)^{3}}\log(r_{T})\right)\right]
\label{eq:deltas}\\
\delta X^{L}_{bb} &=& \frac{g^{2}}{32\pi^{2}}\sin^{2}\theta_{L}\left[f_{1}(x,x')+\cos^{2}\theta_{L}f_{2}(x,x')\right]
\label{eq:delxbb}
\eea

where, $r_{T} = m_{_{T_{p}}}/m_{t},~x = m^{2}_{t}/m^{2}_{W},~ x' = m^{2}_{_{T_{p}}}/m^{2}_{W}$ and,

\bea
f_{1}(x,x') &=& x'-x+3\log\left(\frac{x'}{x}\right) 
\label{eq:f1}\\
f_{2}(x,x') &=& -x'-x+\frac{2x'x}{x'-x}\log\left(\frac{x'}{x}\right) 
\label{eq:f2}
\eea

\section{Partial decay widths}\label{App:B}

In this section we list the formulae for the decay widths of the top partner in various modes.\footnote{We have calculated both the general and model specific expressions for the decay widths independently. The general expressions for the decay widths agree with Ref.~\cite{Buchkremer:2013bha,Kim:2018mks} in various special cases or limits.}  

\begin{itemize}

\item $\bm{T_{p} \to bW}$:

\bea
\mathcal{L}_{T_{p}-b-W} &= &  \overline{T}_{p}\gamma^{\mu} \left(V_{_L}P_{L}\right)b ~W_{\mu}+{\rm H.c.} 
\eea

The exact model independent expression for $\Gamma(T_{p} \to bW)$ is given by 
\bea
\nonumber
\Gamma(T_{p} \to bW) &=& \frac{1}{16\pi}\frac{m^{3}_{_{T_{p}}}}{m^{2}_{_W}}~ \lambda^{1/2}\left(1,\frac{m^{2}_{b}}{m^{2}_{_{T_{p}}}},\frac{m^{2}_{_W}}{m^{2}_{_{T_{p}}}}\right) \\ 
 &&\left[\left(1-\frac{m^{2}_{b}}{m^{2}_{_{T_{p}}}}\right)^{2} +\frac{m^{2}_{_W}}{m^{2}_{_{T_{p}}}}-2 \frac{m^{4}_{_W}}{m^{4}_{T_{p}}}+\frac{m^{2}_{_W}m^{2}_{b}}{m^{2}_{_{T_{p}}}}\right]\left( \frac{|V^{2}_{_L}| }{2}\right)
\label{eq:width_bW}
\eea

where,
\bea
\lambda(a,b,c) &=& a^{2}+b^{2}+c^{2}-2ab-2bc-2ca
\label{eq:lam} 
\eea

In the present model, we have 

\bea
V_{_L} &=& \frac{g\sin\theta_{L}}{\sqrt{2}}
\eea
In the limit $|\sin\theta_{d}|,~|\sin\theta_{S}|,~\varepsilon,~ m_{t}/ m_{_{T_{p}}}\ll 1$  Eq.~(\ref{eq:width_bW})
reduces to 

\bea
\label{eq:width_W}
\Gamma(T_{p}\to bW)& \approx & \frac{1}{16\pi}\frac{m^{3}_{_{T_{p}}}}{v^{2}_{_{EW}}}\sin^{2}\theta_{L}
\eea

\item $\bm{T_{p} \to tV}$: 

\bea
\mathcal{L}_{T_{p}-t-V} &= & -\overline{T}_{p}\gamma^{\mu}\left(V_{_L}P_{_L}+V_{_R}P_{_R}\right)t V_{\mu}+{\rm H.c.}
\eea

The exact model independent expression for $\Gamma(T_{p} \to t V)$ is given by 
\bea
\nonumber
\Gamma(T_{p} \to tV) &=& \frac{1}{16\pi}\frac{m^{3}_{_{T_{p}}}}{m^{2}_{V}}~ \lambda^{1/2}\left(1,\frac{m^{2}_{t}}{m^{2}_{_{T_{p}}}},\frac{m^{2}_{V}}{m^{2}_{_{T_{p}}}}\right) \\ \nonumber &&\left\{\left[\left(1-\frac{m^{2}_{t}}{m^{2}_{_{T_{p}}}}\right)^{2} +\frac{m^{2}_{V}}{m^{2}_{_{T_{p}}}}-2 \frac{m^{4}_{V}}{m^{4}_{T_{p}}}+\frac{m^{2}_{V}m^{2}_{t}}{m^{2}_{_{T_{p}}}}\right] \left(\frac{|V^{2}_{_L}|+|V^{2}_{_R}|}{2}\right) \right. \\ 
&& \left. -3 \frac{m_{t}}{m_{_{T_{p}}}}\frac{m^{2}_{V}}{m^{2}_{_{T_{p}}}}\left(V_{_L}V^{*}_{_R}+V^{*}_{_L}V_{_R}\right)\right\}
\label{eq:width_tV}
\eea


In this model, for $V = Z$, we have 

\bea
V_{_L} &=& \frac{1}{4}\sin 2\theta_{L} \left[\hat{g}_{Z}\left(\cos\theta_{d}+\sin\theta_{d}\hat{t}_{W}\varepsilon \right)+2g_{d}\sin\theta_{d}\right]\\
\nonumber
V_{_R} &=& \frac{1}{2}g_{d}\sin\theta_{d}\sin 2\theta_{R} \\
&=& \frac{1}{2}g_{d}\sin\theta_{d}\sin 2\theta_{L}\frac{m_{_{T_{p}}}m_{t}}{m^{2}_{t}\cos^{2}\theta_{L}+m^{2}_{_{T_{p}}}\sin^{2}\theta_{L}}
\eea

and for $V = \gamma_{d}$, we have 

\bea
V_{_L} &=& \frac{1}{4}\sin 2\theta_{L} \left[\hat{g}_{Z}\left(\sin\theta_{d}-\cos\theta_{d}\hat{t}_{W}\varepsilon \right)-2g_{d}\cos\theta_{d}\right]  \\ 
V_{_R} &=& -\frac{1}{2}g_{d}\cos\theta_{d}\sin 2\theta_{R} \nonumber \\ &=& -\frac{1}{2}g_{d}\cos\theta_{d}\sin 2\theta_{L}\frac{m_{_{T_{p}}}m_{t}}{m^{2}_{t}\cos^{2}\theta_{L}+m^{2}_{_{T_{p}}}\sin^{2}\theta_{L}}
\eea

where $\theta_{d}$ is the mixing angle between the abelian gauge bosons. 
\bea
\hat{t}_{W} &=&\frac{\sin\hat{\theta}_{W}}{\cos\hat{\theta}_{W}}  \\
\sin\hat{\theta}_{W}  &= &\sin\theta_{W} + \mathcal{O}(\varepsilon^{2})\\
\hat{g}_{Z} &=& \frac{g}{\cos\hat{\theta}_{W}}\\
g_{d} &=& \frac{m_{\gamma_{d}}}{v_{d}}+ \mathcal{O}(\varepsilon^{2})\\
g_{d} &=& \frac{g^{\prime}_{d}}{\sqrt{1-\left(\varepsilon^{\prime 2}/\cos^{2}\hat{\theta}_{W}\right)}
}
\eea

In the limit $|\sin\theta_{d}|,~|\sin\theta_{S}|,~\varepsilon,~ m_{t}/ m_{_{T_{p}}}\ll 1$ Eq.~(\ref{eq:width_tV}) simplifies to
the following two equations for $V=Z {~\rm and~} \gamma_d$

\bea
\label{eq:width_Z}
\Gamma(T_{p}\to tZ)& \approx & \frac{1}{32\pi}\frac{m^{3}_{_{T_{p}}}}{v^{2}_{_{EW}}}\sin^{2}\theta_{L}\cos^{2}\theta_{L}
\eea

and,

\bea
\label{eq:width_gamma_d}
\Gamma(T_{p}\to t\gamma_{d})& \approx & \frac{1}{32\pi}\frac{m^{3}_{_{T_{p}}}}{v^{2}_{d}}\sin^{2}\theta_{L}\cos^{2}\theta_{L}\times \left(1+\frac{m^{2}_{_{T_{p}}}m^{2}_{t}}{D^{2}}\right)
\eea

where, $D = m^{2}_{t}\cos^{2}\theta_{L}+m^{2}_{_{T_{p}}}\sin^{2}\theta_{L}$.

\item $\bm{T_{p} \to tS}$:

\bea
\mathcal{L}_{T_{p}-t-S}&=& -\overline{T}_{p}\left(Y_{_L}P_{_L}+Y_{_R}P_{_R}\right)t S+{\rm H.c.}
\eea

The exact model independent expression for $\Gamma(T_{p} \to t S)$ is given by 
\bea
\nonumber
\Gamma(T_{p} \to tS) &=& \frac{1}{16\pi}m_{_{T_{p}}}~ \lambda^{1/2}\left(1,\frac{m^{2}_{t}}{m^{2}_{_{T_{p}}}},\frac{m^{2}_{S}}{m^{2}_{_{T_{p}}}}\right) \\ \nonumber &&\left\{\left[1+\frac{m^{2}_{t}}{m^{2}_{_{T_{p}}}}- \frac{m^{2}_{S}}{m^{2}_{_{T_{p}}}}\right] \left(\frac{|Y^{2}_{_L}|+|Y^{2}_{_R}|}{2}\right) \right. \\ 
 && \left. + \frac{m_{t}}{m_{_{T_{p}}}}\left(Y_{_L}Y^{*}_{_R}+Y^{*}_{_L}Y_{_R}\right)\right\}
\label{eq:width_tS}
\eea


In this model, for $S = h$

\bea
Y_{_L} &=& \frac{1}{\sqrt{2}}\sin\theta_{R} \left(y_{t}\cos\theta_{L}\cos\theta_{S}+\lambda_{T}\sin\theta_{L}\sin\theta_{S}\right) \\
Y_{_R} &=& \frac{1}{\sqrt{2}}\cos\theta_{R} \left(y_{t}\sin\theta_{L}\cos\theta_{S}-\lambda_{T}\cos\theta_{L}\sin\theta_{S}\right)
\eea

and for $S = h_{d}$

\bea
Y_{_L} &=& \frac{1}{\sqrt{2}}\sin\theta_{R} \left(y_{t}\cos\theta_{L}\sin\theta_{S}-\lambda_{T}\sin\theta_{L}\cos\theta_{S}\right) \\
Y_{_R} &=& \frac{1}{\sqrt{2}}\cos\theta_{R} \left(y_{t}\sin\theta_{L}\sin\theta_{S}+\lambda_{T}\cos\theta_{L}\cos\theta_{S}\right)
\eea

where 
\bea
  y_t&=&\sqrt{2}\frac{\sqrt{m_t^2 \cos^2\theta_L + m_{_{T_{p}}}^2 \sin^2\theta_L}}{v_{_{EW}}} \\  
  \label{eq:lam_T}
  \lambda_{T}&=&\frac{(m_{_{T_{p}}}^2-m_t^2)\sin 2\theta_L}{\sqrt{2}v_d\sqrt{m_t^2 \cos^2\theta_L +m_{_{T_{p}}}^2\sin^2\theta_L}} \\
    m_{T}&=&\frac{m_t m_{_{T_{p}}}}{\sqrt{m_t^2 \cos^2\theta_L + m_{_{T_{p}}}^2\sin^2\theta_L}}\\
   \sin\theta_{R}&=& \frac{m_{T}}{m_{t}}\sin \theta_{L}\\
   \cos\theta_{R}&=& \frac{m_{T}}{m_{_{T_{p}}}}\cos \theta_{L} 
  \eea

and $\theta_{S}$ is a mixing angle in the scalar sector.

In the limit $|\sin\theta_{d}|,~|\sin\theta_{S}|,~\varepsilon,~ m_{t}/ m_{_{T_{p}}}\ll 1$  Eq.~(\ref{eq:width_tS}) simplifies to
the following two equations for $S=h {~\rm and~} h_d$

\bea
\label{eq:width_h}
\Gamma(T_{p}\to th)& \approx & \frac{1}{32\pi}\frac{m^{3}_{_{T_{p}}}}{v^{2}_{_{EW}}}\sin^{2}\theta_{L}\cos^{2}\theta_{L}
\eea

and,

\bea
\label{eq:width_hd}
\Gamma(T_{p}\to th_{d})& \approx & \frac{1}{32\pi}\frac{m^{3}_{_{T_{p}}}}{v^{2}_{d}}\sin^{2}\theta_{L}\cos^{2}\theta_{L} \nonumber\\
&&\times\frac{m^{4}_{_{T_{p}}}}{D^{2}} \left(\sin^{4}\theta_{L}+\frac{m^{2}_{t}}{m^{2}_{_{T_{p}}}}\cos^{4}\theta_{L} +4 \frac{m^{2}_{t}}{m^{2}_{_{T_{p}}}}\sin^{2}\theta_{L}\cos^{2}\theta_{L}\right)
\eea


\end{itemize}

\clearpage


\bibliographystyle{JHEP}
\bibliography{References}

\providecommand{\href}[2]{#2}\begingroup\raggedright\begin{thebibliography}{10}

\bibitem{Allanchach:2019wrx}
B.~C. Allanchach, \emph{{Beyond the Standard Model}},
  \href{http://dx.doi.org/10.23730/CYRSP-2019-006.113}{\emph{CERN Yellow Rep.
  School Proc.} {\bf 6} (2019) 113--144}.

\bibitem{Randall:1999ee}
L.~Randall and R.~Sundrum, \emph{{A Large mass hierarchy from a small extra
  dimension}}, \href{http://dx.doi.org/10.1103/PhysRevLett.83.3370}{\emph{Phys.
  Rev. Lett.} {\bf 83} (1999) 3370--3373},
  [\href{https://arxiv.org/abs/hep-ph/9905221}{{\tt hep-ph/9905221}}].

\bibitem{Carena:2007ua}
M.~Carena, E.~Ponton, J.~Santiago and C.~E.~M. Wagner, \emph{{Electroweak
  constraints on warped models with custodial symmetry}},
  \href{http://dx.doi.org/10.1103/PhysRevD.76.035006}{\emph{Phys. Rev. D} {\bf
  76} (2007) 035006}, [\href{https://arxiv.org/abs/hep-ph/0701055}{{\tt
  hep-ph/0701055}}].

\bibitem{Chivukula:2000mb}
R.~S. Chivukula, \emph{{Lectures on technicolor and compositeness}},  in
  \emph{{Theoretical Advanced Study Institute in Elementary Particle Physics
  (TASI 2000): Flavor Physics for the Millennium}}, pp.~731--772, 6, 2000.
\newblock \href{https://arxiv.org/abs/hep-ph/0011264}{{\tt hep-ph/0011264}}.

\bibitem{Kaplan:1983sm}
D.~B. Kaplan, H.~Georgi and S.~Dimopoulos, \emph{{Composite Higgs Scalars}},
  \href{http://dx.doi.org/10.1016/0370-2693(84)91178-X}{\emph{Phys. Lett. B}
  {\bf 136} (1984) 187--190}.

\bibitem{Agashe:2004rs}
K.~Agashe, R.~Contino and A.~Pomarol, \emph{{The Minimal composite Higgs
  model}}, \href{http://dx.doi.org/10.1016/j.nuclphysb.2005.04.035}{\emph{Nucl.
  Phys. B} {\bf 719} (2005) 165--187},
  [\href{https://arxiv.org/abs/hep-ph/0412089}{{\tt hep-ph/0412089}}].

\bibitem{Arkani-Hamed:2002iiv}
N.~Arkani-Hamed, A.~G. Cohen, E.~Katz, A.~E. Nelson, T.~Gregoire and J.~G.
  Wacker, \emph{{The Minimal moose for a little Higgs}},
  \href{http://dx.doi.org/10.1088/1126-6708/2002/08/021}{\emph{JHEP} {\bf 08}
  (2002) 021}, [\href{https://arxiv.org/abs/hep-ph/0206020}{{\tt
  hep-ph/0206020}}].

\bibitem{Chacko:2005pe}
Z.~Chacko, H.-S. Goh and R.~Harnik, \emph{{The Twin Higgs: Natural electroweak
  breaking from mirror symmetry}},
  \href{http://dx.doi.org/10.1103/PhysRevLett.96.231802}{\emph{Phys. Rev.
  Lett.} {\bf 96} (2006) 231802},
  [\href{https://arxiv.org/abs/hep-ph/0506256}{{\tt hep-ph/0506256}}].

\bibitem{Arkani-Hamed:2002ikv}
N.~Arkani-Hamed, A.~G. Cohen, E.~Katz and A.~E. Nelson, \emph{{The Littlest
  Higgs}}, \href{http://dx.doi.org/10.1088/1126-6708/2002/07/034}{\emph{JHEP}
  {\bf 07} (2002) 034}, [\href{https://arxiv.org/abs/hep-ph/0206021}{{\tt
  hep-ph/0206021}}].

\bibitem{Aguilar-Saavedra:2009xmz}
J.~A. Aguilar-Saavedra, \emph{{Identifying top partners at LHC}},
  \href{http://dx.doi.org/10.1088/1126-6708/2009/11/030}{\emph{JHEP} {\bf 11}
  (2009) 030}, [\href{https://arxiv.org/abs/0907.3155}{{\tt 0907.3155}}].

\bibitem{Berger:2012ec}
J.~Berger, J.~Hubisz and M.~Perelstein, \emph{{A Fermionic Top Partner:
  Naturalness and the LHC}},
  \href{http://dx.doi.org/10.1007/JHEP07(2012)016}{\emph{JHEP} {\bf 07} (2012)
  016}, [\href{https://arxiv.org/abs/1205.0013}{{\tt 1205.0013}}].

\bibitem{Buchkremer:2013bha}
M.~Buchkremer, G.~Cacciapaglia, A.~Deandrea and L.~Panizzi, \emph{{Model
  Independent Framework for Searches of Top Partners}},
  \href{http://dx.doi.org/10.1016/j.nuclphysb.2013.08.010}{\emph{Nucl. Phys. B}
  {\bf 876} (2013) 376--417}, [\href{https://arxiv.org/abs/1305.4172}{{\tt
  1305.4172}}].

\bibitem{Kim:2018mks}
J.~H. Kim and I.~M. Lewis, \emph{{Loop Induced Single Top Partner Production
  and Decay at the LHC}},
  \href{http://dx.doi.org/10.1007/JHEP05(2018)095}{\emph{JHEP} {\bf 05} (2018)
  095}, [\href{https://arxiv.org/abs/1803.06351}{{\tt 1803.06351}}].

\bibitem{Alhazmi:2018whk}
H.~Alhazmi, J.~H. Kim, K.~Kong and I.~M. Lewis, \emph{{Shedding Light on Top
  Partner at the LHC}},
  \href{http://dx.doi.org/10.1007/JHEP01(2019)139}{\emph{JHEP} {\bf 01} (2019)
  139}, [\href{https://arxiv.org/abs/1808.03649}{{\tt 1808.03649}}].

\bibitem{Anandakrishnan:2015yfa}
A.~Anandakrishnan, J.~H. Collins, M.~Farina, E.~Kuflik and M.~Perelstein,
  \emph{{Odd Top Partners at the LHC}},
  \href{http://dx.doi.org/10.1103/PhysRevD.93.075009}{\emph{Phys. Rev. D} {\bf
  93} (2016) 075009}, [\href{https://arxiv.org/abs/1506.05130}{{\tt
  1506.05130}}].

\bibitem{Dolan:2016eki}
M.~J. Dolan, J.~L. Hewett, M.~Kr\"amer and T.~G. Rizzo, \emph{{Simplified
  Models for Higgs Physics: Singlet Scalar and Vector-like Quark
  Phenomenology}}, \href{http://dx.doi.org/10.1007/JHEP07(2016)039}{\emph{JHEP}
  {\bf 07} (2016) 039}, [\href{https://arxiv.org/abs/1601.07208}{{\tt
  1601.07208}}].

\bibitem{Chala:2018qdf}
M.~Chala, R.~Gr\"ober and M.~Spannowsky, \emph{{Searches for vector-like quarks
  at future colliders and implications for composite Higgs models with dark
  matter}}, \href{http://dx.doi.org/10.1007/JHEP03(2018)040}{\emph{JHEP} {\bf
  03} (2018) 040}, [\href{https://arxiv.org/abs/1801.06537}{{\tt 1801.06537}}].

\bibitem{Gopalakrishna:2013hua}
S.~Gopalakrishna, T.~Mandal, S.~Mitra and G.~Moreau, \emph{{LHC Signatures of
  Warped-space Vectorlike Quarks}},
  \href{http://dx.doi.org/10.1007/JHEP08(2014)079}{\emph{JHEP} {\bf 08} (2014)
  079}, [\href{https://arxiv.org/abs/1306.2656}{{\tt 1306.2656}}].

\bibitem{Dermisek:2019vkc}
R.~Derm\'\i{}\v{s}ek, E.~Lunghi and S.~Shin, \emph{{Hunting for Vectorlike
  Quarks}}, \href{http://dx.doi.org/10.1007/JHEP04(2019)019}{\emph{JHEP} {\bf
  04} (2019) 019}, [\href{https://arxiv.org/abs/1901.03709}{{\tt 1901.03709}}].

\bibitem{Han:2018hcu}
H.~Han, L.~Huang, T.~Ma, J.~Shu, T.~M.~P. Tait and Y.~Wu, \emph{{Six Top
  Messages of New Physics at the LHC}},
  \href{http://dx.doi.org/10.1007/JHEP10(2019)008}{\emph{JHEP} {\bf 10} (2019)
  008}, [\href{https://arxiv.org/abs/1812.11286}{{\tt 1812.11286}}].

\bibitem{Colucci:2018vxz}
S.~Colucci, B.~Fuks, F.~Giacchino, L.~Lopez~Honorez, M.~H.~G. Tytgat and
  J.~Vandecasteele, \emph{{Top-philic Vector-Like Portal to Scalar Dark
  Matter}}, \href{http://dx.doi.org/10.1103/PhysRevD.98.035002}{\emph{Phys.
  Rev. D} {\bf 98} (2018) 035002},
  [\href{https://arxiv.org/abs/1804.05068}{{\tt 1804.05068}}].

\bibitem{Dobrescu:2016pda}
B.~A. Dobrescu and F.~Yu, \emph{{Exotic Signals of Vectorlike Quarks}},
  \href{http://dx.doi.org/10.1088/1361-6471/aacbfd}{\emph{J. Phys. G} {\bf 45}
  (2018) 08LT01}, [\href{https://arxiv.org/abs/1612.01909}{{\tt 1612.01909}}].

\bibitem{Gopalakrishna:2015wwa}
S.~Gopalakrishna, T.~S. Mukherjee and S.~Sadhukhan, \emph{{Extra neutral
  scalars with vectorlike fermions at the LHC}},
  \href{http://dx.doi.org/10.1103/PhysRevD.93.055004}{\emph{Phys. Rev. D} {\bf
  93} (2016) 055004}, [\href{https://arxiv.org/abs/1504.01074}{{\tt
  1504.01074}}].

\bibitem{Dasgupta:2021fzw}
S.~Dasgupta, R.~Pramanick and T.~S. Ray, \emph{{Broad toplike vector quarks at
  LHC and HL-LHC}},
  \href{http://dx.doi.org/10.1103/PhysRevD.105.035032}{\emph{Phys. Rev. D} {\bf
  105} (2022) 035032}, [\href{https://arxiv.org/abs/2112.03742}{{\tt
  2112.03742}}].

\bibitem{Corcella:2021mdl}
G.~Corcella, A.~Costantini, M.~Ghezzi, L.~Panizzi, G.~M. Pruna and
  J.~\v{S}alko, \emph{{Vector-like quarks decaying into singly and doubly
  charged bosons at LHC}},
  \href{http://dx.doi.org/10.1007/JHEP10(2021)108}{\emph{JHEP} {\bf 10} (2021)
  108}, [\href{https://arxiv.org/abs/2107.07426}{{\tt 2107.07426}}].

\bibitem{Dermisek:2021zjd}
R.~Dermisek, E.~Lunghi, N.~Mcginnis and S.~Shin, \emph{{Tau-jet signatures of
  vectorlike quark decays to heavy charged and neutral Higgs bosons}},
  \href{http://dx.doi.org/10.1007/JHEP08(2021)159}{\emph{JHEP} {\bf 08} (2021)
  159}, [\href{https://arxiv.org/abs/2105.10790}{{\tt 2105.10790}}].

\bibitem{duPlessis:2021xuc}
K.~du~Plessis, M.~M. Flores, D.~Kar, S.~Sinha and H.~van~der Schyf,
  \emph{{Hitting two BSM particles with one lepton-jet: search for a top
  partner decaying to a dark photon, resulting in a lepton-jet}},
  \href{http://dx.doi.org/10.21468/SciPostPhys.13.2.018}{\emph{SciPost Phys.}
  {\bf 13} (2022) 018}, [\href{https://arxiv.org/abs/2112.08425}{{\tt
  2112.08425}}].

\bibitem{PhysRevD.99.115024}
T.~G. Rizzo, \emph{Kinetic mixing and portal matter phenomenology},
  \href{http://dx.doi.org/10.1103/PhysRevD.99.115024}{\emph{Phys. Rev. D} {\bf
  99} (Jun, 2019) 115024}.

\bibitem{Rizzo:2022qan}
T.~G. Rizzo, \emph{{Portal Matter and Dark Sector Phenomenology at Colliders}},
   in \emph{{2022 Snowmass Summer Study}}, 2, 2022.
\newblock \href{https://arxiv.org/abs/2202.02222}{{\tt 2202.02222}}.

\bibitem{Qin:2021cxl}
X.~Qin and J.-F. Shen, \emph{{Search for single production of vector-like $B$
  quark decaying to a Higgs boson and bottom quark at the CLIC}},
  \href{http://dx.doi.org/10.1016/j.nuclphysb.2021.115388}{\emph{Nucl. Phys. B}
  {\bf 966} (2021) 115388}.

\bibitem{Wojcik:2022rtk}
G.~N. Wojcik, \emph{{Kinetic Mixing from Kaluza-Klein Modes: A Simple
  Construction for Portal Matter}},
  \href{https://arxiv.org/abs/2205.11545}{{\tt 2205.11545}}.

\bibitem{Fabbrichesi:2020wbt}
M.~Fabbrichesi, E.~Gabrielli and G.~Lanfranchi, \emph{{The Dark Photon}},
  \href{https://arxiv.org/abs/2005.01515}{{\tt 2005.01515}}.

\bibitem{Foot:2016wvj}
R.~Foot and S.~Vagnozzi, \emph{{Solving the small-scale structure puzzles with
  dissipative dark matter}},
  \href{http://dx.doi.org/10.1088/1475-7516/2016/07/013}{\emph{JCAP} {\bf 07}
  (2016) 013}, [\href{https://arxiv.org/abs/1602.02467}{{\tt 1602.02467}}].

\bibitem{Kim:2019oyh}
J.~H. Kim, S.~D. Lane, H.-S. Lee, I.~M. Lewis and M.~Sullivan, \emph{{Searching
  for Dark Photons with Maverick Top Partners}},
  \href{http://dx.doi.org/10.1103/PhysRevD.101.035041}{\emph{Phys. Rev. D} {\bf
  101} (2020) 035041}, [\href{https://arxiv.org/abs/1904.05893}{{\tt
  1904.05893}}].

\bibitem{ATLAS:2018dyh}
{\scshape ATLAS} collaboration, M.~Aaboud et~al., \emph{{Search for single
  production of vector-like quarks decaying into $Wb$ in $pp$ collisions at
  $\sqrt{s} = 13$ TeV with the ATLAS detector}},
  \href{http://dx.doi.org/10.1007/JHEP05(2019)164}{\emph{JHEP} {\bf 05} (2019)
  164}, [\href{https://arxiv.org/abs/1812.07343}{{\tt 1812.07343}}].

\bibitem{ATLAS:2018uky}
{\scshape ATLAS} collaboration, M.~Aaboud et~al., \emph{{Search for pair
  production of heavy vector-like quarks decaying into hadronic final states in
  $pp$ collisions at $\sqrt{s} = 13$ TeV with the ATLAS detector}},
  \href{http://dx.doi.org/10.1103/PhysRevD.98.092005}{\emph{Phys. Rev. D} {\bf
  98} (2018) 092005}, [\href{https://arxiv.org/abs/1808.01771}{{\tt
  1808.01771}}].

\bibitem{CMS:2016lel}
{\scshape CMS} collaboration, A.~M. Sirunyan et~al., \emph{{Cross section
  measurement of $t$-channel single top quark production in pp collisions at
  $\sqrt s =$ 13 TeV}},
  \href{http://dx.doi.org/10.1016/j.physletb.2017.07.047}{\emph{Phys. Lett. B}
  {\bf 772} (2017) 752--776}, [\href{https://arxiv.org/abs/1610.00678}{{\tt
  1610.00678}}].

\bibitem{CMS:2019too}
{\scshape CMS} collaboration, A.~M. Sirunyan et~al., \emph{{Measurement of top
  quark pair production in association with a Z boson in proton-proton
  collisions at $\sqrt{s}=$ 13 TeV}},
  \href{http://dx.doi.org/10.1007/JHEP03(2020)056}{\emph{JHEP} {\bf 03} (2020)
  056}, [\href{https://arxiv.org/abs/1907.11270}{{\tt 1907.11270}}].

\bibitem{Bizot:2018tds}
N.~Bizot, G.~Cacciapaglia and T.~Flacke, \emph{{Common exotic decays of top
  partners}}, \href{http://dx.doi.org/10.1007/JHEP06(2018)065}{\emph{JHEP} {\bf
  06} (2018) 065}, [\href{https://arxiv.org/abs/1803.00021}{{\tt 1803.00021}}].

\bibitem{Xie:2019gya}
K.-P. Xie, G.~Cacciapaglia and T.~Flacke, \emph{{Exotic decays of top partners
  with charge 5/3: bounds and opportunities}},
  \href{http://dx.doi.org/10.1007/JHEP10(2019)134}{\emph{JHEP} {\bf 10} (2019)
  134}, [\href{https://arxiv.org/abs/1907.05894}{{\tt 1907.05894}}].

\bibitem{Cacciapaglia:2019zmj}
G.~Cacciapaglia, T.~Flacke, M.~Park and M.~Zhang, \emph{{Exotic decays of top
  partners: mind the search gap}},
  \href{http://dx.doi.org/10.1016/j.physletb.2019.135015}{\emph{Phys. Lett. B}
  {\bf 798} (2019) 135015}, [\href{https://arxiv.org/abs/1908.07524}{{\tt
  1908.07524}}].

\bibitem{Chala:2017xgc}
M.~Chala, \emph{{Direct bounds on heavy toplike quarks with standard and exotic
  decays}}, \href{http://dx.doi.org/10.1103/PhysRevD.96.015028}{\emph{Phys.
  Rev. D} {\bf 96} (2017) 015028},
  [\href{https://arxiv.org/abs/1705.03013}{{\tt 1705.03013}}].

\bibitem{Aguilar-Saavedra:2017giu}
J.~A. Aguilar-Saavedra, D.~E. L\'opez-Fogliani and C.~Mu\~noz, \emph{{Novel
  signatures for vector-like quarks}},
  \href{http://dx.doi.org/10.1007/JHEP06(2017)095}{\emph{JHEP} {\bf 06} (2017)
  095}, [\href{https://arxiv.org/abs/1705.02526}{{\tt 1705.02526}}].

\bibitem{Benbrik:2019zdp}
R.~Benbrik et~al., \emph{{Signatures of vector-like top partners decaying into
  new neutral scalar or pseudoscalar bosons}},
  \href{http://dx.doi.org/10.1007/JHEP05(2020)028}{\emph{JHEP} {\bf 05} (2020)
  028}, [\href{https://arxiv.org/abs/1907.05929}{{\tt 1907.05929}}].

\bibitem{Bhardwaj:2022nko}
A.~Bhardwaj, T.~Mandal, S.~Mitra and C.~Neeraj, \emph{{A roadmap to explore the
  vector-like quarks decaying to a new (pseudo)scalar}},
  \href{https://arxiv.org/abs/2203.13753}{{\tt 2203.13753}}.

\bibitem{Das:2018gcr}
K.~Das, T.~Mondal and S.~K. Rai, \emph{{Nonstandard signatures of vectorlike
  quarks in a leptophobic 221 model}},
  \href{http://dx.doi.org/10.1103/PhysRevD.99.115002}{\emph{Phys. Rev. D} {\bf
  99} (2019) 115002}, [\href{https://arxiv.org/abs/1807.08160}{{\tt
  1807.08160}}].

\bibitem{Banerjee:2016wls}
S.~Banerjee, D.~Barducci, G.~B\'elanger and C.~Delaunay, \emph{{Implications of
  a High-Mass Diphoton Resonance for Heavy Quark Searches}},
  \href{http://dx.doi.org/10.1007/JHEP11(2016)154}{\emph{JHEP} {\bf 11} (2016)
  154}, [\href{https://arxiv.org/abs/1606.09013}{{\tt 1606.09013}}].

\bibitem{Banerjee:2022izw}
A.~Banerjee, D.~B. Franzosi and G.~Ferretti, \emph{{Modelling vector-like
  quarks in partial compositeness framework}},
  \href{http://dx.doi.org/10.1007/JHEP03(2022)200}{\emph{JHEP} {\bf 03} (2022)
  200}, [\href{https://arxiv.org/abs/2202.00037}{{\tt 2202.00037}}].

\bibitem{Banerjee:2022xmu}
A.~Banerjee et~al., \emph{{Phenomenological aspects of composite Higgs
  scenarios: exotic scalars and vector-like quarks}},
  \href{https://arxiv.org/abs/2203.07270}{{\tt 2203.07270}}.

\bibitem{Ciuchini:2013pca}
M.~Ciuchini, E.~Franco, S.~Mishima and L.~Silvestrini, \emph{{Electroweak
  Precision Observables, New Physics and the Nature of a 126 GeV Higgs Boson}},
  \href{http://dx.doi.org/10.1007/JHEP08(2013)106}{\emph{JHEP} {\bf 08} (2013)
  106}, [\href{https://arxiv.org/abs/1306.4644}{{\tt 1306.4644}}].

\bibitem{deBlas:2016ojx}
J.~de~Blas, M.~Ciuchini, E.~Franco, S.~Mishima, M.~Pierini, L.~Reina et~al.,
  \emph{{Electroweak precision observables and Higgs-boson signal strengths in
  the Standard Model and beyond: present and future}},
  \href{http://dx.doi.org/10.1007/JHEP12(2016)135}{\emph{JHEP} {\bf 12} (2016)
  135}, [\href{https://arxiv.org/abs/1608.01509}{{\tt 1608.01509}}].

\bibitem{Chen:2017hak}
C.-Y. Chen, S.~Dawson and E.~Furlan, \emph{{Vectorlike fermions and Higgs
  effective field theory revisited}},
  \href{http://dx.doi.org/10.1103/PhysRevD.96.015006}{\emph{Phys. Rev. D} {\bf
  96} (2017) 015006}, [\href{https://arxiv.org/abs/1703.06134}{{\tt
  1703.06134}}].

\bibitem{CMS:2021beq}
{\scshape CMS} collaboration, A.~M. Sirunyan et~al., \emph{{Search for top
  squark production in fully-hadronic final states in proton-proton collisions
  at $\sqrt{s} =$ 13 TeV}},
  \href{http://dx.doi.org/10.1103/PhysRevD.104.052001}{\emph{Phys. Rev. D} {\bf
  104} (2021) 052001}, [\href{https://arxiv.org/abs/2103.01290}{{\tt
  2103.01290}}].

\bibitem{ATLAS:2022vff}
{\scshape ATLAS} collaboration, M.~Aaboud et~al., \emph{{Search for invisible
  particles produced in association with single top quarks in proton–proton
  collisions at $\sqrt{s} =13$ TeV with the ATLAS detector}}, .

\bibitem{Kraml:2016eti}
S.~Kraml, U.~Laa, L.~Panizzi and H.~Prager, \emph{{Scalar versus fermionic top
  partner interpretations of $t\bar t + E_T^{\rm miss}$ searches at the LHC}},
  \href{http://dx.doi.org/10.1007/JHEP11(2016)107}{\emph{JHEP} {\bf 11} (2016)
  107}, [\href{https://arxiv.org/abs/1607.02050}{{\tt 1607.02050}}].

\bibitem{Robens:2015gla}
T.~Robens and T.~Stefaniak, \emph{{Status of the Higgs Singlet Extension of the
  Standard Model after LHC Run 1}},
  \href{http://dx.doi.org/10.1140/epjc/s10052-015-3323-y}{\emph{Eur. Phys. J.
  C} {\bf 75} (2015) 104}, [\href{https://arxiv.org/abs/1501.02234}{{\tt
  1501.02234}}].

\bibitem{Robens:2019ynf}
T.~Robens, \emph{{Investigating extended scalar sectors at current and future
  colliders}}, \href{http://dx.doi.org/10.22323/1.350.0138}{\emph{PoS} {\bf
  LHCP2019} (2019) 138}, [\href{https://arxiv.org/abs/1908.10809}{{\tt
  1908.10809}}].

\bibitem{Robens:2022oue}
T.~Robens, \emph{{More Doublets and Singlets}},  in \emph{{56th Rencontres de
  Moriond on Electroweak Interactions and Unified Theories}}, 5, 2022.
\newblock \href{https://arxiv.org/abs/2205.06295}{{\tt 2205.06295}}.

\bibitem{Adhikari:2022yaa}
S.~Adhikari, S.~D. Lane, I.~M. Lewis and M.~Sullivan, \emph{{Complex Scalar
  Singlet Model Benchmarks for Snowmass}},  in \emph{{Snowmass 2021}}, 3, 2022.
\newblock \href{https://arxiv.org/abs/2203.07455}{{\tt 2203.07455}}.

\bibitem{Robens:2022cun}
T.~Robens, \emph{{Constraining extended scalar sectors at current and future
  colliders - an update}},  in \emph{{8th Workshop on Theory, Phenomenology and
  Experiments in Flavour Physics}: {Neutrinos, Flavor Physics and Beyond}}, 9,
  2022.
\newblock \href{https://arxiv.org/abs/2209.15544}{{\tt 2209.15544}}.

\bibitem{CMS:2022ofi}
{\scshape CMS} collaboration, \emph{{Search for Higgs boson decays to invisible
  particles produced in association with a top-quark pair or a vector boson in
  proton-proton collisions at $\sqrt{s}=13~\mathrm{TeV}$ and combination across
  Higgs production modes, CERN Report No. CMS-PAS-HIG-21-007.}}, .

\bibitem{ATLAS:2023tkt}
{\scshape ATLAS} collaboration, \emph{{Combination of searches for invisible
  decays of the Higgs boson using 139 fb$^{-1}$ of proton-proton collision data
  at $\sqrt{s} = 13$ TeV collected with the ATLAS experiment}},
  \href{http://dx.doi.org/10.1016/j.physletb.2023.137963}{\emph{Phys. Lett. B}
  {\bf 842} (2023) 137963}, [\href{https://arxiv.org/abs/2301.10731}{{\tt
  2301.10731}}].

\bibitem{PDG:2020ssz}
{\scshape Particle Data Group} collaboration, P.~A. Zyla et~al., \emph{{Review
  of Particle Physics}},
  \href{http://dx.doi.org/10.1093/ptep/ptaa104}{\emph{PTEP} {\bf 2020} (2020)
  083C01}.

\bibitem{Jaeckel:2010ni}
J.~Jaeckel and A.~Ringwald, \emph{{The Low-Energy Frontier of Particle
  Physics}},
  \href{http://dx.doi.org/10.1146/annurev.nucl.012809.104433}{\emph{Ann. Rev.
  Nucl. Part. Sci.} {\bf 60} (2010) 405--437},
  [\href{https://arxiv.org/abs/1002.0329}{{\tt 1002.0329}}].

\bibitem{PhysRevD.43.2314}
S.~Davidson, B.~Campbell and D.~Bailey, \emph{Limits on particles of small
  electric charge},
  \href{http://dx.doi.org/10.1103/PhysRevD.43.2314}{\emph{Phys. Rev. D} {\bf
  43} (Apr, 1991) 2314--2321}.

\bibitem{Brockway:1996yr}
J.~W. Brockway, E.~D. Carlson and G.~G. Raffelt, \emph{{SN1987A gamma-ray
  limits on the conversion of pseudoscalars}},
  \href{http://dx.doi.org/10.1016/0370-2693(96)00778-2}{\emph{Phys. Lett. B}
  {\bf 383} (1996) 439--443},
  [\href{https://arxiv.org/abs/astro-ph/9605197}{{\tt astro-ph/9605197}}].

\bibitem{Alwall:2014hca}
J.~Alwall, R.~Frederix, S.~Frixione, V.~Hirschi, F.~Maltoni, O.~Mattelaer
  et~al., \emph{{The automated computation of tree-level and next-to-leading
  order differential cross sections, and their matching to parton shower
  simulations}}, \href{http://dx.doi.org/10.1007/JHEP07(2014)079}{\emph{JHEP}
  {\bf 07} (2014) 079}, [\href{https://arxiv.org/abs/1405.0301}{{\tt
  1405.0301}}].

\bibitem{Alloul:2013bka}
A.~Alloul, N.~D. Christensen, C.~Degrande, C.~Duhr and B.~Fuks,
  \emph{{FeynRules 2.0 - A complete toolbox for tree-level phenomenology}},
  \href{http://dx.doi.org/10.1016/j.cpc.2014.04.012}{\emph{Comput. Phys.
  Commun.} {\bf 185} (2014) 2250--2300},
  [\href{https://arxiv.org/abs/1310.1921}{{\tt 1310.1921}}].

\bibitem{Degrande:2011ua}
C.~Degrande, C.~Duhr, B.~Fuks, D.~Grellscheid, O.~Mattelaer and T.~Reiter,
  \emph{{UFO - The Universal FeynRules Output}},
  \href{http://dx.doi.org/10.1016/j.cpc.2012.01.022}{\emph{Comput. Phys.
  Commun.} {\bf 183} (2012) 1201--1214},
  [\href{https://arxiv.org/abs/1108.2040}{{\tt 1108.2040}}].

\bibitem{Ball:2012cx}
R.~D. Ball et~al., \emph{{Parton distributions with LHC data}},
  \href{http://dx.doi.org/10.1016/j.nuclphysb.2012.10.003}{\emph{Nucl. Phys. B}
  {\bf 867} (2013) 244--289}, [\href{https://arxiv.org/abs/1207.1303}{{\tt
  1207.1303}}].

\bibitem{Bierlich:2022pfr}
C.~Bierlich et~al., \emph{{A comprehensive guide to the physics and usage of
  PYTHIA 8.3}},  \href{https://arxiv.org/abs/2203.11601}{{\tt 2203.11601}}.

\bibitem{Sjostrand:2006za}
T.~Sjostrand, S.~Mrenna and P.~Z. Skands, \emph{{PYTHIA 6.4 Physics and
  Manual}}, \href{http://dx.doi.org/10.1088/1126-6708/2006/05/026}{\emph{JHEP}
  {\bf 05} (2006) 026}, [\href{https://arxiv.org/abs/hep-ph/0603175}{{\tt
  hep-ph/0603175}}].

\bibitem{Park:2642494}
T.~H. Park, \emph{{Jet energy resolution measurement of the ATLAS detector
  using momentum balance}},  2018.

\bibitem{CMS:2016lmd}
{\scshape CMS} collaboration, V.~Khachatryan et~al., \emph{{Jet energy scale
  and resolution in the CMS experiment in pp collisions at 8 TeV}},
  \href{http://dx.doi.org/10.1088/1748-0221/12/02/P02014}{\emph{JINST} {\bf 12}
  (2017) P02014}, [\href{https://arxiv.org/abs/1607.03663}{{\tt 1607.03663}}].

\bibitem{deFavereau:2013fsa}
{\scshape DELPHES 3} collaboration, J.~de~Favereau, C.~Delaere, P.~Demin,
  A.~Giammanco, V.~Lema\^\i{}tre, A.~Mertens et~al., \emph{{DELPHES 3, A
  modular framework for fast simulation of a generic collider experiment}},
  \href{http://dx.doi.org/10.1007/JHEP02(2014)057}{\emph{JHEP} {\bf 02} (2014)
  057}, [\href{https://arxiv.org/abs/1307.6346}{{\tt 1307.6346}}].

\bibitem{Kaplan:2008ie}
D.~E. Kaplan, K.~Rehermann, M.~D. Schwartz and B.~Tweedie, \emph{{Top Tagging:
  A Method for Identifying Boosted Hadronically Decaying Top Quarks}},
  \href{http://dx.doi.org/10.1103/PhysRevLett.101.142001}{\emph{Phys. Rev.
  Lett.} {\bf 101} (2008) 142001}, [\href{https://arxiv.org/abs/0806.0848}{{\tt
  0806.0848}}].

\bibitem{Cacciari:2011ma}
M.~Cacciari, G.~P. Salam and G.~Soyez, \emph{{FastJet User Manual}},
  \href{http://dx.doi.org/10.1140/epjc/s10052-012-1896-2}{\emph{Eur. Phys. J.
  C} {\bf 72} (2012) 1896}, [\href{https://arxiv.org/abs/1111.6097}{{\tt
  1111.6097}}].

\bibitem{Bentvelsen:1998ug}
S.~Bentvelsen and I.~Meyer, \emph{{The Cambridge jet algorithm: Features and
  applications}}, \href{http://dx.doi.org/10.1007/s100520050232}{\emph{Eur.
  Phys. J. C} {\bf 4} (1998) 623--629},
  [\href{https://arxiv.org/abs/hep-ph/9803322}{{\tt hep-ph/9803322}}].

\bibitem{Matsedonskyi:2014mna}
O.~Matsedonskyi, G.~Panico and A.~Wulzer, \emph{{On the Interpretation of Top
  Partners Searches}},
  \href{http://dx.doi.org/10.1007/JHEP12(2014)097}{\emph{JHEP} {\bf 12} (2014)
  097}, [\href{https://arxiv.org/abs/1409.0100}{{\tt 1409.0100}}].

\bibitem{Kidonakis:2022hfa}
N.~Kidonakis, \emph{{Higher-order corrections for $t{\bar t}$ production at
  high energies}},  in \emph{{2022 Snowmass Summer Study}}, 3, 2022.
\newblock \href{https://arxiv.org/abs/2203.03698}{{\tt 2203.03698}}.

\bibitem{Kulesza:2018tqz}
A.~Kulesza, L.~Motyka, D.~Schwartl\"ander, T.~Stebel and V.~Theeuwes,
  \emph{{Associated production of a top quark pair with a heavy electroweak
  gauge boson at NLO$+$NNLL accuracy}},
  \href{http://dx.doi.org/10.1140/epjc/s10052-019-6746-z}{\emph{Eur. Phys. J.
  C} {\bf 79} (2019) 249}, [\href{https://arxiv.org/abs/1812.08622}{{\tt
  1812.08622}}].

\bibitem{Kidonakis:2021vob}
N.~Kidonakis and N.~Yamanaka, \emph{{Higher-order corrections for $tW$
  production at high-energy hadron colliders}},
  \href{http://dx.doi.org/10.1007/JHEP05(2021)278}{\emph{JHEP} {\bf 05} (2021)
  278}, [\href{https://arxiv.org/abs/2102.11300}{{\tt 2102.11300}}].

\bibitem{Mangano:2001xp}
M.~L. Mangano, M.~Moretti and R.~Pittau, \emph{{Multijet matrix elements and
  shower evolution in hadronic collisions: $W b \bar{b}$ + $n$ jets as a case
  study}}, \href{http://dx.doi.org/10.1016/S0550-3213(02)00249-3}{\emph{Nucl.
  Phys. B} {\bf 632} (2002) 343--362},
  [\href{https://arxiv.org/abs/hep-ph/0108069}{{\tt hep-ph/0108069}}].

\bibitem{mt2}
C.~Lester and D.~Summers, \emph{Measuring masses of semi-invisibly decaying
  particle pairs produced at hadron colliders},
  \href{http://dx.doi.org/10.1016/s0370-2693(99)00945-4}{\emph{Physics Letters
  B} {\bf 463} (Sep, 1999) 99–103}.

\bibitem{Maltoni:2012pa}
F.~Maltoni, G.~Ridolfi and M.~Ubiali, \emph{{b-initiated processes at the LHC:
  a reappraisal}}, \href{http://dx.doi.org/10.1007/JHEP04(2013)095}{\emph{JHEP}
  {\bf 07} (2012) 022}, [\href{https://arxiv.org/abs/1203.6393}{{\tt
  1203.6393}}].

\end{thebibliography}\endgroup

\end{document}